\documentclass[12pt]{elsart}
\usepackage{epsfig,graphics,color}
\newcommand{\tr}{\rm tr \,}

\newcommand{\vslash}{\FMslash v}

\def\prt{\partial}

\begin{document}\begin{frontmatter}
\title{Radiative and isospin-violating decays of \\
$D_s$-mesons in the hadrogenesis conjecture}
\author{Matthias F. M. Lutz}
\address{GSI, Planckstrasse 1,  D-64291 Darmstadt, Germany}
\author{Madeleine Soyeur}
\address{DAPNIA/SPhN, CEA/Saclay, F-91191 Gif-sur-Yvette Cedex, France}
\begin{abstract}

The masses and decays of the scalar D$_{s0}^*$(2317) and axial-vector D$_{s1}^*$(2460)
charmed strange mesons are calculated consistently in the hadrogenesis conjecture. These mesons decay either strongly into the isospin-violating $\pi^0$D$_s$ and $\pi^0$D$_s^*$ channels or electromagnetically.
They are ge\-nerated by coupled-channel dynamics based on the lea\-ding order chiral Lagrangian. The effect of
chiral corrections to chiral order Q$_\chi^2$ is investigated. We show that ta\-king into
account large-N$_c$ relations to determine the strength of these correction terms implies a
measurable signal for an exotic axial-vector state in the $\eta \,D^*$ invariant mass distribution.
The one-loop contribution to the electromagnetic decay
amplitudes of scalar and axial-vector states is calculated. The Lagrangian describing electromagnetic interactions is
obtained by gauging the chiral Lagrangian for hadronic interactions and adding gauge-invariant correction terms to chiral order Q$_\chi^2$. In addition the role of
light vector meson degrees of freedom is explored. We confront
our results with measured branching ratios. Once the light vector mesons are included, a natural explanation
of all radiative decay parameters is achieved.

\end{abstract}
\end{frontmatter}

\tableofcontents


\section{Introduction}

The observation \cite{Aubert3,Besson} of two narrow, positive parity, charmed strange mesons at masses lower than
expected in quark models \cite{Godfrey,Cahn}
may provide new insight into the way hadrons are generated. The properties of these two mesons, the scalar
D$_{s0}^*$(2317)$^\pm$ and the axial-vector D$_{s1}^*$(2460)$^{\pm}$, appear indeed sensitive to the degrees
of freedom building up hadronic excitations and to strong interaction
symmetries \cite{Swanson,Zhu}. It is therefore of interest to compute the properties of these states in effective field theories involving hadronic degrees of freedom and constrained by specific symmetries.

The importance of symmetries in constructing effective actions describing heavy-light mesons was emphasized a long time ago. As early as 1993, Nowak, Rho and Zahed
\cite{Nowak1} derived an effective action combining chiral and heavy-quark symmetries and predicted that the pseudoscalar (0$^-$) and vector (1$^-$) ground states should have chiral partners as a consequence of the spontaneous breaking of chiral symmetry. The splitting between the two sets of states was found to be rather small (of the order of the constituent quark mass). It was also noted that the large N$_c$ limit appeared compatible with the heavy-quark limit in the heavy-light meson sector. Later work by the same authors \cite{Nowak2} addressed specifically the observed D$_{s0}^*$(2317) and D$_{s1}^*$(2460) mesons. Similar considerations linking chiral and heavy-quark symmetries to parity-doubled heavy-light mesons systems were published at the same time by Bardeen and Hill
\cite{Bardeen}. These papers rely on the chiral quark model which predicts the heavy-light $0^+$ and $1^+$ states as chiral partners of the heavy-light
$0^-$ and $1^-$  antitriplet states. There is no prediction of that kind for non-linear realizations of the chiral SU(3) group.

 The discovery of the D$_{s0}^*$(2317) and D$_{s1}^*$(2460) mesons in 2003 and the difficulty of conventional approaches to reproduce their masses motivated a large number of theoretical studies involving different descriptions based on multiquark states, molecular pictures or dynamical generation of resonances (see \cite{Swanson,Zhu} for a broad set of references).

In order to get more insight into the nature of the D$_{s0}^*$(2317) and D$_{s1}^*$(2460) mesons, it was
pointed out by Mehen and Springer \cite{Mehen-Springer-2004} that the electromagnetic decays of these states could be most helpful in distinguishing among models. This property was illustrated by a specific comparison of leading order heavy-hadron chiral perturbation theory predictions for these decays to a corresponding calculation in the molecular picture.

The purpose of our paper is to study consistently the
masses, electromagnetic and strong decays of the
D$_{s0}^*$(2317) and D$_{s1}^*$(2460) mesons in the hadrogenesis conjecture where earlier
work \cite{Lutz-Kolomeitsev-2004,Kolomeitsev-Lutz-2004,Hofmann-Lutz-2004} showed that such states can
be produced at their observed masses.
This approach for heavy-light mesons exploits both heavy-quark and spontaneously broken chiral symmetries
and
ge\-nerates D$_s$-mesons through relativistic coupled-channel dynamics.
Goldstone bosons are scattered off heavy-light pseudoscalar and vector D-meson ground states.
For the
D$_{s0}^*$(2317)$^+$ meson, the calculation involves the
$\eta D_s^+$, $K^0 D^+$ and $K^+ D^0$ channels coupled further to the $\pi^0 D_s^+$ channel through an
isospin-mixing parameter. Analogously we consider for the D$_{s1}^*$(2460)$^{+}$ meson the
 $\eta D_s^{*+}$, $K^0 D^{*+}$, $K^+ D^{*0}$ and $\pi^0 D_s^{*+}$ channels. The strong, parity-violating
 decays of the D$_{s0}^*$(2317) and D$_{s1}^*$(2460) mesons give the main part of their total widths
 and reflect both the coupled-channel dynamics and the treatment of isospin-mixing effects. The electromagnetic widths are very sensitive to the details of the coupled-channel dynamics and to the vector degrees of freedom treated explicitly in the interaction Lagrangian.

A series of papers have addressed specific issues related to the contents of our work. The D$_{s0}^*$(2317) and D$_{s1}^*$(2460) mesons were dynamically generated based on the leading order heavy chiral Lagrangian by Guo et al. \cite{Guo-2006,Guo-2007}. They also computed the strong decays of the D$_{s0}^*$(2317) and D$_{s1}^*$(2460) mesons.
As this calculation is quite comparable to our first step leading order derivation, we will compare our results
for the strong widths and comment on the diffe\-rences.
Open-charm meson systems and their decays were also studied from a Lagrangian based on the SU(4) flavour symmetry
by Gamermann et al. \cite{Gamermann1,Gamermann2,Gamermann3}. The starting SU(4) degeneracy in this approach is quite different from the light-quark chiral symmetry and charmed quark heavy-quark symmetry underlying our work. The SU(4) symmetry is not an approximate symmetry in the presence of a heavy quark and needs to be largely broken by phenomenological interactions \cite{Gamermann1}.
It is therefore expected that the properties of heavy-light mesons will differ significantly in both approaches even though specific effects in radiative decays can be similar. We will illustrate this point in the discussion of the electromagnetic decay of the D$_{s0}^*$(2317) state.
 The D$_{s0}^*$(2317) and D$_{s1}^*$(2460) mesons were interpreted as hadronic molecules by Faessler et al.
\cite{Gutsche1,Gutsche2}. It is assumed that the D$_{s0}^*$(2317) is a strong bound state of K and D mesons while the D$_{s1}^*$(2460) is treated as a bound state of K* and D mesons. The strong and radiative decays were calculated using a phenomenological Lagrangian. While this phenomenological approach is quite far from our description, we find some common features in the strong decay properties. We will point them out.

We summarize now the experimental data available on the charmed strange mesons.
We display in Fig.  1 the D$_s^{\pm}$-meson spectrum as presently known \cite{PDG:2006,Aubert1}. We also
indicate the DK threshold whose closeness to the D$_{s0}^*$(2317) and D$_{s1}^*$(2460) mesons
influences the dynamics of these states.
The spin and parity of the D$_s^{\pm}$-mesons are well-established for the ground state
and for the D$_{s1}^*$(2460)$^{\pm}$. The spin and parity of the other states need confirmation.
We have quoted their most probable values.
\vskip 0.4true cm
\begin{figure}[h]
\noindent
\begin{center}
\mbox{\epsfig{file=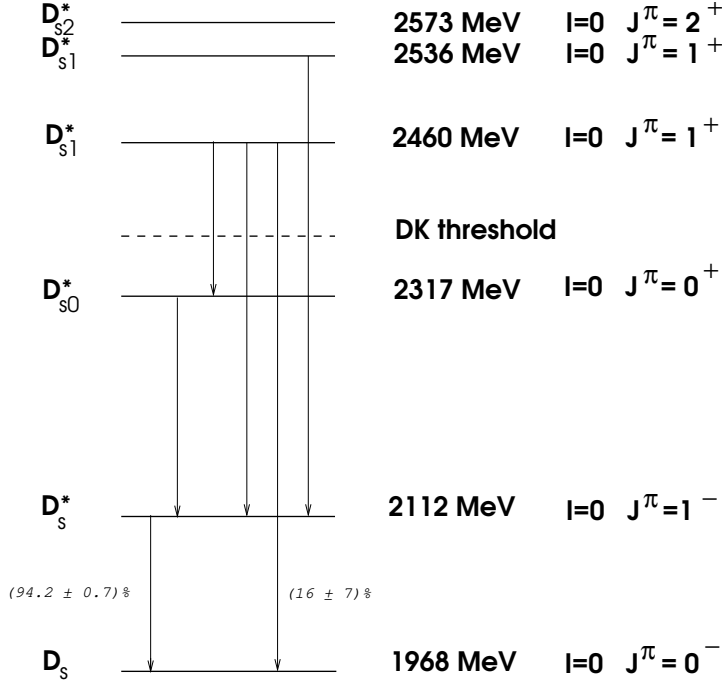, height= 9 truecm}}
\end{center}
\vskip 0.4 true cm
\caption{The D$_s^{\pm}$-meson spectrum with the most probable spin-parity assignments together
with the radiative transitions on which there is experimental information \cite{PDG:2006,Aubert1}.
The arrows without number indicate that there
are experimental constraints on these decays but no absolute branching ratios. The dashed line shows
the DK threshold.}
\label{f1_spectrum}
\end{figure}
\vglue 0.4truecm

The D$_{s}^*$(2112)$^\pm$ and D$_{s0}^*$(2317)$^\pm$ mesons
lie below the DK threshold and are therefore expected to be very narrow states.
They can decay either electromagnetically
or into the isospin-violating D$_{s}^\pm\,\pi^0$ channel.

The D$_{s}^*$(2112) has a width $\Gamma < 1.9$ MeV and
decays dominantly by a radiative transition to the scalar ground state with a probability of (94.2 $\pm$ 0.7) $\%$ \cite{PDG:2006}.
Its decay probability to the $D_s \pi^0$ channel is therefore (5.8 $\pm$ 0.7) $\%$.

The most stringent upper limit obtained for the  D$_{s0}^*$(2317) width is $\Gamma < 3.8$ MeV \cite{Aubert2}.
The  D$_{s0}^*$(2317) was first observed through its
decay into the D$_{s}(1968)\,\pi^0$ channel \cite{Aubert3}. Its radiative decay to the D$_{s}(1968)$ has never been seen.
The upper limits available
on the ratio of the radiative to pionic decay widths of the D$_{s0}^*$(2317) to the ground state
and to the D$_{s}^*$(2112) are \cite{PDG:2006}
\begin{eqnarray}
\frac{\Gamma \,[D_{s0}^*(2317) \, \rightarrow \,  D_{s}(1968)\, \gamma]}
{\Gamma\,[D_{s0}^*(2317)\, \rightarrow \, D_{s}(1968)\, \pi^0]}
<\, 0.05
\label{Dstar0widthtoDsfirst}
\end{eqnarray}
\begin{eqnarray}
\frac{\Gamma \,[D_{s0}^*(2317) \, \rightarrow \,  D_{s}^*(2112)\, \gamma]}
{\Gamma\,[D_{s0}^*(2317)\, \rightarrow \, D_{s}(1968)\, \pi^0]}
<\, 0.059.
\label{Dstar0widthtoDstarfirst}
\end{eqnarray}

The D$_{s1}^*$(2460) meson is located above the DK threshold but appears nevertheless very narrow:
its total width was found to be less than 3.5 MeV
\cite{Aubert2}. Constraints on its radiative decays to the D$_{s}$(1968), to the D$_{s}^*$(2112)
 and to the D$_{s0}^*$(2317) are as follows \cite{PDG:2006},
\begin{eqnarray}
\frac{\Gamma \,[D_{s1}^*(2460) \, \rightarrow \,  D_{s}(1968)\, \gamma]}
{\Gamma\,[D_{s1}^*(2460) \, \rightarrow \, D_{s}^*(2112)\, \pi^0]}
=0.31 \pm 0.06,
\label{Dstar1widthtoDsfirst}
\end{eqnarray}
\begin{eqnarray}
\frac{\Gamma \,[D_{s1}^*(2460) \, \rightarrow \,  D_{s}^*(2112)\, \gamma]}
{\Gamma\,[D_{s1}^*(2460) \, \rightarrow \, D_{s}^*(2112)\, \pi^0]}
<\,0.16,
\label{Dstar1widthtoDstarfirst}
\end{eqnarray}
\begin{eqnarray}
\frac{\Gamma \,[D_{s1}^*(2460) \, \rightarrow \,  D_{s0}^*(2317)\, \gamma]}
{\Gamma\,[D_{s1}^*(2460) \, \rightarrow \, D_{s}^*(2112)\, \pi^0]}
<\,0.22.
\label{Dstar1widthtoDstar0first}
\end{eqnarray}

An absolute measurement of the decay probability of the D$_{s1}^*$(2460)$^-$
to the $D_{s}(1968)^-\, \gamma$ channel gave recently (16 $\pm$ 7)$\%$ \cite{Aubert1}.
The same data also provided the branching fraction
B$(D_{s1}^*(2460)^-\,\rightarrow \, D_{s}^*(2112)^- \pi^0$) = (56 $\pm$ 22)$\%$.
The ratio is in agreement with the value quoted in (\ref{Dstar1widthtoDsfirst}).

The D$_{s1}^*$(2536) and D$_{s2}^*$(2573) widths were recently measured to be
(1.03$\pm$0.017) MeV and (27.1 $\pm$6.2) MeV respectively \cite{Aubert4,Aubert5}.
The radiative decay widths of these states are not known.

We restrict our calculations to the radiative and isospin-violating $\pi^0$ decays of the
D$_{s0}^*$(2317)$^+$ and D$_{s1}^*$(2460)$^{+}$ states on which there are fragmentary but
significant data. We note that the presence of a 2$^+$ state about 100 MeV above the D$_{s1}^*$(2460)$^{+}$ state is an indication of physics outside our scheme around that energy scale which should guide future
investigations of higher-lying states.
In view of the uncertainties in the measured widths of the D$_{s0}^*$ and D$_{s1}^*$ states, our main concern will be to check the ability
of the hadrogenesis conjecture to provide a consistent picture of the main features of the decay scheme. We aim at
identifying the important contributions to the dynamics of these decays, determining characteristic ranges for the
parameters involved and making predictions able
to test further the structure of the D$_s$-mesons and the specific conjecture on which this work relies.

This paper is organized as follows. Section 2 is devoted to the generation of
the D$_{s0}^*$(2317)$^+$ and D$_{s1}^*$(2460)$^{+}$ mesons and to the calculation of their strong isospin-violating decay widths. We present first a calculation based
on the leading order chiral Lagrangian in which massive
vector particles are described in terms of antisymmetric tensor fields.
We introduce subsequently chiral correction terms to chiral order Q$_\chi^2$ to take into account s- and u-channel D-meson
exchange processes and local two-body counter terms.
Section 3 deals with the coupling of the electromagnetic field to the hadrons. In a first step we gauge the hadronic
interactions introduced in Section 2. We add gauge-invariant interaction terms of chiral order Q$_\chi^2$.
We consider also interaction
vertices probed when including the light vector mesons as explicit degrees of freedom. We comment on the values of
the parameters associated with these terms in relation to QCD symmetries and discuss the renormalization of the
ultraviolet singularities. The explicit expressions of the electromagnetic decays are derived in Section 4 for
the scalar state
D$_{s0}^*$(2317) and in Section 5 for the axial-vector state D$_{s1}^*$(2460). Our numerical results are presented
in Section 6 and compared to the available data. We discuss the role of the different contributions and the
constraints expected on the range of values for the coupling constants of specific interaction terms.
We conclude in Section 7. We relegate lengthy derivations in seven appendices (A-G).


\section{Generation and strong isospin-violating decays of molecules}

This section deals with the generation of the scalar $D^*_{s0}(2317)$ and isovector $D^*_{s1}(2460)$ mesons
in the coupled-channel framework of  \cite{Kolomeitsev-Lutz-2004,Hofmann-Lutz-2004}.
This des\-cription is based on the scattering of Goldstone bosons
off heavy-light $0^-$ and $1^-$ mesons respectively.
Isospin-breaking effects arise from the diffe\-rence between the up and down quark masses which leads to
isospin-violating strong decay
amplitudes, $D^*_{s0}(2317) \to \pi^0\,D_s$ and $D^*_{s1}(2460) \to \pi^0\,D^*_s$. These isospin-breaking effects were not included in earlier work based on the assumption of perfect isospin symmetry \cite{Kolomeitsev-Lutz-2004,Hofmann-Lutz-2004}.
We consider scalar and axial-vector states successively. In the latter case, we reformulate the derivation of \cite{Kolomeitsev-Lutz-2004,Hofmann-Lutz-2004} in terms of massive 1$^-$ fields represented by antisymmetric tensors. This parti\-cular development is needed later to arrive at expressions which are
gauge-invariant in a transparent manner.
For both states, we consider first the hadronic interactions resulting from the leading order chiral Lagrangian and treat afterwards chiral corrections to chiral order Q$_\chi^2$.

We emphasize the importance of studying scalar and axial-vector mesons in
the open-charm sector on equal footing. The properties of spin 0 and spin 1 heavy-light mesons are indeed closely related by the heavy-quark symmetry of QCD \cite{Wise92,YCCLLY92,BD92,Jenkins94,Casalbuoni}. Even though the charm quark mass is much larger that the light (u,d,s) quark masses, the limit in which the mass of the charm quark goes to infinity is an approximation which may require significant corrections \cite{Lu,Dosch}. Rather than applying a formalism where scalar and vector fields are fully degenerate and grouped together in one field as implied by exact heavy-quark symmetry, we use separate scalar and vector D-meson fields to allow for the observed mass difference between spin multiplets.

\subsection{The scalar state $D^*_{s0}(2317)$ }

The open-charm $D^*_{s0}(2317)$ state has been shown to be dynamically
generated as a direct consequence of the leading order chiral Lagrangian density \cite{Kolomeitsev-Lutz-2004,Hofmann-Lutz-2004},
\begin{eqnarray}
{\mathcal L} &=& \frac{1}{4} \,{\tr } (\partial_\mu \Phi)\,(\partial^\mu \Phi) -\frac{1}{4}\,\tr
\chi_0\,\Phi^2+\ (\partial_\mu D) \, (\partial^\mu \bar D)- D\,M^2_{0^-}   \,\bar D
\nonumber\\
&+& \frac{1}{8\,f^2}\,\Big\{ (\partial^\mu D)\,
\,[\Phi  , (\partial_\mu \Phi)]_-\,\bar D -D\,
\,[\Phi  , (\partial_\mu \Phi)]_-\,(\partial^\mu \bar D )
 \Big\} \,,
 \label{WT-term-scalar}
\end{eqnarray}
where $\Phi$ and $D $ are the pseudoscalar octet and triplet fields. We use the notation
$\bar D = D^\dagger$. In the particle representation the
Goldstone and ground state open-charm meson fields are
\begin{eqnarray}
&&  \Phi =\left(\begin{array}{ccc}
\pi^0+\frac{1}{\sqrt{3}}\,\eta &\sqrt{2}\,\pi^+&\sqrt{2}\,K^+\\
\sqrt{2}\,\pi^-&-\pi^0+\frac{1}{\sqrt{3}}\,\eta&\sqrt{2}\,K^0\\
\sqrt{2}\,K^- &\sqrt{2}\,\bar{K}^0&-\frac{2}{\sqrt{3}}\,\eta
\end{array}\right) \,,\qquad D=(D^0, -D^+, D_s^+)  \,.
\label{def-fields-scalar}
\end{eqnarray}
The Weinberg-Tomozawa term in (\ref{WT-term-scalar}), which is proportional to $f^{-2}$, is obtained by
chirally gauging the kinetic term of the D-mesons. It is of chiral order Q$_\chi$ as it involves a single derivative of the light Goldstone fields.
The parameter $f \simeq f_\pi =92.4$ MeV in (\ref{WT-term-scalar}) is the octet meson decay constant. It
defines the scale of chiral symmetry breaking and is approximatively known from the weak decay of the
charged pions. A precise determination of $f$  requires a chiral SU(3) extrapolation
of some data set. In \cite{Lutz:Kolomeitsev:2002} the value $f \simeq 90$ MeV was obtained from a detailed
study of pion- and kaon-nucleon scattering data.  We will use $f = 90$ MeV throughout this work unless specified otherwise.
This parameter determines the leading s-wave interaction of the Goldstone bosons with the
open-charm meson fields. We note that our approach is consistent
with heavy-hadron chiral perturbation theory \cite{Mehen-Springer-2004} but at variance with the work of  \cite{Gamermann1,Gamermann2,Gamermann3} where the large breaking of the assumed SU(4) flavour symmetry leads to
different interactions.
The ground state scalar D-meson mass matrix is denoted by $M_{0^-}$. The mass term of the Goldstone bosons is proportional to the quark-mass matrix
\begin{eqnarray}
\chi_0 &=& 2\,B_0 \, \left(
\begin{array}{ccc}
m_u & 0 & 0\\
0 & m_d & 0 \\
0 & 0 & m_s
\end{array}
\right) =
 \frac{1}{3} \left(
m_\pi^2+2\,m_K^2 \right)\,1
+\frac{2}{\sqrt{3}}\,\left(m_\pi^2-m_K^2\right)
\lambda_8 \,.
\label{GMOR}
\end{eqnarray}
At leading order the latter can be expressed in terms of the pion and kaon masses as indicated in (\ref{GMOR}).

If we admit isospin-breaking effects, i.e. $m_u \neq m_d$,
there is a term in (\ref{WT-term-scalar}) proportional
to $(m_u-m_d)\,\pi^0\,\eta$, inducing $\pi^0\,\eta$ mixing. A unitary transformation is required such that the transformed fields $\tilde \pi^0$ and $\tilde \eta$ defined by
\begin{eqnarray}
&& \pi^0 = \tilde \pi^0\,\cos \epsilon - \tilde \eta \,\sin \epsilon
\,, \qquad \eta = \tilde \pi^0\,\sin \epsilon + \tilde \eta \,\cos \,
\epsilon \,,
\label{def-mixing}
\end{eqnarray}
decouple. The Lagrangian density (\ref{WT-term-scalar}), when written in terms of the new fields, does
not show a $\tilde\pi^0\,\tilde\eta$ term if and only if
\begin{eqnarray}
\frac{\sin (2\,\epsilon)}{\cos (2 \,\epsilon)} =
\sqrt{3}\,\frac{m_d-m_u}{2\,m_s-m_u-m_d} \,.
\label{eps-def}
\end{eqnarray}
According to \cite{Gasser-Leutwyler-1985} the ratio of quark masses relevant in  (\ref{eps-def}) takes the value
\begin{eqnarray}
\frac{m_d-m_u}{m_s-(m_u+m_d)/2} = \frac{1}{43.7 \pm 2.7} \,,
\end{eqnarray}
which implies the mixing angle
\begin{eqnarray}
\epsilon = 0.010 \pm 0.001 \,.
\label{value-epsilon}
\end{eqnarray}

Heavy-light meson resonances with quantum numbers $J^P\!=\!0^+$  manifest
themselves as poles in the s-wave scattering amplitude $T(s)$. We consider the
four isospin states $\langle K\,D, I |, \langle \pi \,D_s, 1 | $ and $\langle \eta \,D_s, 0 |$.
In the presence of isospin mi\-xing all channels couple. The mixing of the
two isospin sectors is of order $\epsilon$. Using the phase convention of \cite{Kolomeitsev-Lutz-2004,Hofmann-Lutz-2004}, we define the four states
\begin{eqnarray}
&&\langle 1|=\langle \,\tilde \pi^0  \,\,\!D_s^+ | =
\cos \epsilon \,\langle  \pi^0  \,\,\! D_s,
1 | + \sin \epsilon \,\langle \eta \,\,\!D_s ,0 |
\,, \qquad \nonumber\\
&& \langle 2|=\langle  \, \tilde \eta \;\,\,D_s^+ | =  \cos \epsilon\,\langle  \eta
\,\,\!D_s ,0 |
- \sin \epsilon \,\langle  \pi^0  \,\,\! D_s, 1 |\,, \nonumber\\
&& \langle 3|=\langle K^0 \,\!D^+ | = +{\textstyle{1\over \sqrt{2}}}\,
\big(\langle K\,D, 0 |- \langle K\,D, 1 | \big)\,,
\nonumber\\
&& \langle 4|=\langle K^+ \,\!D^0 | = -{\textstyle{1\over
\sqrt{2}}}\,\big(\langle K\,D, 0 |+ \langle K\,D, 1 |\big)\,.
\label{particle-isospin}
\end{eqnarray}

The Weinberg-Tomozawa interaction (\ref{WT-term-scalar})
implies a scattering amplitude of the simple form
\cite{Kolomeitsev-Lutz-2004,Hofmann-Lutz-2004}
\begin{eqnarray}
&&  T(s) = \Big[ 1- V(s)\,J(s)\Big]^{-1}\,
V(s)\,.
\label{final-t}
\end{eqnarray}
The matrix of loop functions, $J(s)$, is diagonal and given by \cite{Kolomeitsev-Lutz-2004,Hofmann-Lutz-2004}
\begin{eqnarray}
&& J(s) = I(s)-I(\mu_M^2)\,,
\nonumber\\
&& I(s)=\frac{1}{16\,\pi^2}
\left( \frac{p_{cm}}{\sqrt{s}}\,
\left( \ln \left(1-\frac{s-2\,p_{cm}\,\sqrt{s}}{m^2+M^2} \right)
-\ln \left(1-\frac{s+2\,p_{cm}\sqrt{s}}{m^2+M^2} \right)\right)
\right.
\nonumber\\
&&\qquad \qquad + \left.
\left(\frac{1}{2}\,\frac{m^2+M^2}{m^2-M^2}
-\frac{m^2-M^2}{2\,s}
\right)
\,\ln \left( \frac{m^2}{M^2}\right) +1 \right)+I(0)\;,
\label{i-def}
\end{eqnarray}
where $\sqrt{s}= \sqrt{M^2+p_{cm}^2}+ \sqrt{m^2+p_{cm}^2}$ and $p_{cm}$ is the center of mass momentum of the scattering particles. Each diagonal element depends on the masses of the
light (Goldstone) and heavy (open-charm) mesons $m$ and $M$ defining the coupled-channel state.
The matching scale $\mu_M$ in (\ref{i-def}) should be identified
with the ground state mass of the $D_s$-meson, i.e. $\mu_M \simeq 1968 $ MeV \cite{Lutz:Kolomeitsev:2002}. For such a value, s- and u-channel unitarized scattering amplitudes may be smoothly matched around the matching scale so
as to define a full scattering amplitude that is crossing symmetric by construction. The determination of the matching scale and the induced approximate crossing symmetry parallels the derivation published for kaon-nucleon scattering in \cite{Lutz:Kolomeitsev:2002}. The minimal critical point needed to open the matching window for $\pi$D$_s$ or KD scattering is typically of the order of $\sqrt {m_{\pi}^2+M_{D_s}^2}$ $\simeq$
$\sqrt {m{_K}^2+M_{D}^2}$ $\simeq$ $M_{D_s}$.
One may vary the matching scale slightly around its natural value. As shown
in \cite{Lutz-Kolomeitsev-2004} the resulting effects are small for reasonable variations. A large variation is excluded
since it would make the matching of u- and s-channel unitarized amplitudes possible only at the price of introducing
a strong discontinuity, which is at odds with causality.

The coupled-channel interaction kernel $V_{ij}(s)$ in (\ref{final-t}) is determined by
the leading order chiral SU(3) Lagrangian (\ref{WT-term-scalar}) to be \cite{Kolomeitsev-Lutz-2004},
\begin{eqnarray}
&& V^{(0^+)}_{WT,ij}(s) = \frac{C_{ij}}{8\,f^2}\,
\nonumber\\
&&\times \Big(
3\,s-M^2-\bar M^2-m^2-\bar m^2
 -\frac{M^2-m^2}{s}\,(\bar M^2-\bar m^2)\Big) \,,
\label{VWT}
\end{eqnarray}
where $(m,M)$ and $(\bar m, \bar M)$ are the masses of initial and final mesons and the indices $i$ and $j$ refer to the states defined in  (\ref{particle-isospin}).
The 4$\times$4 matrix $C_{ij}$, whose elements characterize the interaction strength in
a given channel, can be expressed in terms of
the mixing angle $\epsilon$ and the isospin zero $C^{(0)}_{ij}$ and
isospin one $C^{(1)}_{ij}$  coupling matrices of \cite{Kolomeitsev-Lutz-2004}. For the channels of positive strangeness considered in  (\ref{particle-isospin}), we have
\begin{eqnarray}
&& C_{11} = C^{(0)}_{22}\,\sin^2 \epsilon + C^{(1)}_{11}\,\cos^2 \epsilon \, \qquad
 \mkern 25 mu C_{12} = (C^{(0)}_{22}-C^{(1)}_{11})\,\sin\epsilon \,\cos \epsilon \, \nonumber\\
&& C_{22} = C^{(0)}_{22}\,\cos^2 \epsilon + C^{(1)}_{11}\,\sin^2 \epsilon \, \qquad
 \mkern 25 mu C_{13} = {\textstyle{1\over \sqrt{2}}}\,\Big(
C^{(0)}_{12}\,\sin \epsilon -\cos \epsilon \,C^{(1)}_{12} \Big)\, \nonumber\\
&& C_{23} = {\textstyle{1\over \sqrt{2}}}\,\Big(
\cos \epsilon \,C^{(0)}_{12}+\sin \epsilon\,C^{(1)}_{12} \Big)\, \qquad
C_{33} = {\textstyle{1\over 2}}\,\Big(C^{(0)}_{11}+
C^{(1)}_{22} \Big)\,\nonumber\\
&& C_{14} = {\textstyle{-1\over \sqrt{2}}}\,\Big(
C^{(0)}_{12}\,\sin \epsilon+C^{(1)}_{12}\,\cos \epsilon  \Big)\, \qquad
\mkern -2 mu C_{24} = {\textstyle{-1\over \sqrt{2}}}\,\Big(
C^{(0)}_{12}\,\cos \epsilon -C^{(1)}_{12}\,\sin \epsilon  \Big)\, \nonumber\\
&& C_{34} = {\textstyle{-1\over 2}}\,\Big(C^{(0)}_{11}-
C^{(1)}_{22} \Big)\, \qquad
 \mkern 78 mu C_{44} = {\textstyle{1\over 2}}\,\Big(C^{(0)}_{11}+
C^{(1)}_{22} \Big)\,
\nonumber\\
&&C^{(0)}_{11} =2 = 2\,C_{12}^{(1)}\, \qquad C_{12}^{(0)} = \sqrt{3}\, \qquad\mkern 6 mu
C^{(0)}_{22} =C^{(1)}_{11} = C^{(1)}_{22} =0 \,.
\label{def-44}
\end{eqnarray}

Given the coupled-channel scattering amplitude (\ref{final-t}) with the effective interaction (\ref{VWT}),
it is straightforward to determine the mass and width of possible resonances. The Weinberg-Tomozawa
interaction is strongly attractive in the isospin strangeness (I,S) = (0,1) sector,
leading to the formation of  a scalar resonance of mass $M_{0^+}$. The latter manifests itself as a pole in the
scattering amplitude which factorizes close to the pole at $s=M_{0^+}^2$,
\begin{eqnarray}
T_{ij}(s) \simeq - \frac{2\,g_i^*\,M^2_{0^+}g_j}{s-M_{0^+}^2+i\,\Gamma_{0^+}\,M_{0^+}}\,,
\label{def-gi-0plus}
\end{eqnarray}
with the coupling constants $g_i$ and the width $\Gamma_{0^+}$. We do not consider the possibility of background terms as we are dealing with a narrow resonance.

The numerical results obtained at leading order should be viewed as qualitative. At the end of this section  more quantitative results including chiral corrections will be presented.

To reproduce the empirical mass of 2317.6 MeV at leading order for the $0^+$ state requires
using an effective parameter $f_{eff}= 95.5$ MeV in  (\ref{VWT}). Taking $\epsilon=0.010$, the
coupling constants are
\begin{eqnarray}
&&g_{\,\eta \;D_s} \;\,\simeq 1.95\,, \qquad \qquad \quad  g_{\pi^0 D_s} \;\simeq 0.056,
\nonumber\\
&& g_{K^0 D^+} \simeq 2.25 \,, \qquad \qquad \;\;\;g_{K^+ D^0} \simeq -2.25 \,.
 \label{value-0plus-couplings}
\end{eqnarray}
Isospin-breaking effects in the KD coupling constants are found to
be negligible, i.e. $ g_{K^0 D^+} \simeq
-g_{K^+ D^0}  $ holds quite accurately. It is therefore meaningful to work with the isospin coupling constants
\begin{eqnarray}
g^{(0+)}_{K D} = \sqrt{2}\,g_{K^0 D^+}\simeq 3.18 \,,\qquad \qquad g^{(0+)}_{\eta \,D_s} =g_{\,\eta \,D_s}\simeq 1.95\,.
\label{isospin-coupling-0plus}
\end{eqnarray}
The
flavour SU(3) limit suggests $\sqrt{3}\,g^{(0+)}_{\eta D_s} = g^{(0+)}_{K
D}$, a result quite compatible with the values given in
(\ref{value-0plus-couplings}, \ref{isospin-coupling-0plus}).

The strong width of the D$_{s0}^*$(2317)
is related to $g_{\pi^0 D_s}$ by
\begin{eqnarray}
&&\Gamma_{D^*_{s0}(2317) \to \pi^0 D_s} = |g_{\pi^0 D_s}|^2\,\frac{p_{cm}}{4\,\pi }
\simeq |g_{\pi^0 D_s}|^2\,23.7\, {\rm MeV} \simeq 76 \,{\rm keV}
 \,,
 \label{def-total-width-0plus}
\end{eqnarray}
where $p_{cm}$ is the pion momentum in the center of mass frame. Our value of 76 keV is almost an order of magnitude
larger than the value given in \cite{Guo-2006} based on the same interaction.

To trace the origin of this difference, it is useful to understand the physics underlying $g_{\pi^0 D_s}$.
At linear order in isospin breaking, it is the sum of two terms,
\begin{eqnarray}
g_{\pi^0 D_s} = \tilde \epsilon\,g^{(0+)}_{K D}+\epsilon\,g^{(0+)}_{\eta D_s}\,.
\label{def-eff-0plus}
\end{eqnarray}
The contribution proportional to $g^{(0+)}_{\eta D_s}$ is unambiguously determined by the angle $\epsilon$, as
a direct consequence of the $\pi^0\eta $ mixing defined in (\ref{def-mixing}), and linked to the finite value of ($m_u - m_d$). If it were the only contribution to $g_{\pi^0 D_s}$, the corresponding strong width $\Gamma_{D^*_{s0}(2317) \to \pi^0 D_s}$ would be $\sim$ 9 keV, a value compatible with the 8.7 keV obtained in
\cite{Guo-2006} based on $\pi^0 \eta $ mixing.

The contribution proportional to $g^{(0+)}_{K D}$ appears at the same level and is not included in  \cite{Guo-2006}. It is induced by the
mass difference between the neutral and charged kaons
and between the neutral and charged D-mesons. Like the
$\pi^0\eta$ mixing phenomenon, it couples the two isospin sectors and is related to the finite value of ($m_u - m_d$). It is clearly not there if one assumes $m_{K^+}=m_{K^0}$ and $M_{D^+}= M_{D^0}$.
This effect in chiral coupled-channel dynamics is analogous to the mixing phenomenon recently pointed out
 in the mole\-cular picture \cite{Gutsche1,Gutsche2} where it is induced by the exchange of vector K$^*$- and D$^*$-mesons.
The parameter $\tilde \epsilon$ in  (\ref{def-eff-0plus}) must have the form
\begin{eqnarray}
\tilde \epsilon = \tilde \epsilon_1+\tilde \epsilon_2 \,, \qquad \tilde \epsilon_1  \sim
( m_{ K^+}-m_{K^0}) \,,
\qquad \tilde \epsilon_2  \sim ( M_{D^+}-M_{D^0}) \,,
\label{def-tilde-epsilon}
\end{eqnarray}
where the proportionality factors depend on the details of the coupled-channel dynamics.
At leading order we predict
$\tilde \epsilon \simeq 0.012$ with $\tilde \epsilon_1 \simeq 0.004$ and
$\tilde \epsilon_2 \simeq 0.008$. The term $\tilde \epsilon \,g^{(0+)}_{K D}$ in
$g_{\pi^0 D_s}$ is roughly twice larger than $\epsilon\,g^{(0+)}_{\eta D_s}$ and appears dominated
by the isospin-breaking induced by $M_{ D^+}\neq M_{D^0}$. These results agree qualitatively with the findings of  \cite{Gutsche1,Gutsche2}.

We mention two sources of uncertainty in the value derived above for the strong width of the $D^*_{s0}$(2317). First we use the physical masses of the neutral and charged kaons and D-meson including the electromagnetic contribution to these masses. The latter should in principle be generated by the coupling of the hadronic Lagrangian to the electromagnetic field. The  splitting of hadronic and electromagnetic interactions is a nontrivial issue \cite{GasserRS-2003,BijnensPrades} which we do not address here. The use of physical masses in the unitarization loop function (\ref{i-def}) is in
line with the scheme developed in \cite{Lutz:Kolomeitsev:2002}. Loop corrections implied by photon-exchange processes enter the effective coupled-channel interaction $V(s)$ in (\ref{final-t}) at order $Q^3_\chi$. Another
 source of uncertainty lies in the coupling constants $g_{\eta D_s}$ and $g_{K D}$. It was shown in  \cite{Hofmann-Lutz-2004} that chiral correction terms of order $Q^2_\chi $ modify these couplings and lead to the values $g_{\eta D_s} \simeq 3.7$ and
$g_{K D} \simeq 3.7$.
The width of $76$ keV quoted above is therefore most likely a lower limit as suggested by inserting the coupling
constants of \cite{Hofmann-Lutz-2004} into (\ref{def-total-width-0plus}) and (\ref{def-eff-0plus}). We will return to this issue in the final part of this section.

\subsection{The axial-vector state $D^*_{s1}(2460)$}

We turn to the generation and strong decay of the axial-vector meson $D^*_{s1}(2460)$. This calculation involves the same procedure as followed for the scalar state except that we now build the axial-vector state
by scattering Goldstone bosons off vector D-mesons.

Since we aim at predicting electromagnetic decay amplitudes, we have to construct gauge-invariant
expressions. Our Lagrangian involves
massive scalar and vector D-meson particles. We are faced with a serious complication, namely, the mixing of scalar and vector modes.
It is a quite cumbersome enterprize to arrive at gauge-invariant expressions in the presence of such mixing
phenomena \cite{Nowakowski:1993,Veltman:1994,Kaloshin:1996}. This is a known and non-trivial
complication of the standard model where the Higgs boson may mix with the longitudinal component of the $Z$ boson \cite{Atwood:1994}. A solution to this problem is to represent the $1^-$ D-mesons
in terms of antisymmetric tensor fields \cite{Jones:1962,Kyriakopoulos:1971,Kyriakopoulos:1972,Ecker:1989}.
The massive vector field is proportional to the divergence of the antisymmetric tensor.

To proceed with this particular representation we demonstrate first that the results of \cite{Kolomeitsev-Lutz-2004}, which
were obtained using the conventional vector field representation, can be recovered
with the tensor field representation.

We start with the Lagrangian density,
\begin{eqnarray}
{\mathcal L}&=&  -(\partial_\mu D^{\mu \alpha})   \,(\partial^\nu \bar D_{\nu \alpha})
+ \frac{1}{2}\,D^{\mu \alpha}\,M^2_{1^-}  \,\bar D_{\mu \alpha}
\nonumber\\
&-& \frac{1}{8\,f^2}\,\Big\{ (\partial^\nu D_{\nu \alpha })\,
\,[\Phi  , (\partial_\mu \Phi)]_-\,\bar D^{\mu \alpha }  - D_{\nu \alpha }\,
\,[\Phi  , (\partial_\nu \Phi)]_-\,(\partial_\mu \bar D^{\mu \alpha } ) \Big\} \,,
 \label{WT-term-tensor-axial}
\end{eqnarray}
involving the kinetic term and its associated Weinberg-Tomozawa interaction. The antisymmetric triplet fields $D_{\mu \nu}=-D_{\nu \mu}$ and $\bar D_{\mu \nu}= D_{\mu \nu}^\dagger$ with
$D_{\mu \nu} =(D^0_{\mu \nu}, -D^+_{\mu \nu}, D^+_{s, \mu \nu})$ describe
the heavy-quark multiplet partners of the fields $D$ introduced in (\ref{def-fields-scalar}). $M_{1^-}$ denotes their mass matrix.
Since the tensor field
representation is not frequently used in the literature, we recall the definitions of the propagator and of the associated wave function,
\begin{eqnarray}
&&\langle 0| \,T\,\bar D_{\mu \nu}(x)\,D_{\alpha \beta}(y)\,| 0 \rangle = -\frac{i}{M_{1^-}^2}\,
\int \frac{d^4 k}{(2\pi)^4} \, \frac{e^{-i\,k \cdot (x-y)}}{k^2-M_{1^-}^2+i\,\epsilon}\,
\nonumber\\
&& \qquad \times \Bigg[ (M_{1^-}^2-k^2)\,g_{\mu \alpha}\,g_{\nu \beta} + g_{\mu \alpha}\,k_\nu\,k_\beta-
g_{\mu \beta}\,k_\nu\,k_\alpha- (\mu \leftrightarrow \nu)\Bigg]\,
\label{def-propagator}
\end{eqnarray}
and
\begin{eqnarray}
&&\langle 0| D_{\mu \nu}(0)\,| D(p,\lambda) \rangle =
\epsilon_{\mu \nu}(p,\lambda)=\frac{i}{M_{1^-}}\,\Big[ p_\mu \,\epsilon_\nu(p,\lambda)-p_\nu \,\epsilon_\mu(p,\lambda) \Big]
\,, \qquad
\nonumber\\
&&  \sum_{\lambda=1}^3 \epsilon^\dagger_\mu(p,\lambda)\,\epsilon_\nu(p,\lambda) = -g_{\mu \nu}+ \frac{p_\mu\,p_\nu}{M_{1^-}^2}\,,
\label{def-wave-function}
\end{eqnarray}
where the wave function is expressed most economically in terms of the conventional wave
function $\epsilon_\mu(p, \lambda)$ of
a vector particle in the vector
representation.

To derive the on-shell scattering amplitude based on the interaction (\ref{WT-term-tensor-axial}), we have to reformulate with the tensor representation the technique developed in \cite{Lutz-Kolomeitsev-2004} using
the vector representation of spin one fields. The on-shell part of
the scattering amplitude in the vector representation takes the simple form,
\begin{eqnarray}
&& T^{\rm on-shell}_{\mu \nu}  = \sum_{J,P}\,M^{(J P)}(s)\,
{\mathcal Y}^{(J P)}_{\mu \nu}(\bar q, q;w) \,,
\label{T-on-shell}
\end{eqnarray}
where the projectors ${\mathcal Y}^{(J P)}_{\mu \nu}(\bar q, q;w)$ are constructed to carry well-defined
total angular momentum $J$ and parity $P$.
The projectors are polynomials in the initial and final 4-momenta of the
Goldstone bosons, $q_\mu$ and $\bar q_\mu$, as well as in the total 4-momentum $w_\mu$ ($w^2=s$). We are
interested only in the $J^P=1^+$ sector to generate the $D^*_{s1}(2460)$. We recall the appropriate projector
\begin{eqnarray}
{\mathcal Y}^{(1+)}_{\mu \nu} = \frac{3}{2}\,\Big(\frac{w^\mu\,w^\nu}{w^2}-g^{\mu \nu} \Big)\,.
\label{def-1plus-proj}
\end{eqnarray}
The merit of the
projectors is their property of solving the Bethe-Salpeter coupled-channel equation analytically for quasi-local interactions. The partial-wave amplitudes $M^{(J P)}(s)$ are Lorentz invariant.
They can be computed in terms of an effective interaction $V^{(J P)}(s)$ and loop functions $J^{(J P)}(s)$,
\begin{eqnarray}
&& M^{(J P)}(s) = \Big[ 1- V^{(J P)}(s)\,J^{(J P)}(s)\Big]^{-1}\,
V^{(J P)}(s)\,,
\nonumber\\
&& J^{(J P)}(s) = N^{(J P)}(s)\,\Big\{I(s)-I(\mu_M^2) \Big\}\,,
\label{M-generic}
\end{eqnarray}
where the factors $N^{(J P)}(s)$ reflect the presence of spin and angular momentum. The latter, if multiplied by
an appropriate factor $s^n$, are polynomials in $s$ and the masses of the intermediate states.
The universal integral $I(s)$ is defined in (\ref{i-def}).
The matching scale $\mu_M$ is taken to be the mass of the vector-meson ground state, $\mu_M = 2012$ MeV, following the same reasoning as given in Section 2.1. for the scalar case.

We seek a set of tensor projectors,
${\mathcal Y}^{(J P)}_{\mu \nu, \alpha \beta }(\bar q, q;w)$,
with  properties analogous to those of ${\mathcal Y}^{(J P)}_{\mu \nu }(\bar q, q;w)$.
They are defined in terms of the previous ones by
\begin{eqnarray}
&&{\mathcal Y}^{(J P)}_{\mu \nu, \alpha \beta }(\bar q, q;w) =
{\textstyle{1\over 4}}\,\bar p_\nu\,{\mathcal Y}^{(J P)}_{\mu \alpha  }(\bar q, q;w)\,p_\beta
-{\textstyle{1\over 4}}\,\bar p_\nu\,{\mathcal Y}^{(J P)}_{\mu \beta  }(\bar q, q;w)\,p_\alpha
\nonumber\\
&& \qquad \qquad \qquad \;-\,{\textstyle{1\over 4}}\,\bar p_\mu\,{\mathcal Y}^{(J P)}_{\nu \alpha  }(\bar q, q;w)\,p_\beta
+{\textstyle{1\over 4}}\,\bar p_\mu\,{\mathcal Y}^{(J P)}_{\nu \beta  }(\bar q, q;w)\,p_\alpha \,,
\label{def-projectors-tensors}
\end{eqnarray}
where $p_\mu =w_\mu-q_\mu$ and $\bar p_\mu =w_\mu- \bar q_\mu$. By construction we have
\begin{eqnarray}
\epsilon^{\dagger,\mu \nu}(\bar p)\,{\mathcal Y}^{(J P)}_{\mu \nu, \alpha \beta }(\bar q, q;w) \,
\epsilon^{\alpha \beta}(p) =
\sqrt{\bar p^2\,p^2}\,\epsilon^{\dagger,\mu }(\bar p)\,{\mathcal Y}^{(J P)}_{\mu\nu }(\bar q, q;w) \,
\epsilon^{\nu}(p)\,,
\label{matching-vector-tensor}
\end{eqnarray}
where we made use of  (\ref{def-wave-function}).
The identity (\ref{matching-vector-tensor}) provides the relation we are after for the scattering
amplitude,
\begin{eqnarray}
&& T^{\rm on-shell}_{\mu \nu, \alpha \beta}  = \sum_{J,P}\,M^{(J P)}(s)\,
{\mathcal Y}^{(J P)}_{\mu \nu, \alpha \beta}(\bar q, q;w) \,,
\label{scattering amplitude-tensor}
\end{eqnarray}
where the invariant partial-wave amplitudes are given by an equation of the form (\ref{M-generic}).
The only modification compared to the expressions of \cite{Lutz-Kolomeitsev-2004}
is a rescaling of the loop functions by a factor $M^2$. The normalization factor reads
\begin{eqnarray}
N^{(1+)}(s) = \frac{3}{2}\,M^2+ \frac{p_{cm}^2}{2} \,,\qquad \sqrt{s}= \sqrt{M^2+p_{cm}^2}+ \sqrt{m^2+p_{cm}^2}\,.
\end{eqnarray}

Using
a convention for the coupled-channel states analogous to (\ref{particle-isospin}),
the effective interaction $V^{(1+)}_{ij}(s)$ implied by (\ref{WT-term-tensor-axial})
is obtained by a straightforward application of \cite{Lutz-Kolomeitsev-2004},
\begin{eqnarray}
V^{(1+)}_{WT,ij}(s) = \frac{\bar M^2+M^2}{3\,\bar M^2\,M^2} \,V^{(0^+)}_{WT,ij}(s)
+ \frac{(\bar M^2-M^2)}{12\,f^2\,\bar M^2\,M^2}\,(\bar m^2-m^2)\,C_{ij}\,,
\label{def-Vij-1plus}
\end{eqnarray}
in terms of the matrix $V^{(0^+)}_{WT,ij}(s)$ and the $C_{ij}$ coefficients defined in (\ref{def-44}).
The parameters $M$, $\bar M$, $m$ and
$\bar m$ have the same meaning as in  (\ref{VWT}).

In the limit $\bar M =M$, we recover the expressions obtained in
\cite{Kolomeitsev-Lutz-2004}. The invariant amplitude $M^{(1+)}(s)$ is identical to that of  \cite{Kolomeitsev-Lutz-2004} within a factor $2\,M^2/3$. This implies in particular that
the predictions for the axial-vector spectrum are consistent with the expectation of the heavy-quark symmetry.

The $D^*_{s1}(2460)$ state is generated dynamically from the interaction Lagrangian (\ref{WT-term-tensor-axial}).
The scattering amplitude develops a pole at $s=M_{1^+}^2$. Close to the pole it has the form
\begin{eqnarray}
M^{(1+)}_{ij}(s) \simeq - \frac{2}{3\,\bar M\,M}\,\frac{2\,g_i^*\,M^2_{1^+}g_j}{s-M_{1^+}^2+i\,\Gamma_{1^+}\,M_{1^+}}\,,
\label{def-gi-1plus}
\end{eqnarray}
with the coupling constants $g_i$ and the width parameter $\Gamma_{1^+}$. For the same reason as in  (\ref{def-gi-0plus}),
we do not consider background terms. The normalization of the coupling constants,
$g_i$, is such that in the heavy-quark limit they are identical to those of (\ref{def-gi-0plus}). The values discussed in
the following can be compared directly to those given in \cite{Hofmann-Lutz-2004}.

To reproduce the empirical mass of 2459 MeV at leading order for the $1^+$ state requires using an effective parameter $f_{eff}=97.2$ MeV. The coupling constants are then
\begin{eqnarray}
&& g_{\,\eta\, D^*_s}\;\, \simeq 1.95\,, \qquad \qquad \quad g_{\pi^0 D^*_s} \,\simeq 0.049\,,\nonumber\\
&&  g_{K^0 D^*_+}  \simeq 2.25 \,, \qquad \qquad \;\;\;g_{K^+ D^*_0} \simeq - 2.25\,.
\label{value-1plus-couplings}
\end{eqnarray}
Isospin-breaking effects in the coupling constants are again found negligible.
For a quantitative study that considers chiral correction terms we refer to the end of this section.
The values (\ref{value-1plus-couplings}) are identical (or very similar for $g^{(1+)}_{\pi^0 D^*_s}$)
to those given
in (\ref{value-0plus-couplings}). This approximate degeneracy is expected from heavy-quark symmetry.
The strong width of the $D^*_{s1}(2460)$ is determined by $g_{\pi^0 D^*_s}$. We get 55 keV using $\epsilon=0.010$. Our result for the
width parameter is five times larger than the value obtained in \cite{Guo-2007} based on the same
interaction. The reason for such a discrepancy is again the neglect of
isospin-breaking effects in the kaon and D*-meson masses in \cite{Guo-2007}. The uncertainties in the strong width discussed for the $D^*_{s0}(2317)$ apply equally for the $D^*_{s1}(2460)$.

\subsection{Chiral correction terms}

\begin{figure}[ht]
\noindent
\begin{center}
\mbox{\epsfig{file=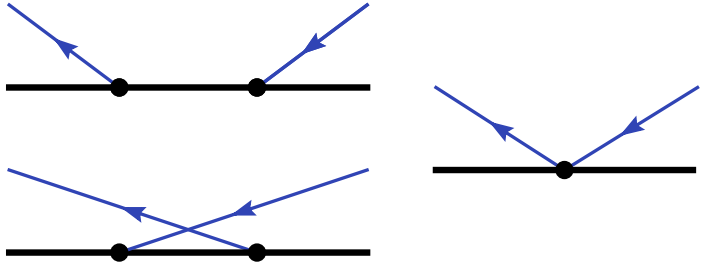, height= 3.5 truecm}}
\end{center}
\vskip 0.4 true cm
\caption{Chiral correction terms from s- and u-channel exchange processes of the D-meson (left) and
local 2-body counter terms (right). The thin and thick lines represent Goldstone and heavy-light mesons respectively.}
\label{chiral-corrections}
\end{figure}

It was shown in  \cite{Hofmann-Lutz-2004} that chiral
correction terms at subleading order (Q$_\chi^2$) provide additional mass shifts
for the  $D^*_{s0}$ and $D^*_{s1}$ states, such that their masses are consistent with the observed
values when using $f= 90$ MeV rather than the effective decay parameters $f_{eff}$ of the
previous sections.
We construct these chiral corrections to the leading order interactions
(\ref{WT-term-scalar}) and  (\ref{WT-term-tensor-axial}) by adjusting the expressions of  \cite{Hofmann-Lutz-2004} to the
tensor representation of  massive vector particles.
There will be two types of contributions for s-wave scattering. On the one hand we include
s- and u-channel exchanges of the D-meson ground states based on the leading order vertices involving a Goldstone boson and two D-mesons. On the other hand
local 2-body counter terms (breaking chiral symmetry and chiral symmetric respectively) will be constructed. These different contributions are shown
in Fig.  \ref{chiral-corrections}.
\vskip 0.4true cm
\vglue 0.4truecm
The
s- and u-channel exchanges of the D-meson ground states at chiral order Q$_\chi^2$ are generated by leading order (Q$_\chi$) 3-point vertices,

\begin{eqnarray}
{\mathcal L} &=& i\,\frac{g_P}{f}\,\Big\{D_{\mu \nu}\,(\partial^\mu
\Phi )\,(\partial^\nu \bar D)
 - (\partial^\nu D )\,(\partial^\mu
\Phi)\,\bar D_{\mu \nu} \Big\}
\nonumber\\
&+& \frac{\tilde g_P}{4\,f}\,\epsilon^{\mu \nu \alpha \beta}\,\Big\{
D_{\mu \nu}\,(\partial_\alpha \Phi)\,
(\partial^\tau \bar D_{\tau \beta} )
+ (\partial^\tau D_{\tau \beta})\,(\partial_\alpha
\Phi)\,\bar D_{\mu \nu}) \Big\} \,.
\label{def-gP}
\end{eqnarray}
As derived in Appendix A, the decay of the charged $D^*$-mesons  implies
\begin{eqnarray}
|g_P| = 0.57 \pm 0.07 \,.
\label{value-gP}
\end{eqnarray}

The parameter $\tilde g_P$ in (\ref{def-gP}) can not be extracted from empirical data directly.  In the absence of an accurate evaluation
within unquenched lattice QCD, the size of $\tilde g_P$
can be estimated using the heavy-quark symmetry of QCD. As discussed in Appendix B, one expects at leading order
\begin{eqnarray}
\tilde g_P = g_P \,.
\label{gP-hqs}
\end{eqnarray}

The coupling constant $g_P$ contributes to the effective interaction $V^{(0^+)}(s)$ and $V^{(1^+)}(s)$
via s- and u-channel exchange processes of the D-meson ground states.
The specific processes involved are linked to the choice of representation for the
massive vector particles.
In the vector representation, the propagator contains a non-propagating 0$^+$ component. In the tensor representation, it contains a non-propagating 1$^+$ component.
We derive the relevant contributions applying
the on-shell reduction scheme of \cite{Lutz-Kolomeitsev-2004}.

In the scalar sector, only the u-channel
exchange of the $J^P=1^-$ contributes as a consequence of the tensor representation.
We have
\begin{eqnarray}
&& V^{(0^+)}_{u-ch}(s) =  g_P^2\,\frac{C_{u-ch}}{4\,s\,f^2}\,
 \Big(s-\bar M^2+\bar m^2\Big)\,\Big(s-M^2+m^2\Big)
\nonumber\\
&& \qquad +\,g_P^2\, \frac{C_{u-ch}}{f^2}\,\int_{-1}^1 \frac{d x}{2}
\frac{(\bar q\cdot q) \,\mu_{1^-}^2- (\bar m^2 - \bar q\cdot p )\,
(m^2- \bar p\cdot q )}{M^2+\bar M^2 -\mu_{1^-}^2 -s +2\,\bar q\cdot q} \,,
\nonumber\\ \nonumber\\
&& \bar q \cdot q = \sqrt{\bar m^2+\bar p_{\rm cm}^2}\,\sqrt{m^2+p_{\rm cm}^2}
-\bar p_{\rm cm}\,p_{\rm cm}\,x \,,
\nonumber\\
&& \bar q\cdot p = -\bar q\cdot q  + \frac{s-\bar M^2+\bar m^2}{2}\,,
\qquad \bar p \cdot q = -\bar q\cdot q  + \frac{s- M^2+m^2}{2}\,,
\nonumber\\
&& \sqrt{s}= \sqrt{M^2+p_{cm}^2}+ \sqrt{m^2+p_{cm}^2}=
\sqrt{\bar M^2+\bar p_{cm}^2}+ \sqrt{\bar m^2+\bar p_{cm}^2} \,,
\label{u-channel-0plus}
\end{eqnarray}
where $(m,M)$ and $(\bar m, \bar M)$ are the masses of the initial and final mesons. The parameter
$\mu_{1^-} \simeq 2059$ MeV is taken to be the average of the vector D-meson masses. The coefficients
$C^{(I,S)}_{u-ch}$ are given in Table \ref{tab:coeff} for the channels
relevant for the formation and decay of the $D^*_{s0}(2317)$.
\vglue0.2truecm
\begin{table}[h]
\tabcolsep=1.0mm
\begin{center}
\begin{tabular}{||c||c||c|c|c|c|c|c|c|c|c||}
\hline
\hline
(I,S) & Channel & $C^{(I,S)}$ & $C^{(I,S)}_{\pi,0}$ & $C^{(I,S)}_{\pi,1}$ & $C^{(I,S)}_{K,0}$ & $C^{(I,S)}_{K,1}$ & $C^{(I,S)}_2$ & $C^{(I,S)}_3$ & $C^{(I,S)}_{s-ch}$ & $C^{(I,S)}_{u-ch}$ \\
\hline
\hline
(0,+1)      & 11    &2      &0  &0      &4  & 0     &2  &0     & 4& 0 \\
\hline
        & 12    &$\sqrt3$   &0  &$-\frac{\sqrt3}2$&0 & $\frac5{2\sqrt3}$&0   &$-\frac1{2\sqrt3}$ &$\frac{4}{\sqrt{3}}$& $-\frac{2}{\sqrt{3}}$\\
\hline
        & 22    &0      &$-\frac43$& $-\frac{2}{3}$  & $\frac{16}{3}$  & 0 & 2   &$\frac13$  &$\frac{4}{3}$& $\frac{4}{3}$ \\
\hline
\hline
(1,+1)      & 11    &0      &4  &$-2$     &0  &0      &2  &1     &0 & 0 \\
\hline
        & 12    &1      &0  &$-\frac12$  &0  &$-\frac12$  &0  &$\frac12$  &0& $-2$\\
\hline
        & 22    &0      &0  &0      &4  &$-2$     &2  &1    &0& 0  \\
\hline
\hline
\end{tabular}
\vglue 0.25 true cm
\caption{The coefficients $C^{(I,S)}$ that characterize the interaction of Goldstone bosons with
heavy-meson fields for
given isospin (I) and strangeness (S). The channel indices 1 and 2 refer to the isospin basis
used in  \cite{Hofmann-Lutz-2004}}.
\label{tab:coeff}
\end{center}
\end{table}
\vglue0.2truecm

As already noted in \cite{Hofmann-Lutz-2004} the influence of the u-channel process is of very minor importance
for the formation of the $D^*_{s0}(2317)$. At $f=90$ MeV and $g_P=0$ we obtain a mass of 2304 MeV
which is pulled down by 1 MeV only if we switch on $g_P=0.57$.

The situation is different for the axial-vector state
$D^*_{s1}(2460)$. Three processes contribute: the s-channel exchange of the $1^-$ state and the
u-channel exchanges of the $0^-$ and $1^-$ charmed mesons. The two u-channel contributions give
\begin{eqnarray}
&& V^{(1^+)}_{u-ch}( s) = - \frac{C_{u-ch}}{f^2}\,\int_{-1}^1 \frac{d x}{4}\,
\Big\{ A_1\,(1+x^2) + \bar p_{cm}\,p_{cm}\,x\,(1-x^2)\,A_5 \Big\}\,,
\nonumber\\
&& A_1 = \frac{\tilde g_P^2}{16\,\bar M^2\,M^2\,\mu_{1^-}^2}\,\Big\{-\Big(M^4 + (3\,\mu_{1^-}^2 -
u)\,M^2 + \mu_{1^-}^2\,u \Big)\,\bar M^4
\nonumber\\
&& \qquad +\,  \Big((u - 3\,\mu_{1^-}^2)\,M^4 + (\mu_{1^-}^2\,(2\,s - u) - u^2)\,M^2
\nonumber\\
&& \qquad \quad +\, m^2\,(M^2 + \mu_{1^-}^2)\,(M^2 - u) +
        2\,\mu_{1^-}^2\,s\,u\Big )\,\bar M^2
\nonumber\\
&& \qquad +\,\bar m^2\,(\bar M^2 - m^2 - u)\,\Big((M^2 + \mu_{1^-}^2)\,\bar M^2 + \mu_{1^-}^2\,(M^2 + u)\Big)
\nonumber\\
&& \qquad +\, \mu_{1^-}^2\,(M^2 + u)\,\Big((M^2 - u)\,m^2 + u\,(-M^2 + 2\,s + u)\Big)\Big\}
\,,
\nonumber\\
&& A_5= \frac{g^2_P}{u-\mu_{0^-}^2}
+ \tilde g_P^2\,\Bigg(1+ \frac{(M^2+\bar M^2+ u)\, \mu_{1^-}^2}{M^2\,\bar M^2 }\Bigg)\,
\frac{M^2 + \bar M^2 - s - u }{8\,\mu_{1^-}^2\,(u-\mu_{1^-}^2)}
\,,
\nonumber\\ \nonumber\\
&& \qquad \qquad u= \bar M^2+M^2 -s +2\,\bar q\cdot q\,,
\label{u-channel-1plus}
\end{eqnarray}
where the kinematics is given by (\ref{u-channel-0plus}). We use $\mu_{0^-}= 1918$ MeV as the average mass of
the $0^-$ charmed mesons. The influence of
the u-channel exchange interaction (\ref{u-channel-1plus}) on the formation of the $D^*_{s1}(2460)$ state
is somewhat more important than it is in the scalar sector.  For $f=90$ MeV and $g_P=\tilde g_P=0$, we
obtain 2441 MeV, a mass which is pushed up by 5 MeV upon incorporation of (\ref{u-channel-1plus})
with $g_P=\tilde g_P=0.57$. The effect is dominated largely by the exchange of the $1^-$ state.

We turn to the s-channel exchange.
In contrast to \cite{Hofmann-Lutz-2004} based on the
vector-field representation of the spin-one D-mesons, there is a contribution from the s-channel exchange
of the $1^-$ state within the tensor-field approach,
\begin{eqnarray}
&& V^{(1^+)}_{s-ch}( s) = \frac{2\,\tilde g_P^2}{3\,\mu^2_{1^-}}\,\frac{C_{s-ch}}{4\,s\,f^2}\,
 \Big(s-\bar M^2+\bar m^2\Big)\,\Big(s-M^2+m^2\Big) \,,
\label{s-channel-1plus}
\end{eqnarray}
where the coefficients $C^{(I,S)}_{s-ch}$ are given in Table \ref{tab:coeff}.
The combined effect of (\ref{u-channel-1plus}) and (\ref{s-channel-1plus}) yields a resonance mass of 2432 MeV
for $f =90$ MeV and $g_P=\tilde g_P=0.57$.

We turn to the local 2-body interaction terms, considering successively the chiral symmetry breaking and the chiral symmetric terms introduced in \cite{Hofmann-Lutz-2004}.

At chiral order $Q^2_\chi$, the following terms break chiral symmetry explicitly,
\begin{eqnarray}
&& {\mathcal L}_{\chi-SB} = -2\,c_1\, D \,\chi_0\,\bar D -(4\,c_0-2\,c_1 )\, (D\,\bar D)\,\tr \chi_0
\nonumber\\
&& \qquad + \frac{2\,c_0-c_1}{f^2}\,D\,\bar D\,{\tr } \Big(\Phi \,\chi_0\, \Phi\Big)
 + \frac{c_1}{4\,f^2}\,D\, \Big\{\Phi, \Big\{ \Phi\,,\chi_0 \Big\} \Big\}\, \bar D
\nonumber\\
&& \qquad + \,\tilde c_1\, D_{\alpha \beta} \,\chi_0\,\bar D^{\alpha \beta} +(2\,\tilde c_0-\tilde c_1 )\,
(D_{\alpha \beta}\,\bar D^{\alpha \beta})\,\tr \chi_0
\nonumber\\
&& \qquad - \frac{2\,\tilde c_0-\tilde c_1}{2\,f^2}\,D_{\alpha \beta}\,\bar D^{\alpha \beta}\,{\tr } \Big(\Phi \,\chi_0\, \Phi\Big)
 - \frac{\tilde c_1}{8\,f^2}\,D_{\alpha \beta}\, \Big\{\Phi, \Big\{ \Phi\,,\chi_0 \Big\} \Big\}\, \bar D^{\alpha \beta}
\,,
\label{explicit-csb}
\end{eqnarray}
where the matrix $\chi_0$ is defined in (\ref{GMOR}). The parameters $c_1$ and $\tilde c_1$
are determined by the empirical mass differences of the
$J^P=0^-$ and $J^P=1^-$ charmed mesons. According to \cite{Hofmann-Lutz-2004} we have
\begin{eqnarray}
c_1 \simeq 0.44 \,, \qquad \qquad \tilde c_1 \simeq 0.47\,.
\label{value-c1}
\end{eqnarray}
The parameters $c_0$ and $\tilde c_0$ could in principle be determined by unquenched lattice QCD simulation
upon studying the pion-  and kaon-mass dependence of the D-meson ground states. So far they are unknown.
We construct the effective interaction

\begin{eqnarray}
&& V^{(0^+)}_{\chi-SB}(s) =  2\,\frac{m_\pi^2}{f^2}\,\Big( c_0\,C^{}_{\pi,0}+c_1\,C^{(I,S)}_{\pi,1}\Big)
+ 2\,\frac{m_K^2}{f^2}\,\Big( c_0\,C^{}_{K,0}+c_1\,C^{}_{K,1}\Big)
 \,,
\nonumber\\
&& V^{(1^+)}_{\chi-SB}(s) =  \frac{2}{3\,\bar M\,M}\, V^{(0^+)}_{\chi-SB}(s)\Big|_{c_i \to \tilde c_i} + \cdots
 \,,
\label{chSB}
\end{eqnarray}
where the dots represent higher order terms that we neglect.
The coefficients $C^{(I,S)}_{\pi,0}, C^{(I,S)}_{\pi,1}, C^{(I,S)}_{K,0}$ and $C^{(I,S)}_{K,1}$
are given in Table \ref{tab:coeff}.

The chiral symmetric terms of order-$Q^2_\chi$ contributing to s-wave scattering are
\begin{eqnarray}
&& {\mathcal L}_{CT} =  \frac{2\,c_2+c_3}{f^2}\,D\,\bar D\,{\tr } \Big((\partial_\mu \Phi)\,(\partial^\mu \Phi)\Big)
-\frac{c_3}{f^2}\,D \,(\partial_\mu \Phi)\,(\partial^\mu \Phi)\,\bar D
\nonumber\\
&& \quad - \frac{2\,\tilde c_2+\tilde c_3}{2\,f^2}\,D_{\alpha \beta}\,\bar D^{\alpha \beta}\,{\tr } \Big((\partial_\mu \Phi)\,(\partial^\mu \Phi)\Big)
 +\frac{\tilde c_3}{2\,f^2}\,D_{\alpha \beta} \,(\partial_\mu \Phi)\,(\partial^\mu \Phi)\,\bar D^{\alpha \beta}
\,.
 \label{def-cs}
\end{eqnarray}
They imply the effective s-wave interactions
\begin{eqnarray}
&& V^{(0^+)}_{CT}(s) =
\Big( c_2\,\frac{C^{}_2}{s\,f^2} +c_3\,\frac{C^{}_3}{s\,f^2} \Big)
\,\Big(s-\bar M^2+\bar m^2\Big)\,\Big(s-M^2+m^2\Big) \,,
\nonumber\\
&&V^{(1^+)}_{CT}(s) =\frac{2}{3\,\bar M\,M}\,V^{(0^+)}_{CT}(s) + \cdots \,,
\label{CT}
\end{eqnarray}
where the dots represent again higher order terms.
The coefficients $C^{(I,S)}_{2}$ and $C^{(I,S)}_{3}$
are given in Table \ref{tab:coeff}. The full effective interactions $V^{(0^+)}(s)$ and
$V^{(1^+)}(s)$ are then defined as the sums
\begin{eqnarray}
V (s) = V_{WT}(s)+ V_{u-ch}(s)+V_{s-ch}(s) + V_{\chi-SB}(s)+ V_{CT}(s).
\label{eff-interaction-complete}
\end{eqnarray}
The number of unknown parameters, $c_{0,2,3}$ and $\tilde c_{0,2,3}$ appears large at first. A free fit
to the masses of the $D^*_{s0}(2317)$ and $D^*_{s1}(2460)$  states only would not be significant.
Additional constraints from QCD should be used. According to
\cite{Hofmann-Lutz-2004}, the parameters $c_i$ and $\tilde c_i$ are degenerate
in the heavy-quark mass limit, i.e. we expect
\begin{eqnarray}
c_i \simeq  \tilde c_i \,.
\label{hqs-cs}
\end{eqnarray}
It is reassuring that the values for $c_1$ and $\tilde c_1$ given in (\ref{value-c1}) are
quite compatible with the expectation of the heavy-quark symmetry relations (\ref{hqs-cs}).
We consider further constraints from QCD as they arise in the limit of a large number of colors $N_c$ \cite{large-N_c-reference}.
Since at leading order in a $1/N_c$ expansion single-flavour trace interactions are dominant, the corresponding couplings should go to zero in the
$N_c\rightarrow \infty $ limit, suggesting
\begin{eqnarray}
c_0 \simeq \frac{c_1}{2} \,, \qquad \qquad c_2 \simeq -\frac{c_3}{2}\,, \qquad \qquad
\tilde c_0 \simeq \frac{\tilde c_1}{2} \,, \qquad \qquad \tilde c_2 \simeq -\frac{\tilde c_3}{2}\,.
\label{large-Nc}
\end{eqnarray}
In the combined heavy-quark and large-$N_c$ limit (\ref{hqs-cs}, \ref{large-Nc}),
we are left with one free parameter only,
$c_3=\tilde c_3$. The optimal value $c_3=\tilde c_3 =1.2 $ (together with the parameters
$c_1$ and $\tilde c_1$ given by (\ref{value-c1}), $g_P=\tilde g_P=0.57$ and $f=90$ MeV)  predicts
2330 MeV and 2448 MeV for the masses of the  $D^*_{s0}(2317)$ and $D^*_{s1}(2460)$ states respectively.
\begin{table}
\begin{center}
\begin{tabular}{|c||c|c|c||c|c|c|}
\hline
J$^P_{(I,S)}$ &$0^+_{(0,1)}$&$0^+_{(0,1)}$ &$0^+_{(\frac12,0)}$& $1^+_{(0,1)}$&$1^+_{(0,1)}$&$1^+_{(\frac12,0)}$\\
\hline
\hline
M$_R$ [MeV]  & 2317.6   & 2317.6 & 2410.5     & 2459.2  & 2459.2  & 2568   \\
\hline
$\Gamma$ [MeV]& 0.14    & 0.25   & 2.18       & 0.14    & 0.25    & 18     \\
\hline
$|g_1|$       & 3.27    & 3.27   & 0.24       & 2.97    & 2.97    &  $<$0.05    \\
\hline
$|g_2|$       & 2.50    & 2.50   & 1.37       & 2.42    & 2.42    & 1.4        \\
\hline
$|g_3|$       &-        & -      & 2.12       &-        & -       & 2.5        \\
\hline \hline
 $\epsilon$   &  0.01   & 0.02   &  0         &  0.01   & 0.02    & 0   \\
\hline
\end{tabular}
\vglue 0.3 truecm
\caption{Masses, widths and coupling constants for dynamically generated $0^+$ and $1^+$ states. The coupling constants
$g_i$ corresponds to specific decay channels labeled in the isospin basis used in  \cite{Hofmann-Lutz-2004}.
We use $f=90$ MeV,
$g_P=\tilde g_P =0.57 $, $c_3=1.0$, $\tilde c_3 =1.4 $. The remaining parameters are
implied by the relations (\ref{value-c1}, \ref{hqs-cs}, \ref{large-Nc}).}
\vglue 0.4 truecm
\label{tab:hadronic-decay}
\end{center}
\end{table}

For the determination of the
width parameters, it is important to reproduce the masses accurately. We allow therefore for small variations of the
parameters around the heavy-quark scenario, leaving the large-$N_c$ relations (\ref{large-Nc}) untouched.
A precise reproduction of the scalar and axial-vector state masses is  achieved with $c_3 =1.0$ and $\tilde c_3 = 1.4$.
Detailed results are collected in Table \ref{tab:hadronic-decay}. For the mixing angle $\epsilon=0.01$,
the strong decay widths of the $D^*_{s0}(2317)$ and $D^*_{s1}(2460)$ are both 0.14 MeV. These widths are
significantly larger than
the leading order estimates discussed above. This effect is mainly a consequence of somewhat larger coupling
constants $g_{\eta D_s}^{(0^+)} = 2.50 $ and $g_{\eta D^*_s}^{(1^+)}=2.42$
given in Table  \ref{tab:hadronic-decay} (where they are denoted by $g_2$).
The sensitivity to the mixing angle $\epsilon$ is illustrated by the 3rd and 6th row. Taking $\epsilon = 0.02$ instead of $\epsilon = 0.01$ leads to a strong width of 0.25 MeV rather than 0.14 MeV.

We briefly comment on the previous results of \cite{Hofmann-Lutz-2004}. In that work a different scenario
was investigated. Using the conventional vector-field representation of the $1^-$  charmed mesons, it was assumed that
the axial-vector resonance $D^*_1(2420)$ was a member of the exotic sextet, predicted at leading order
by chiral coupled-channel dynamics \cite{Kolomeitsev-Lutz-2004,Hofmann-Lutz-2004}. The crucial question is whether chiral correction terms reduce the weak attraction predicted at leading order or possibly enhance it. This is clearly sensitive to the implementation of chiral symmetry. For example, in  \cite{Gamermann1,Gamermann2} based on the broken SU(4) flavour symmetry, the
sextet states are weakly bound only and therefore quite broad.
In \cite{Hofmann-Lutz-2004} chiral correction terms were tuned in such a way as to
pull down the exotic axial state with (I,S) = $(\frac{1}{2},0)$ to match the properties of the $D^*_1(2420)$. The invariant $\pi D$ and $\pi D^*$ mass distributions as measured by the BELLE collaboration \cite{BELLE} were used as an additional constraint. It was argued that the scalar heavy-quark partner of the exotic axial state decouples from the $\pi D$ channel and therefore is not seen in the data. Based on the large-$N_c$ relations, that were not
considered in \cite{Hofmann-Lutz-2004}, we would deem this scenario unlikely.
In order to compare to the large-$N_c$
scenario advocated in  the present work we included in  Table \ref{tab:hadronic-decay} the rows 4 and 7, which give the
characteristics of the exotic $(\frac{1}{2},0)$ states. Like in \cite{Hofmann-Lutz-2004} the scalar state is
quite narrow with a mass below the $\eta D$ threshold. We obtain a mass and a width of 2411 MeV and 2.2 MeV respectively.
As seen from the coupling constants in Table \ref{tab:hadronic-decay}, the state couples most strongly to the
$\eta D$ and $K \,D_s$ channels.

\begin{figure}[t]
\begin{center}
\includegraphics[width=14cm,clip=true]{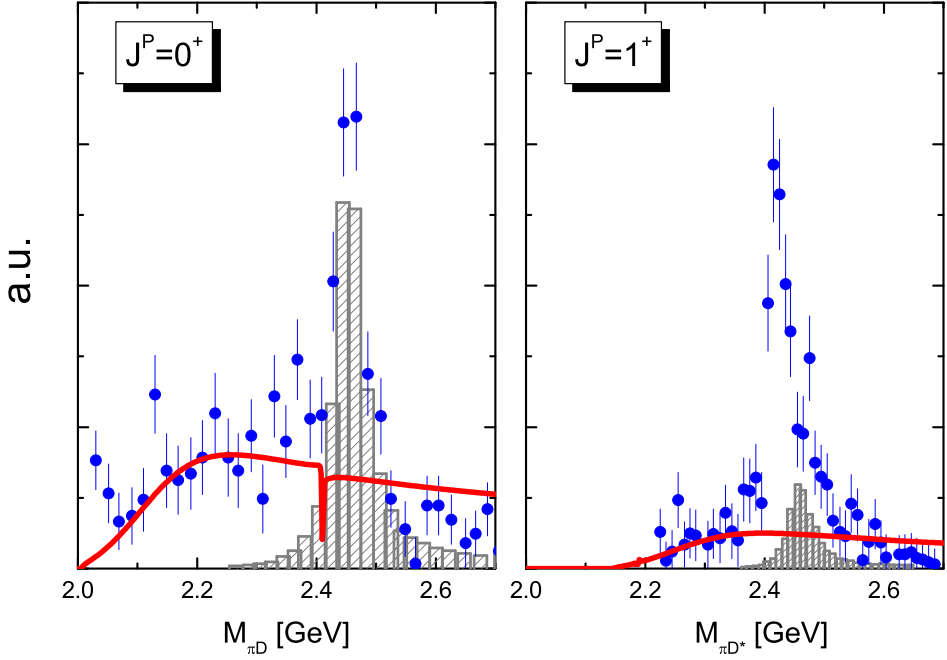}
\end{center}
\caption{Mass spectra of the $(\frac{1}{2},0)$-resonances as seen in the
$ \pi \,D(1867)$-channel (l.h. panel $J^P=0^+$) and $ \pi \,D^*(2008)$-channel (r.h. panel $J^P=1^+$).
The solid lines show the theoretical mass distributions. The data are obtained from the $B \to \pi \,D(1867)$  and $B \to  \pi \,D^*(2008)$
decays \cite{BELLE}. The histograms indicate the contribution from the
$J=2$ resonances $D_2^*(2460)$ as given in \cite{BELLE}. }
\label{fig:Dpi-distribution}
\end{figure}

In Fig.  \ref{fig:Dpi-distribution} we confront the imaginary part of the
$\pi D \to \pi D$ amplitude with the invariant $\pi D$ mass distribution measured by the BELLE collaboration \cite{BELLE}. Such a comparison is approximate since it does not resolve the structure of the initial state. The empirical distribution is dominated by the broad $(\frac{1}{2},0)$ state, a member of an antitriplet like the $D^*_{s0}$(2317), and the tensor state $D_2^*(2460)$. The contribution of this state is
illustrated by the histograms. The possible presence of a narrow $(\frac{1}{2},0)$
state is not excluded by the present data. It is interesting to observe that the exotic state leads to a dip
in the mass distribution rather than a peak. This is  a consequence of the nearby $\eta D$ channel that couples strongly
to that state. With the exception of a strong cusp effect at the $\bar K D$ threshold
in the $(0,-1)$ sector, there is no further strong signal of any sextet state in this scenario.

A striking prediction of the large-$N_c$ scenario is a clear measurable signal of an exotic axial state
in the $\eta D^*$ invariant mass distribution with a mass of
2568 MeV. It lies above the $\eta D^*$ threshold and has a width of about 18 MeV. In Fig.  \ref{fig:Dpi-distribution}
we confront the imaginary part of the $\pi D^* \to \pi D^*$ amplitude with the invariant $\pi D^*$ mass
distribution measured by the BELLE collaboration \cite{BELLE}. The empirical distributions show two axial and
one tensor states. The contribution of the tensor state is again illustrated by the histograms. In conventional approaches
the tensor state $D_2^*(2460)$ is grouped together with the $D^*_1(2420)$ state to form a heavy-quark multiplet. Within the
hadrogenesis conjecture we would expect to generate that multiplet dynamically via coupled-channel effects once the light
vector mesons are considered as additional and explicit degrees of freedom. The theoretical amplitude of the present work describes
only the broad state, which has a width of about 300 MeV. In contrast to the $\pi D$ distribution
shown in the left panel of Fig.  \ref{fig:Dpi-distribution}, we do not predict any significant signal in the
$\pi D^*$ distribution that one may use to discover the exotic axial state. This reflects a coupling constant of that
state to the $\pi D^*$ channel that is almost compatible with zero.
Nonetheless, the exotic axial state could be discovered by ongoing experiments once the
invariant $\eta D^*$ mass distribution is analyzed. The discovery of
the scalar state would require a measurement of the $\pi D$  invariant mass with an
energy resolution of a few MeV as may be possible with the PANDA experiment at FAIR.

We close this section with the remark that very similar results are obtained with the vector-field representation
once the parameters are chosen in accordance with the expectation of large-$N_c$ QCD rather than with the fitting procedure recalled above \cite{Hofmann-Lutz-2004,private-communication-Hofmann}.


\section{Electromagnetic interactions}

To compute the radiative decay of the scalar and axial-vector states gene\-rated in
the last section we need to couple the photon to the hadronic fields, ensuring consistency with the U(1) gauge symmetry of
electromagnetic interactions. We consider first the electromagnetic interactions obtained
by gauging the hadronic interactions of Section 2. We add
terms involving Goldstone and D-meson fields which are separately gauge-invariant and of chiral order $Q_\chi^2$. Such terms are not sufficient to reproduce the available data satisfactorily.
We introduce an additional 4-point vertex of order $Q_\chi^3$ to evaluate the need to include higher chiral orders.
With such a term, the data on the radiative decays of the $D^*_{s0}(2317)$ and $D^*_{s1}(2460)$
mesons can be understood at the expense of an unnaturally large value for the coupling constant
of the term of order $Q_\chi^3$. We expect this result to
 reflect the effectiveness of the underlying theory and more precisely the absence of light vector mesons in the dyna\-mics of radiative transitions. We introduce them as additional explicit degrees of freedom and will indeed find that this leads to natural values of all the para\-meters, justifying a posteriori this extension of the theory.
 Light vector mesons are not included in the coupled-channel part of this work but should become part of that scheme in a later stage to study
the formation of tensor states, in particular the low-lying 2$^+$ state expected at 2573 MeV.

We gauge the hadronic Lagrangians (\ref{WT-term-scalar}, \ref{WT-term-tensor-axial}, \ref{def-gP}) using the covariant derivatives
\begin{eqnarray}
\partial_\mu \Phi + i\,e \,[Q, \Phi]\,A_\mu \,, \qquad \qquad \partial_\mu D +i\,e \,D \,Q'\,A_\mu \,,
\label{gauge-covariant}
\end{eqnarray}
where $e=|e|$ and the charge matrices are defined by
\begin{eqnarray}
 Q = \left( \begin{array}{ccc} \frac{2}{3} & 0 & 0 \\
0 & -\frac{1}{3} & 0 \\
 0 & 0 & - \frac{1}{3}
\end{array}
\right) \,,\qquad \quad
 Q' = \left( \begin{array}{ccc} 0 & 0 & 0 \\
0 & 1 & 0 \\
 0 & 0 & 1
\end{array}
\right)\,.
\label{charge-matrices}
\end{eqnarray}

The radiative decays of the D$_{s0}^*$(2317) and D$_{s1}^*$(2460) mesons will involve only 3-point and 4-point vertices. Gauging 4-point vertices, such as the Weinberg-Tomozawa
terms in (\ref{WT-term-scalar}) and (\ref{WT-term-tensor-axial}) or the chiral correction terms (\ref{def-cs}),
would lead to 5-point vertices of no relevance.
The kinetic terms in (\ref{WT-term-scalar}) and (\ref{WT-term-tensor-axial}) imply the 3-point
couplings
\begin{eqnarray}
&&{\mathcal L}_{\rm e.m.} = i\,\frac{e}{2}\,A^\mu\, {\tr} \,
\Big((\partial_\mu \Phi)\,[Q,\,\Phi]_-\Big) +\,
i\,e\,A^\mu\, \Big( D \,Q' (\partial_\mu\,\bar D ) -
(\partial_\mu\, D)\,Q'\,\bar D\Big)
\nonumber\\
&& \qquad
-\,i\,e\,A_\mu\,D_{\mu \nu} \,Q'\,
(\partial^\beta  \bar D_{\beta \nu } )
+i\,e \,A_\mu\,  ( \partial_\alpha D^{\alpha \nu})\,Q'\,\bar D_{\mu \nu } \,.
\label{em-kinetic}
\end{eqnarray}
Given the interactions
(\ref{WT-term-scalar}, \ref{WT-term-tensor-axial}, \ref{em-kinetic}) only, the electromagnetic
decay amplitudes $0^+ \to \gamma \, 1^-$ and $1^+ \to \gamma \,0^-$ or $\gamma\,1^-$ are
zero identically as already noted in \cite{Hai-Yang-Cheng-1993}.
These decay processes probe on the other hand the 3-point hadronic vertices introduced in (\ref{def-gP}).
The importance of these terms in determining decays contrasts with the minor
role they play for the formation of the $D^*_{s0}(2317)$ and $D^*_{s1}(2460)$ states.
The gauging of the interaction (\ref{def-gP}) yields the terms
\begin{eqnarray}
{\mathcal L}_{e.m.} &=& -\,\frac{e\,g_P}{f}\,\Big\{D_{\mu \nu}\,[Q, \Phi]\,A_\mu \,(\partial^\nu \bar D)
 - (\partial^\nu D )\,[Q, \Phi]\,A^\mu \,\bar D_{\mu \nu} \Big\}
\nonumber\\
&+&\frac{e\,g_P}{f}\,\Big\{D_{\mu \nu}\,(\partial^\mu
\Phi )\,A^\nu \,Q'\,\bar D
 + A^\nu D\,Q' \,(\partial^\mu
\Phi)\,\bar D_{\mu \nu} \Big\}
\nonumber\\
&+& i\,\frac{e\,\tilde g_P}{4\,f}\,\epsilon^{\mu \nu \alpha \beta}\,\Big\{
D_{\mu \nu}\,[Q, \Phi]\,A_\alpha \,
(\partial^\tau \bar D_{\tau \beta} )
+ (\partial^\tau D_{\tau \beta})\,[Q, \Phi]\,A_\alpha\,\bar D_{\mu \nu} \Big\}
\nonumber\\
&-&i\, \frac{\tilde g_P}{4\,f}\,\epsilon^{\mu \nu \alpha \beta}\,\Big\{
D_{\mu \nu}\,(\partial_\alpha \Phi)\,
A^\tau \,Q'\bar D_{\tau \beta}
- A^\tau\, D_{\tau \beta}\,Q'\,(\partial_\alpha
\Phi)\,\bar D_{\mu \nu} \Big\}
\,.
\label{gP-gauging}
\end{eqnarray}
These terms are part of the gauge-invariant vertex associated with a leading chiral
power $Q_\chi$, hence showing that the standard counting scheme requires the photon field $A_\mu$
to be of order $Q_\chi$.

We construct the chiral correction terms of order $Q^2_\chi$. They are gauge-invariant separately as they involve the electromagnetic field strength tensor $F_{\mu \nu}$ (of chiral order $Q^2_\chi$). Terms of a similar structure were discussed previously, for example in  \cite{Hai-Yang-Cheng-1993}, using a Lagrangian where massive spin 1 fields were described in the vector representation. Clearly the Lorentz structure of these terms will be different in our case.

Chiral correction terms of order $Q^2_\chi$
describe anomalous processes. A photon hitting a pseudoscalar charmed meson may
convert the latter into its heavy-quark partner, a charmed vector meson.
In the absence of such processes the
decay amplitudes $D^*\to \gamma \,D$ and $D^*_s\to \gamma \,D_s$ would vanish identically, contrary to experiment. The leading order anomalous vertex should be of the form
\begin{eqnarray}
&& {\mathcal L}_{\rm e.m.} =
\frac{1}{2\,M^2_V} \,F^{\mu \nu} \,\epsilon_{\mu \nu \alpha \beta}\,\Big\{
(\partial_\tau D^{\tau \beta} )\,\Big(e_C+e_Q\,Q\Big)\,(\partial^\alpha  \bar D )
\nonumber\\
&& \qquad \qquad \qquad \qquad \quad \;\; +\,(\partial^\alpha   D )\Big(e_C+e_Q\,Q\Big)\,(\partial_\tau \bar D^{\tau \beta} ) \Big\}
\,.
\label{def-eC-eQ}
\end{eqnarray}
The vertices (\ref{def-eC-eQ}) carry
the leading chiral power $Q^2_\chi$ \cite{Hai-Yang-Cheng-1993,Cho-Georgi-1992,Stewart-1998}.
The radiative decay properties of the charmed vector mesons suggest (see Appendix A)
\begin{eqnarray}
e_Q  =0.91 \pm 0.10\,, \qquad \quad e_C =0.13 \pm 0.05   \,.
\label{value-eQ-eC}
\end{eqnarray}
The values of $e_Q$ and $e_C$ given in (\ref{value-eQ-eC})
reproduce the empirical branching ratios of the  $D^{*-}\to D^-\,\gamma $ and $D^{*0}\to D^0\,\gamma $ decays.
The data on the $D^{*}_s\to D_s\,\gamma $ and
$D^{*}_s\to D_s\,\pi^0 $ suggest the smaller value $e_Q  \simeq 0.52$ (see Appendix A).
The anomalous interaction (\ref{def-eC-eQ}) by itself
is at odds with the heavy-quark symmetry of QCD which relates the interactions of pseudoscalar and vector $D$-mesons.
Additional terms are required for consistency to
parameterize the magnetic moments of the charmed vector mesons. They read
\begin{eqnarray}
&& {\mathcal L}_{\rm e.m.} =\frac{i}{M_V^2} \,F^{\mu \nu} \,(\partial^\alpha D_{\alpha \mu} )\,\Big( e\,Q' -\tilde e_C+\tilde e_Q\,Q\Big)\,
(\partial^\beta  \bar D_{\beta \nu } )\,.
\label{def-tilde-eC-eQ}
\end{eqnarray}
The parameters $\tilde e_C$ and $\tilde e_Q$ are related to the magnetic moments of the charmed vector mesons
by \cite{Jones:1962,Kyriakopoulos:1971,Kyriakopoulos:1972}
\begin{eqnarray}
&& \mu_{D^*_0} =   \frac{M_{D^*_0}}{2\,M_V^2}\,\Big(\tilde e_C- \frac{2}{3}\,\tilde e_Q\Big)\,, \qquad \quad
 \mu_{D^*_+} =    \frac{M_{D^*_+}}{2\,M_V^2}\,\Big(\tilde e_C+ \frac{1}{3}\, \tilde e_Q\Big) \,,
\nonumber\\
&& \mu_{D^*_s} =  \frac{M_{D^*_s}}{2\,M_V^2}\,\Big(\tilde e_C+ \frac{1}{3}\, \tilde e_Q\Big) \,.
\label{results-magnetic-moments}
\end{eqnarray}
At present there is no empirical information on these
magnetic moments of help to determine the values of $\tilde e_C$ and $\tilde e_Q$.
Note that the term proportional to $Q'$ in (\ref{def-tilde-eC-eQ}) cancels a corresponding contribution which
is implied by minimally gauging the kinetic term (\ref{em-kinetic}).

It is instructive to interpret the result (\ref{results-magnetic-moments})
in terms of the constituent quark model. The contribution from $\tilde e_C$
reflects the magnetic moment of the charm quark. It is
SU(3) flavour-blind. In the heavy-quark mass limit, the parameter
$\tilde e_C$ approaches a constant. In contrast, the term proportional to $\tilde e_Q$ models the contribution of the
magnetic moment of the light quark. It is flavour-dependent, being proportional to the charge
matrix of the light quarks $Q$. In the heavy-quark mass limit,
 $\tilde e_Q$ scales linearly with the charm quark mass. To leading order, the parameters
 $\tilde e_C$ and $\tilde e_Q$ are related to $e_C$ and $e_Q$ by (see Appendix B)
\begin{eqnarray}
e_Q = \tilde e_Q \,, \qquad \quad e_C = \tilde e_C \,.
\label{result-eC-eQ}
\end{eqnarray}

We emphasize that the two sets of terms (\ref{def-eC-eQ}) and (\ref{def-tilde-eC-eQ})
are the only chiral corrections to the leading order Lagrangian at order $Q^2_\chi$. It does not appear possible to fit the data
on the $D^*_{s0}(2317)$ and $D^*_{s1}(2460)$ radiative decays with these electromagnetic terms together with the Lagrangians (\ref{em-kinetic}) and (\ref{gP-gauging})
 (see Section 6 for details). This suggests that higher order terms should be consi\-dered. We introduce
an additional term involving one Goldstone boson field and proportional to $F_{\mu \nu}$
and to the charge matrix $Q$ of the light quarks (expected to be dominant in the heavy-quark mass limit
\cite{Hai-Yang-Cheng-1993}),
\begin{eqnarray}
{\mathcal L}_{\rm e.m.} &=& \frac{e_P}{f\,m_V^2}\,F^{\mu \nu}\,\Big\{D_{\mu \alpha}\,[(\partial_\nu
\Phi ),Q]\,(\partial^\alpha \bar D)
 - (\partial^\alpha D )\,[(\partial_\nu
\Phi),Q]\,\bar D_{\mu \alpha} \Big\}
\nonumber\\
&-& i\,\frac{\tilde e_P}{4\,f\,m_V^2}\,F_{\mu \nu}\,\epsilon^{\sigma \tau \mu \beta}\,\Big\{
D_{\sigma \tau}\,[(\partial^\nu
\Phi ),Q]\,
(\partial^\alpha \bar D_{\alpha \beta} )
\nonumber\\
& & \qquad \qquad \qquad \quad +\, (\partial^\alpha D_{\alpha \beta})\,[(\partial^\nu
\Phi ),Q]\,\,\bar D_{\sigma \tau}) \Big\} \,.
\label{def-eP}
\end{eqnarray}
The heavy-quark symmetry predicts the relation (see Appendix B)
\begin{eqnarray}
e_P = \tilde e_P\,.
\label{eP-hqs}
\end{eqnarray}
One expects $|e_P| \sim e\simeq 0.303$.
According to standard counting rules the vertices (\ref{def-eP}) carry the leading chiral power $Q^3_\chi$.
It should be emphasized, however, that the counting rules depend on a
naturalness assumption, i.e. that the dimensional parameters of the effective Lagrangian density scale with
appropriate power of the chiral symmetry breaking scale $4 \pi \,f \simeq 1131$ MeV (or more pragmatically with
the mass $m_V$ of the lightest degree of freedom that is integrated out). This is the rationale behind the
particular representation of the vertex (\ref{def-eP}) in terms of the dimensionless parameters $e_P$ and $\tilde e_P$.
One may assign the
vertex (\ref{def-eP}) the order $Q^2_\chi$. On a formal level that would justify to consider the effect
of (\ref{def-eP}) while neglecting additional hadronic vertices of chiral
order $Q^3_\chi$, such as the SU(3) flavour breaking effects in the coupling of the Goldstone bosons to the
D mesons (see (\ref{def-gP})).

There is no empirical
estimate available of the size of $e_P$ and $\tilde e_P$, in particular no data on
the three-body decay process $D^{*+} \to \gamma\,\pi^+\, D^0 $ sensitive to $e_P$. These parameters have therefore to be determined indirectly.
We shall see in the section on numerical results
that the magnitude of the parameter $e_P$ is
larger than expected from the naive naturalness assumption if we try to reproduce the data
with the effective Lagrangian derived above. Such a phenomenon is not unusual in effective
field theories, though asking for a physical explanation.
We assume that the size of $e_P$ is a manifestation of the need to incorporate additional degrees
of freedom, the most natural ones being the light vector mesons. Once they are incorporated one expects
the fitted value of $e_P$ to reduce significantly. We take this as a motivation
to explore the role played by the light vector mesons in the radiative decays of the $D^*_{s0}(2317)$ and $D^*_{s1}(2460)$ states.


\subsection{Light vector mesons}

For the light vector mesons, there is no well-defined chiral power counting
scheme to the authors' knowledge, at least for the case of virtual vector mesons.
The treatment of these degrees of freedom is therefore more phenomenological than the chiral expansion approach used for Goldstone bosons and heavy-light mesons.
The 3-point electromagnetic vertex probed when considering the light vector mesons is the anomalous process analogous
to (\ref{def-eC-eQ}) where a photon excites a Goldstone boson into a vector meson. It reads
\begin{eqnarray}
&& {\mathcal L}_{\rm e.m.} =
\frac{e_A}{8\,f\,m_V}\,F^{\mu \nu}\,\epsilon_{\mu \nu \alpha \beta}\,{\rm tr}\Big((\prt^\alpha
\Phi)\,[Q,\,\partial_\tau V^{\tau \beta} ]_+\Big) \,,
\label{def-eA}
\end{eqnarray}
where we use the representation
\begin{eqnarray}
V_{\mu \nu} = \left(\begin{array}{ccc}
\rho^0_{\mu \nu}+\omega_{\mu \nu} &\sqrt{2}\,\rho_{\mu \nu}^+&\sqrt{2}\,K_{\mu \nu}^+\\
\sqrt{2}\,\rho_{\mu \nu}^-&-\rho_{\mu \nu}^0+\omega_{\mu \nu}&\sqrt{2}\,K_{\mu \nu}^0\\
\sqrt{2}\,K_{\mu \nu}^- &\sqrt{2}\,\bar{K}_{\mu \nu}^0&\sqrt{2}\,\phi_{\mu \nu}
\end{array}\right)\,.
\label{def-Vmumnu}
\end{eqnarray}
The particular form (\ref{def-eA}) is equivalent to the anomalous photon coupling used for example in  \cite{Klingl}.
From the radiative decays of the light vector mesons considered in Appendix A, one derives conflicting values for the
parameter $e_A$. A quantitative description requires the inclusion of SU(3) flavour breaking effects.
Only three processes,
$K^{*0} \to K^0 \gamma, K^{*\pm} \to K^\pm \gamma $ and $\phi\to \eta\gamma $,
are relevant for the radiative decays of the $D^*_{s0}$ and $D^*_{s1}$ states. We will include SU(3) breaking by using the phenomenological couplings obtained in Appendix A for each of these processes,
\begin{eqnarray}
&&|e^{(K^{*}_0 \to K_0 \gamma)}_A | =  0.119 \pm 0.006 \,, \qquad \quad
|e^{(K^{*}_\pm \to K_\pm  \gamma)}_A | =0.090 \pm 0.004 \,,
\nonumber\\
&& |e^{(\;\,\phi \;\,\to \;\eta\; \gamma\;)}_A | = 0.053\pm 0.001 \,.
\label{value-eA}
\end{eqnarray}
The anomaly vertex (\ref{def-eA}) can contribute to the radiative decays of the
$D^*_s$-mesons only in the presence of additional hadronic 3-point vertices. The light vector meson that is created
by the photon must be absorbed by a heavy meson. The corresponding interaction is constructed by analogy with (\ref{def-gP}),
\begin{eqnarray}
{\mathcal L} &=&  i\,\frac{g_V}{2\,f}\, \Big\{ D \, (\partial_\alpha V^{\alpha \mu}) \,
(\partial_\mu \bar D) - (\partial_\mu D)\,(\partial_\alpha V^{\alpha \mu}) \,\bar D\Big\}
\nonumber\\
&-& i\,\frac{\tilde g_V}{2\,f}\, \Big\{D^{ \mu \nu} \,(\partial^\alpha V_{\alpha \mu}) \,
(\partial^\beta \bar D_{\beta \nu} ) -
(\partial^\beta D_{\beta \nu} ) \,(\partial^\alpha V_{\alpha \mu}) \bar D^{ \mu \nu} \,
\Big\}\,
\nonumber\\
&+& \frac{\tilde g_T}{4\,f}\,\epsilon^{\mu \nu \alpha \beta}\,\Big\{
(\partial_\alpha D)\,V_{\mu \nu } \,(\partial^\tau  \bar D_{\tau \beta})+
(\partial^\tau  D_{\tau \beta})\,V_{\mu \nu } \,(\partial_\alpha \bar D)
 \Big\}\,
 \nonumber\\
&+& i\,\frac{g_T}{2\,f}\,(\partial^\alpha D_{\alpha \mu} )\,V^{\mu \nu} \,(\partial^\beta \bar D_{\beta \nu} )
\nonumber\\
&+& \frac{g_E}{4\,f}\,\epsilon_{\mu \nu \alpha \beta}\,\Big\{D^{\mu \nu }\,
(\partial_\tau  V^{\tau \beta})\,(\partial^\alpha  \bar D)
+(\partial^\alpha  D)\,(\partial_\tau  V^{\tau \beta})\,\bar D^{\mu \nu } \Big\}
\,.
 \label{def-gV-gT}
\end{eqnarray}
The electromagnetic terms implied by the covariant derivatives (\ref{gauge-covariant}) should be added.
The particular form of (\ref{def-gV-gT})
was guided by the Lagrangian (\ref{def-gP}) and the flavour SU(4) limit. We complement (\ref{def-gV-gT}) by additional electromagnetic interactions
analogous to (\ref{def-eP}),
\begin{eqnarray}
{\mathcal L} &=&  \frac{e_V}{2\,f\,m_V^2}\,F^{\mu \nu}\, \Big\{ D \, \big[(\partial_\alpha V^{\alpha \nu}),\,Q\big] \,
(\partial_\mu \bar D) - (\partial_\mu D)\,\big[(\partial_\alpha V^{\alpha \nu}),\,Q\big] \,\bar D\Big\}
\nonumber\\
&-& \frac{\tilde e_V}{2\,f\,m_V^2}\, F^{\mu \nu }\,\Big\{D_{ \mu \tau} \,\big[(\partial_\alpha V^{\alpha \nu}),\,Q\big] \,
(\partial_\beta \bar D^{\beta \tau} )
- (\partial_\beta D^{\beta \tau} ) \,\big[(\partial_\alpha V^{\alpha \nu}),\,Q\big]\, \bar D_{ \mu \tau} \,
\Big\}\,
\nonumber\\
&-& i\,\frac{\tilde e_T}{4\,f\,m_V^2}\,F^{\mu \nu}\,\epsilon_{\mu \sigma \alpha \beta}\,\Big\{
(\partial^\alpha D)\,\big[V_{\nu}^{\;\; \sigma } ,\,Q\big]\,(\partial_\tau  \bar D^{\tau \beta})
 \nonumber\\
&& \qquad \qquad + (\partial_\tau  D^{\tau \beta})\,\big[V_{\nu}^{\;\, \sigma },\,Q\big] \,(\partial^\alpha \bar D)
 \Big\}\,
  \nonumber\\
&+& \frac{e_T}{2\,f\,m_V^2}\,F^{\mu \nu }\,(\partial^\alpha D_{\alpha \mu} )\,
\big[V^{\nu \tau},\,Q\big] \,(\partial^\beta \bar D_{\beta \tau} )
\nonumber\\
&-&\frac{e_E}{f\,m_V^2}\,F^{\mu \nu }\,(\partial_\alpha D )\,
\big\{ V_{\mu \nu},\,Q\big\} \,(\partial^\alpha \bar D )
\nonumber\\
&+&\frac{\tilde e_E}{f\,m_V^2}\,F^{\mu \nu }\,(\partial_\alpha D^{\alpha \tau} )\,
\big\{ V_{\mu \nu},\,Q\big\} \,(\partial^\beta \bar D_{\beta \tau} )\,.
\label{def-eV-eT-eE}
\end{eqnarray}

The hadronic vertices (\ref{def-gP}, \ref{def-gV-gT}) should also have some
impact on the formation of the molecules based on the
leading orders coupled-channel interaction.
The influence of the parameters $g_P$ and $\tilde g_P$ was
studied in \cite{Hofmann-Lutz-2004} and in Section 2.3.
Their effects on the mass were found to be rather minor,
confirming the expectation that the corresponding interactions are of subleading order in a chiral expansion.
Nevertheless, this issue deserves further studies. In particular, the role played by additional
inelastic channels involving the light vector mesons should be worked out. We anticipate that they
are responsible for the formation of tensor molecules.
We will return to this issue in the next section.

The various parameters in (\ref{def-gV-gT}, \ref{def-eV-eT-eE}) are correlated by the heavy-quark symmetry of QCD.
As shown in Appendix B, one expects
\begin{eqnarray}
&& \tilde g_V = g_V \,, \qquad \qquad \tilde g_T = g_T\,,
\nonumber\\
&& e_V = \tilde e_V \,,\qquad \qquad e_T = \tilde e_T \,, \qquad \qquad e_E = \tilde e_E\,,
\label{gV-gT-hqs}
\end{eqnarray}
at leading order. The size of $g_V$ and $\tilde g_V$ can be estimated by the phenomenological
assumption of universally coupled light vector mesons and the KSFR relation to be
\begin{eqnarray}
g_V =\tilde g_V = \frac{g\,f}{m_V} = \frac{\pm1}{\sqrt{2}} \simeq \pm 0.71 \,,
\label{def-universal-g}
\end{eqnarray}
with the universal vector coupling constant $|g | \simeq 6$ \cite{Bando}. The phase of the coupling
constant $g_V$ is not determined by this assumption.

The estimate of the size of the coupling constants
$g_T$ and $g_E$ is more difficult. While lacking QCD lattice simulations, we may use
a flavour SU(4) ansatz, admittedly a rather rough and questionable tool.
In order to do so, we introduce flavour SU(4) multiplet fields
\begin{eqnarray}
&& V^{\mu \nu}_{[16]} =
\left(
\begin{array}{cc}
V^{\mu \nu} & 0 \\
0 & 0
\end{array} \right)
+\left(\begin{array}{cc}
0 & \sqrt{2}\,\bar D^{\mu \nu} \\
\sqrt{2}\,D^{\mu \nu} & J/\psi
\end{array} \right) \,,
\nonumber\\
&& \Phi_{[16]} =
\left(
\begin{array}{cc}
\Phi & 0 \\
0 & 0
\end{array} \right)
+\left(\begin{array}{cc}
0 & \sqrt{2}\,\bar D \\
\sqrt{2}\,D & \eta_c
\end{array} \right) \,,
\label{def-SU4-multiplet}
\end{eqnarray}
in terms of the SU(3) multiplet fields $\Phi, D$ and $V_{\mu \nu}, D_{\mu \nu}$.
We construct a minimal SU(4) invariant Lagrangian density that comprises the interaction terms introduced
in (\ref{def-gP}, \ref{def-gV-gT}),
\begin{eqnarray}
{\mathcal L}_{SU(4)}=&& \frac{g_1}{4\,f}\,\epsilon_{\mu \nu \alpha \beta}\, \tr\,
(\partial^\alpha \Phi_{[16]})\,\Big\{V^{\mu \nu }_{[16]} \,(\partial_\tau  V^{\tau \beta}_{[16]})
+(\partial_\tau  V^{\tau \beta}_{[16]})\,V^{\mu \nu }_{[16]} \Big\}
\nonumber\\
+&& \frac{i\,g_2}{4\,f}\,{\tr }\,(\partial_\alpha V_{[16]}^{\alpha \mu})\,V_{\mu \nu }^{[16]} \,(\partial_\beta  V^{\beta \nu}_{[16]})
-\frac{i\,g_3}{2\,f}\,\tr (\partial_\mu \Phi_{[16]})\,V^{\mu \nu}_{[16]}\,(\partial_\nu \Phi_{[16]}) \,.
\label{SU4-ansatz}
\end{eqnarray}
The ansatz (\ref{SU4-ansatz}) suggests the identification
\begin{eqnarray}
 \tilde g_T =g_E= \tilde g_P =g_1\,, \qquad \qquad \tilde g_V = g_T=g_2 \,, \qquad \qquad g_V =g_P=g_3 \,.
\label{SU4-result}
\end{eqnarray}
The result (\ref{SU4-result}) deserves and requires some discussion.
We observe that according to (\ref{gP-hqs}, \ref{gV-gT-hqs}) the coupling constants
$g_i$ with $i=1,2,3$ in (\ref{SU4-ansatz}) should approach a universal value in the heavy-quark mass limit, for instance the parameters $g_P$ and $g_V$. Confronted
with the values (\ref{value-gP}) and (\ref{def-universal-g}), this expectation appears verified within less
than 20$\%$. However, the result (\ref{SU4-result}) would predict also a common value for $g_P$ and $g_T$.
We point out that this implication is troublesome: whereas $g_P$ approaches a finite value, the parameter $g_T$ must vanish
in the heavy-quark mass limit. This is an immediate consequence of the fact that
the QCD action is linear in the charm quark mass. While the term proportional to $g_P$ involves one
derivative acting on a heavy field, the term proportional to $g_T$ involves two. It is therefore clear that
we must not use the relations (\ref{SU4-result}) as they stand. From (\ref{SU4-result}) we retain only the predicted
phase relations for the coupling constants, i.e. we assume all coupling constants to take positive values.
In addition, for the unknown parameter $g_E $, we would anticipate the range
\begin{eqnarray}
g_P  \leq g_E \leq g_V\,,
\label{result-gE}
\end{eqnarray}
a conjecture consistent with the expected scaling behaviour at large charm-quark masses.
The estimate of the remaining parameters $g_T \simeq \tilde g_T$ is most uncertain. From (\ref{SU4-result}) we
would expect the range
\begin{eqnarray}
0 \leq g_T   \leq \frac{m_V}{M_V}\,g_P \,,
\end{eqnarray}
with $m_V =776$ MeV and $M_V=$ 2000 MeV representing the typical masses of light and heavy vector meson.


\subsection{Radiative decay of molecules}

\begin{figure}[b]
\begin{center}
\includegraphics[width=10cm,clip=true]{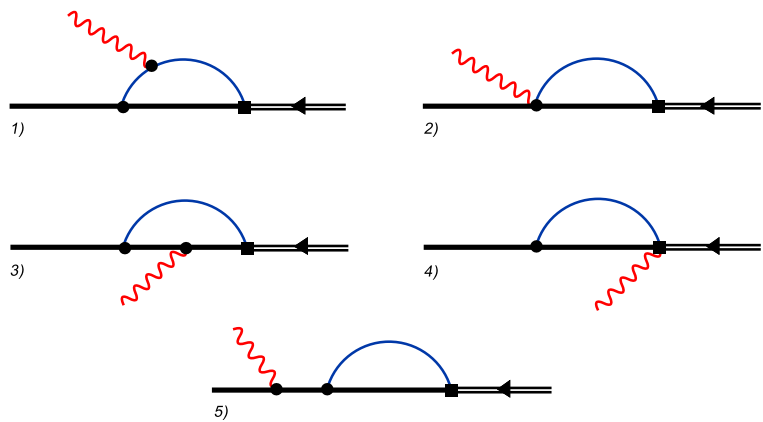}
\end{center}
\caption{Single-loop diagrams contributing to the decay amplitude of a scalar or axial-vector molecule (see text). }
\label{fig:generic-decay-a}
\end{figure}

We display in Fig. \ref{fig:generic-decay-a} the various graphs contributing to the radiative decay of the
$D^*_{s0}(2317)$ and $D^*_{s1}(2460)$ states which are linear in the hadronic three-point vertices. The $D^*_{s0}$ and $D^*_{s1}$ mesons generated by coupled-channel dyna\-mics appear as blocks, the sum of the bubble diagrams defining the effective propagator of the resonance state.
It is drawn as a double line. The dynamics of the decay is contained in a single loop. Solid lines represent the propagation of
the pseudoscalar or vector mesons. The thin lines stand for the light mesons and the thick lines for the heavy
mesons. The wavy line is the photon.

\begin{figure}[t]
\begin{center}
\includegraphics[width=13cm,clip=true]{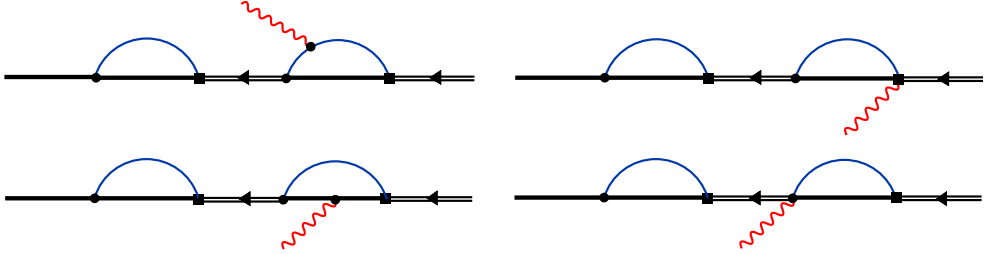}
\end{center}
\caption{Two-loop diagrams contributing to the decay width of scalar and axial-vector molecules where the photon couples to the resonance directly (see text). These terms vanish identically in
the tensor representation of massive vector particles. }
\label{fig:generic-decay-b}
\end{figure}

We display in Fig. \ref{fig:generic-decay-b} two-loop diagrams which, at first sight, should be relevant for
the radiative decay width of scalar and axial-vector molecules. The figure shows contributions where the photon couples to the resonance directly. We do not resolve the structure of the vertex since these diagrams
vanish identically. This is an immediate consequence of using the tensor-field representation of the spin one
particles. Consider for instance the decay of a scalar molecule, where the final state carries
$J^P=1^-$ quantum numbers. All diagrams in Fig. \ref{fig:generic-decay-b} factorize into two contributions.
Since the left part is contracted with the antisymmetric
wave function of the $1^-$ state, it vanishes identically. This contribution carries indeed the
two indices $\alpha$ and $\beta$  and depends on one 4-momentum only, the 4-momentum of the final state.
It must be symmetric in the indices $\alpha$ and $\beta$ and therefore vanishes if contracted with
the wave function. Analogous arguments hold for the decay modes of axial-vector molecules.
There are two possibilities.
If the final state carries $J^P=0^-$ quantum numbers, the antisymmetric and outgoing
indices of the intermediate axial-vector propagator are contracted necessarily with the 4-momentum of the
final state. If the final state carries $J^P=1^-$ quantum numbers, the final 3-point vertex involves the
antisymmetric tensor $\epsilon_{\mu \nu \alpha \beta}$ due to parity conservation. At least two indices must be
saturated by the 4-momentum of the $1^-$ D$_s$-meson.

A striking conclusion of our discussion is that the radiative decay of a scalar or axial-vector
molecule is determined fully by an effective hadronic resonance vertex. The detailed structure of the resonance
propagator is not probed. We emphasize that
this is a consequence of using the tensor-field representation of spin-one particles. The
application of the vector-field representation would require a detailed study of the
resonance propagator. This could be quite cumbersome. In particular, enforcing gauge invariance
is highly non-trivial. Our observation implies that even in the presence
of the additional coupled-channel interactions (\ref{def-gP}, \ref{def-gV-gT}), the formal evaluation of the diagrams of Fig. \ref{fig:generic-decay-a} would be unchanged.

Because of the particular structure of the decay diagrams, it is useful to introduce flavour triplet fields,
$R$ and $R_{\mu \nu }$, which interpolate the scalar and axial-vector molecules. All what is needed are the
coupling strengths of the molecules to the hadronic final states. We introduce the effective coupling constants
$g_R$ and $\tilde g_R$,
\begin{eqnarray}
&& {\mathcal L}_{R} =-\frac{g_R}{2\,f}\,\Big\{
(\partial_\mu D) \,(\partial^\mu \Phi) \,\bar R +  R\,(\partial^\mu \Phi) \, (\partial_\mu \bar D) \Big\}
\nonumber\\
&& \quad \;\; -\,i\,\frac{\tilde g_R}{2\,f}\,\,\Big\{
 D^{\mu \nu} \,(\partial_\mu \Phi) \,(\partial^\tau \bar R_{\tau \nu} )
- (\partial^\tau R_{\tau \nu} )\,(\partial^\mu \Phi) \,  \bar D^{\mu \nu} \Big\} \,,
\label{def-gR}
\end{eqnarray}
where we do not write explicitly terms linear in the photon field that are required by gauge invariance.
We may add terms involving the electromagnetic
field strength tensor $F_{\mu \nu}$ that are gauge-invariant separately. Like the parameters $g_R$ and $\tilde g_R$,
the structure and size of these terms have to be extracted from a coupled-channel computation. As will become
clear in the subsequent section, the role of such terms is very minor. Within our renormalization scheme such
contributions vanish identically for the particular choice where the matching scale $\mu_M$ and the mass of the
hadronic final state are degenerate. This degeneracy is almost realized.
For on-shell conditions, the first vertex of (\ref{def-gR}) is
equivalent to a vertex of the generic form $  D \,\Phi \,\bar R $ as implied by the
projector technique described in Section 2.1. This follows from the replacement
\begin{eqnarray}
2\,(\partial_\mu D)\,(\partial^\mu \Phi)\,\bar R \to
D\,\Phi \,(\partial^\mu \partial_\mu \bar R ) - ( \partial^\mu \partial_\mu D)\,\Phi\,\bar  R
- D\,(\partial^\mu \partial_\mu \Phi ) \,\bar R  \,.
\label{eq:75}
\end{eqnarray}
In general the two vertices would give different results in one-loop diagrams.
Within our renormalization scheme discussed in the next subsection, the vertices are equivalent.
The corresponding decay amplitudes coincide up to terms that may be generated by effective molecule vertices
involving $F_{\mu \nu}$. As argued above such terms
vanish identically for the particular choice where the matching scale $\mu_M$ is
identified with the mass of the hadronic final state.
An ana\-logous argument shows the equivalence of the second vertex of (\ref{def-gR}) with the
generic form $ ( \partial_\alpha D^{\alpha \tau}) \,\Phi \,( \partial^\beta \,\bar R_{\beta \tau}) $
implied in Section 2.2. This vertex leads to the form (\ref{T-on-shell}) of the scattering amplitude
 for $J^P=1^+$.

At leading order the magnitude of the parameters $g_R$ and $\tilde g_R$ are
determined by the coupling constants as
extracted from the pole structure of the scattering amplitude (see (\ref{def-gi-0plus}, \ref{def-gi-1plus})).
We identify
\begin{eqnarray}
&& g_R= \frac{4\,\sqrt{2}\,f\,M_{0^+}}{M_{0+}^2-M_D^2\,\,-m_K^2}\,g^{(0+)}_{ K D}\,
= \frac{4\,\sqrt{6}\,f\,M_{0^+}}{M_{0+}^2-M_{D_s}^2-m_\eta^2}\,g^{(0+)}_{\eta D_s}\,,
\nonumber\\
&& \tilde g_R = \frac{4\,\sqrt{2}\,f\,M_{D^*}}{M_{1+}^2-M_{D^*}^2-m_K^2}\, g^{(1+)}_{ K D^*}
 = \frac{4\,\sqrt{6}\,f\,M_{D_s^*}}{M_{1+}^2-M_{D^*_s}^2-m_\eta^2}\,g^{(1+)}_{\eta D^*_s} \,.
\label{identify-gR}
\end{eqnarray}
As detailed in Appendix B, heavy-quark symmetry predicts
\begin{eqnarray}
g_R = \tilde g_R \,,
\label{hqs-gR}
\end{eqnarray}
at leading order. While (\ref{hqs-gR}) is realized quite accurately,
the SU(3) flavour breaking effects in the coupling constant $g_R$ and $\tilde g_R$
are sizeable as can be inferred from the values given in Table \ref{tab:hadronic-decay}.
The relation (\ref{hqs-gR}) is satisfied at the 10 $\%$ level
\begin{eqnarray}
\frac{\tilde g_R}{g_R} \simeq 0.986 \;\frac{g^{(1+)}_{ K D^*}}{g^{(0+)}_{ K D}}
 \simeq 0.989 \;\frac{g^{(1^+)}_{\eta D_s^*}}{g^{(0+)}_{\eta D_s}} \,.
\end{eqnarray}
In contrast, the SU(3) relations
\begin{eqnarray}
g_R \simeq 0.720 \;g^{(0+)}_{K D} \simeq 1.710\;g^{(0+)}_{\eta D_s}\,, \qquad
\tilde g_R \simeq 0.710 \;g^{(1+)}_{K D^*} \simeq 1.691\;g^{(1+)}_{\eta D^*_s}\,,
\end{eqnarray}
are violated at the $95 \%$ level. The value $g_R \simeq 4.28$ derived from the
$\eta D_s$ channel with $f=90$ MeV is almost a factor of two larger than the value obtained from the $K D$ channel.

Before turning to the renormalization issue,  we discuss additional resonance vertices involving
light vector mesons. The latter arise necessarily in a coupled-channel computation once
any of the parameters $g_V, \tilde g_V, g_T , \tilde g_T$ or $g_E$ is non-vanishing.
Though we are not presenting results of such a computation, we study the
relevance of light vector meson in radiative decay processes of scalar and axial-vector molecules.
We introduce the effective resonance vertices (lea\-ving out additional terms linear in the photon field required by gauge invariance)
\begin{eqnarray}
&& {\mathcal L}_{R} = \frac{g_H}{2\,f}\,\Big\{
 R\,(\partial^\tau V_{\tau \nu }) \, (\partial_\mu \bar D^{\mu \nu})
 +(\partial_\mu  D^{\mu \nu }) \,(\partial^\tau V_{\tau \nu }) \,\bar R  \Big\}
\nonumber\\
&& \quad \;\; +\,i\,\frac{ \tilde g_H}{2\,f}\,\,\Big\{
 (\partial^\mu R_{\mu \nu} )\,(\partial_\tau V^{\tau \nu }) \, \bar D
 - D \,(\partial_\tau V^{\tau \nu }) \,( \partial^\mu \bar R_{\mu \nu} ) \Big\}
\nonumber\\
&& \quad \;\; -\,\,\frac{\hat g_H}{4\,f}\,\epsilon^{\alpha \beta \mu \nu }\,\Big\{
(\partial^\sigma  R_{\sigma \mu})\,(\partial^\tau V_{\tau \nu }) \, \bar D_{\alpha \beta}
 + D_{\alpha \beta} \,(\partial^\tau V_{\tau \nu }) \,(\partial^\sigma \bar R_{\sigma \mu}) \Big\}\,,
\label{def-gH}
\end{eqnarray}
where the parameters $g_H, \tilde g_H$ and $\hat g_H$ are unknown at this stage and degenerate in the heavy-quark limit. As worked out in Appendix B, we have indeed
\begin{eqnarray}
g_H = \tilde g_H = \hat  g_H \,.
\label{}
\end{eqnarray}
For the axial-vector molecules, additional vertices that are not on-shell equi\-valent to those of
(\ref{def-gH}) may be constructed. Such terms are suppressed by their d-wave phase-space behaviour.
The presence of the vertices (\ref{def-gH}) leads to additional diagrams contributing to the
radiative decay amplitudes of scalar and axial-vector molecules. Such diagrams are part of
Fig. \ref{fig:generic-decay-a} given the meaning of the solid lines which represent the propagation of
either pseudoscalar or vector mesons.


\subsection{Renormalization}

We finally turn to the renormalization issue.
The decay diagrams are ultraviolet divergent. Applying the Passareno-Veltman
reduction \cite{Passarino:Veltman:1979}, which is rigourously justified within dimensional regularization,
the integrals of Fig.~\ref{fig:generic-decay-a} may be expressed in terms of a set of scalar
integrals of the form
\begin{eqnarray}
&& I_a = -i\,\int
\frac{d^4l}{(2\pi)^4}\,S_a (l) \,, \qquad \qquad \qquad \;S_a(l)= \frac{1}{l^2-m_a^2+i\,\epsilon}\,,
\nonumber\\
&&  I_{ab}= -i\,\int
\frac{d^4l}{(2\pi)^4}\,S_a (l)\,S^{}_{b}(l+p)\,, \qquad  \bar I_{ab}= -i\,\int
\frac{d^4l}{(2\pi)^4}\,S_a (l)\,S^{}_{b}(l+\bar p)\,,
\nonumber\\
&& J_{abc}= +i\,\int
\frac{d^4l}{(2\pi)^4}\,S_a (l)\,S^{}_{b}(l+q)\,S^{}_{c}(l+p)\,,
\nonumber\\
&& \bar J_{abc}= +i\,\int
\frac{d^4l}{(2\pi)^4}\,S_a (l)\,S^{}_{b}(l+\bar p)\,S^{}_{c}(l+p)\,,
\label{def-master-loops}
\end{eqnarray}
where we focus on the physical limit with space-time dimension four. In our convention the
4-momentum $p_\mu$ characterizes the decaying molecule,
$q_\mu $ is the 4-momentum of the photon and $\bar p_\mu = p_\mu-q_\mu$ is the 4-momentum of the hadronic final
state. The quantities defined in (\ref{def-master-loops})are evaluated in Appendix C.

Divergent structures arise only from the tadpole
integral $I_a$ and the two-propagator integrals $I_{ab}$ and $\bar I_{ab}$. Since the combinations
\begin{eqnarray}
&& I_{ab}-\bar I_{ab} \,, \qquad \qquad
I_{ab} - \frac{I_b-I_a}{m_a^2-m_b^2} \,,
\label{finite-combinations}
\end{eqnarray}
are finite as space-time dimension approaches four,
one may take the viewpoint that all divergent structures are caused by the tadpole integral $I_a$ (see also \cite{Semke-Lutz-2006}).

\begin{figure}[b]
\begin{center}
\includegraphics[width=11cm,clip=true]{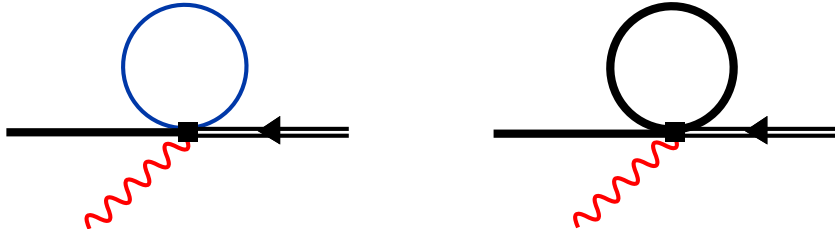}
\end{center}
\caption{Examples of counter loop contributions that renormalize the decay amplitude of molecules. }
\label{fig:renormalization}
\end{figure}

Any divergent contribution can not be renorma\-lized
by simply adding a counter vertex where the photon couples directly to the resonance. Since the resonance is formed
by coupled-channel dynamics its radiative decays amplitude must be renormalized by a loop subtraction me\-chanism similar
to the one introduced in \cite{Lutz:2000}. This is illustrated in Fig.  \ref{fig:renormalization}. Typically the counter
terms needed are appropriate 5-point vertices as indicated in  Fig.  \ref{fig:renormalization} by an solid square. A
contribution to the radiative decay amplitude is generated by contracting two identical lines giving
rise to a tadpole-type contribution. The latter have to cancel the tadpole contributions from Fig.  \ref{fig:generic-decay-a} as discussed above.

There is yet another important observation to make. In Fig.  \ref{fig:generic-decay-a} the  integral
arising from the second and fourth diagrams have a structure similar to the loop functions that build
up the resonance state. Therefore the renormalization of the coupled-channel
dynamics and of the radiative-decay amplitude are necessarily related. Since the coupled-channel dynamics
involves an infinite number of Feynman diagrams, one has to leave the well-trotted path of perturbative
renormalization, i.e. infinite sets of counter terms present in effective field theories
have to be introduced at each order.  We emphasize that one should
carefully discriminate two issues. First, the renormalization scale independence of a scattering amplitude and second,
a possible scheme dependence of how the infinite number of diagrams are treated. We point out that the application of the
on-shell reduction formalism developed in \cite{Lutz:2000,Lutz:Kolomeitsev:2002,Lutz-Kolomeitsev-2004},
as is implied by the unique existence of  an algebra of covariant projectors, may be considered as a scheme dependence.
The additional ingredient, the requirement of a smooth matching of scattering amplitudes unitarized in different
channels, can be considered again as a scheme dependence. It is an economical, though not unique, procedure
to build crossing symmetry into the scheme. From this point of view the matching parameter $\mu_M$ in
(\ref{final-t}, \ref{M-generic}) reflects a scheme dependence, not a
renormalization scale dependence.

As a consequence of the on-shell reduction formalism our coupled-channel dynamics is based on,  neither the effective potential $V^{(JP)}(s)$ nor the loop functions $J^{(JP)}(s)$ gain any contribution at leading order
from a tadpole-type loop integral $I_a$. Using the leading order two-body interaction
as the driving force of a Bethe-Salpeter equation would introduce plenty of tadpole contributions
\cite{Lutz:Kolomeitsev:2002,Lutz-Kolomeitsev-2004}. As was discussed in great
detail in \cite{Lutz:Kolomeitsev:2002,Lutz-Kolomeitsev-2004}, the contribution of reduced tadpoles can be trusted only
at a level where one computes one-loop corrections to the Bethe-Salpeter interaction kernel, i.e. for two-particle irreducible
diagrams. While for the latter conventional power-counting arguments are
applicable, they are not for the reducible contributions, i.e. those summed by the unitarization.

Closing our chain of arguments, we arrive at the result that reduced tadpole contributions in the decay amplitude
should be dropped in our leading order computation. In addition a finite renormalization is applied to the
integral $I_{ab}$. It is identified with the expression introduced in (\ref{i-def}) with
\begin{eqnarray}
I_{ab}=I(p^2)-I(\mu_M^2)\,, \qquad \qquad m_a=M \,, \qquad m_b = m \,,
\end{eqnarray}
and  the matching scale $\mu_M$ used in the coupled-channel computation.


\section{Radiative decay of scalar molecules: $D^*_{s0}(2317)\to \gamma\,D^*_{s}(2112)$}

We express the partial decay width $\Gamma_{0^+\to \gamma \,1^-}$ of the $D^*_{s0}(2317)$-meson
in terms of the transition amplitudes
$M^{\alpha \beta, \mu }_{0^+ \to \gamma
\,1^-}$. Due to gauge invariance, the transition tensor is characterized  by
a single number $d_{0^+\to \gamma
\,1^-}$ defined by
\begin{eqnarray}
&&\epsilon^\dagger_\mu(q, \lambda_q)\,\epsilon^\dagger_{\alpha \beta}(\bar p, \lambda_{\bar p})\,
M^{\alpha \beta, \mu }_{0^+ \to \gamma
\,1^-}
\nonumber\\
&& \qquad = d_{0^+\to \gamma
\,1^-}\;\epsilon^\dagger_\mu(q,\lambda_q)\,
\epsilon^\dagger_{\alpha \beta}(\bar p, \lambda_{\bar p})\,
\Big\{ g^{\mu \alpha}\,(p \cdot q)-p^\mu\,q^\alpha \Big\}\,\frac{i\, \bar p^\beta}{\sqrt{\bar p^2}}\,,
\label{def-transition-tensor-scalar}
\end{eqnarray}
where $\epsilon_\mu(q, \lambda_q)$ and $\epsilon_{\alpha \beta}(\bar p, \lambda_{\bar p})$
denote the wave functions of the outgoing photon and vector meson and $p_\mu=\bar p_\mu+q_\mu$ the decaying  resonance momentum. The radiative width reads
\begin{eqnarray}
\Gamma_{0^+\to \,\gamma \,1^-} = \frac{|d_{0^+\to \gamma
\,1^-}|^2}{4\pi}\, \left(\frac{M_{0^+}^2-M_{1^-}^2}{2\,M_{0^+}} \right)^3\,,
\label{result:width-0plus}
\end{eqnarray}
with $M_{1^-}^2 = \bar p^2 $ and $M_{0^+}^2=p^2$.
The decay parameter
can be obtained by the projection formula,
\begin{eqnarray}
&& d_{0^+ \to \,\gamma\,1^-} =
\frac{-i}{(p \cdot q)\,M_{1^-}}\,\Bigg\{
g_{\mu \alpha}- \frac{p_\mu\,q_\alpha}{(p \cdot q)} \Bigg\}\,\bar p_\beta\,M^{\alpha \beta , \mu}_{0^+ \to \,\gamma\,1^-}  \,,
\label{result-projection-0plus}
\end{eqnarray}
since the decay amplitude is transverse with respect to
the photon momentum and antisymmetric in the indices $\alpha \leftrightarrow \beta$.
This implies that the off-shell amplitude $ \bar p_\beta \,M^{\alpha \beta ,\mu}$ is orthogonal to $q_\mu$ and
$\bar p_\alpha$, i.e. is characterized by two parameters only. With (\ref{result-projection-0plus}) we
project onto the relevant component.

There are 5 classes of contributions as shown in Fig.  \ref{fig:generic-decay-a}. All terms have the topology of  a one-loop Feynman diagram. A given contribution is either proportional to the coupling constant $g_R$ or $g_H$. The thin line attached to the resonance vertex represents the
propagation of a pseudoscalar ($g_R$) or a vector meson ($g_H$) and the thick line the propagation of a heavy meson. The outgoing line represents a vector $D_s^*$ meson.

It is convenient to group together diagrams that are gauge-invariant separately. We discuss the various
contributions proportional to the  coupling constant $g_R$. We consider first the terms of class 2) in Fig. \ref{fig:generic-decay-a}. They are proportional to $e\,g_P/f $ and involve two types of tensors. We form two gauge-invariant combinations with the
transverse tensors $A^{\alpha \beta, \mu}(\bar p,p)$ and $B^{\alpha \beta, \mu}(\bar p,p)$,
receiving respectively contributions from diagrams of classes 1) + 2) and 2) + 3) of Fig.  \ref{fig:generic-decay-a}. The tensors $A^{\alpha \beta, \mu}(\bar p,p)$ and $B^{\alpha \beta, \mu}(\bar p,p)$ describe
the processes where the photon couples to the charge of the light or of the heavy pseudoscalar meson,
\begin{eqnarray}
&& A^{\alpha \beta, \mu}_{ab}(\bar p, p) =-i\,\int
\frac{d^4l}{(2\pi)^4}\,S_a(l)\,S_{b}(p+l)\,
\nonumber\\
&& \qquad  \quad \times
\Big\{ S_a(l+q)\,(q+2\,l)^\mu \, \bar p^\alpha\,(p+l)^\beta
+ g^{\mu \alpha}\,(p+l)^\beta
\Big\}\,,
\nonumber\\ \nonumber\\
&& B^{\alpha \beta, \mu}_{ab}(\bar p, p) =+i\, \int
\frac{d^4l}{(2\pi)^4}\,S_a (l)\,S_{b}(p+l)\,
\nonumber\\
&& \qquad \quad \times
\Big\{ S_b(\bar p+l)\,(\bar p+p+2\,l)^\mu  \, \bar p^\alpha\,(\bar p+l)^\beta
 + g^{\mu \beta}\, l^\alpha \Big\}\,,
 \label{def-AB-0plus}
\end{eqnarray}
with
\begin{eqnarray}
&&S_a(p) = \frac{1}{p^2-m_a^2} \,.
 \label{def-Sa}
\end{eqnarray}
We note that for technical simplicity the
tensors (\ref{def-AB-0plus}) are constructed with an effective molecule
vertex of the generic form $  D \,\Phi \,\bar R   $ as implied by the on-shell reduction
technique applied in the coupled-channel computation of Section 2.1. This is the reason why
diagrams of class 4) do not contribute to the expressions (\ref{def-AB-0plus}).
The latter are associated with a process where the photon
is emitted from a hadronic 3-point vertex involving the resonance field.

The contributions implied by the interactions involving the electromagnetic field
strength tensor $F_{\mu \nu}$ are proportional to the anomalous coupling strengths
$e_A, e_Q, e_C$  of (\ref{def-eA}, \ref{def-eC-eQ}) or the parameter $e_P$ of (\ref{def-eP}).
The terms proportional to $e_P/(f\,m_V^2)$ probe the class 2) of Fig.  \ref{fig:generic-decay-a} only.
Their effect is encoded into the transverse tensor $\bar A^{\alpha \beta, \mu}(\bar p,p)$,
\begin{eqnarray}
&& \bar A^{\alpha \beta, \mu}_{ab}(\bar p, p) =-i\,\int
\frac{d^4l}{(2\pi)^4}\,S_a(l)\,S_{b}(p+l)\,\Big\{(q \cdot l)\, g^{\mu \alpha} -l^\mu\,q^\alpha \Big\}\,(p+l)^\beta\,.
\label{def-barA-0plus}
\end{eqnarray}
Additional tensors
$C^{\alpha \beta, \mu}(\bar p,p)$ and $\bar C^{\alpha \beta, \mu}(\bar p,p)$ describe the processes belonging to the diagrams of class 1) where the
Goldstone boson, which emits the photon, is converted into a light vector meson.
They are proportional to the parameter combinations $e_A\,\tilde g_T/f^2$ and $e_A\,\tilde g_E/f^2$ respectively.
We introduce
\begin{eqnarray}
&&C_{abc}^{\alpha \beta, \mu}(\bar p, p) =-i\, \int
\frac{d^4l}{(2\pi)^4}\,S_a (l)\,S_{c}(p+l)\,
\nonumber\\
&& \qquad \quad \times 2\,q_\rho\,l_\sigma\,\epsilon^{\rho \mu \sigma}_{\quad \; \tau }\,
\bar p^\alpha\,(p+l)^{\bar \rho}\, \epsilon_{ \bar \sigma \bar \tau \bar \rho}^{\quad \;\,\beta}\,S^{\bar \sigma\bar \tau ,\tau}_b(l+q)\,,
\nonumber\\ \nonumber\\
&&\bar C_{abc}^{\alpha \beta, \mu}(\bar p, p) =-i\, \int
\frac{d^4l}{(2\pi)^4}\,S_a (l)\,S_{c}(p+l)\,
\nonumber\\
&& \qquad \quad \times 2\,q_\rho\,l_\sigma\,\epsilon^{\rho \mu \sigma}_{\quad \, \tau }
(p+l)_{\bar \rho}\, \epsilon^{ \alpha \beta}_{\;\;\;\; \bar \rho \bar \sigma}\,S^{ \bar \sigma \tau}_b(l+q)\,,
\label{def-C-0plus}
\end{eqnarray}
where
\begin{eqnarray}
&&S^{\mu \nu}_a(p) =  p_\alpha \,S^{\alpha \mu, \beta \nu}_a(p)\,p_\beta \,,
\nonumber\\
&& S^{\mu \nu, \alpha \beta }_a(p)=-\frac{1}{m_a^2}\,\frac{1}{p^2-m_a^2+i\,\epsilon}\,
\Bigg[ (m_a^2-p^2)\,g_{\mu \alpha}\,g_{\nu \beta}
\nonumber\\
&& \qquad \qquad \qquad  + g_{\mu \alpha}\,p_\nu\,p_\beta-
g_{\mu \beta}\,p_\nu\,p_\alpha- (\mu \leftrightarrow \nu)\Bigg] \,,
\nonumber\\
&& S^{\mu ,\beta \nu}_a(p) =  p_\alpha \,S^{\alpha \mu, \beta \nu}_a(p)\,, \qquad \qquad
 S^{\alpha \mu ,\nu}_a(p) = S^{\alpha \mu, \beta \nu}_a(p)\,p_\beta \,.
 \label{def-Sb}
\end{eqnarray}
Anomalous processes analogous to those described by the tensors (\ref{def-C-0plus}) are driven by the
parameters $e_Q$ and $ e_C$. In this case the photon is emitted by a pseudoscalar D-meson, which is converted
into a vector D-meson. The contribution is included in class 3) of Fig.  \ref{fig:generic-decay-a}.
We define the corresponding loop tensor
\begin{eqnarray}
&&D_{abc}^{\alpha \beta, \mu}(\bar p, p) =+i\, \int
\frac{d^4l}{(2\pi)^4}\,S_a (l)\,S_{c}(p+l)\,
\label{def-D-0plus}\\
&& \qquad \quad \times 2\,q_\rho\,(l+p)_\sigma\,\epsilon^{\rho \mu \sigma}_{\quad \;\, \tau }
\Big\{l_{\bar \rho}\, \epsilon^{\alpha \beta \bar \rho}_{\quad \;\, \bar \tau}\,S^{\bar \tau \tau}_b(\bar p+l)\,
-\bar p^\alpha\,l^{\bar \rho}\, \epsilon_{ \bar \sigma \bar \tau \bar \rho}^{\quad \;\, \beta}\,S^{\bar \sigma  \bar \tau ,\tau}_b(\bar p+l)\,
\Big\}\,. \nonumber
\end{eqnarray}
We emphasize that each of the tensor integrals introduced in
(\ref{def-AB-0plus}, \ref{def-barA-0plus}, \ref{def-C-0plus}, \ref{def-D-0plus})
is gauge-invariant separately, i.e. vanishes identically if contracted with $q_\mu$. This was checked
by explicit calculations.

Collecting all contributions from the  $K D$ and $\eta \,D_s$ channels,
 we arrive at the following decay amplitude,
\begin{eqnarray}
&& i\,M^{\alpha \beta, \mu }_{0^+ \to \gamma \,1^-} =
-\frac{e\,g_P}{f}\,\,h^{(0+)}_{K D}\, \Big\{ A_{KD}^{ \alpha \beta, \mu }(\bar p, p)
+B_{KD}^{\alpha \beta,\mu }(\bar p, p) \Big\}
\nonumber\\
&& \qquad  \;\;\;
-\frac{e_P}{f\,m_V^2}\,\,h^{(0+)}_{K D}\, \bar A_{KD}^{ \alpha \beta, \mu }(\bar p, p)
-2\,\frac{e\,g_P}{\sqrt{3}\,f}\,\,h^{(0+)}_{\eta D_s}\, B_{\eta D_s}^{ \alpha \beta, \mu }(\bar p, p)
\nonumber\\
&& \qquad  \;\;\;
+\frac{e_A}{48\,f^2\,m_V}\,h^{(0+)}_{K D}\, \Big\{ \tilde g_T\,C_{K K^*D}^{\alpha \beta, \mu }(\bar p, p)
+ g_E\,\bar C_{K K^*D}^{\alpha \beta, \mu }(\bar p, p) \Big\}
\nonumber\\
&& \qquad  \;\;\;
+\frac{e_A}{\sqrt{3}\,12\,f^2\,m_V}\,h^{(0+)}_{\eta D_s}\,
\Big\{
\tilde g_T\,C_{\eta  \phi D_s}^{\alpha \beta, \mu}(\bar p, p)
+ g_E\,\bar C_{\eta  \phi D_s}^{\alpha \beta, \mu}(\bar p, p) \Big\}
\nonumber\\
&& \qquad  \;\;\;
+\frac{\tilde g_P}{4\,f\,M^2_V}\,\Big\{ \frac{2}{\sqrt{3}}\,\frac{3\,e_C-e_Q}{6}\,h^{(0+)}_{\eta D_s}\,
D_{\eta  D_s^* D_s}^{\alpha \beta, \mu}(\bar p, p)
\nonumber\\
&& \qquad \qquad  \qquad \qquad \quad +\frac{6\,e_C+e_Q}{6}\,h^{(0+)}_{K D}\,
D_{K  D^* D}^{\alpha \beta, \mu}(\bar p, p)\Big\}  \,,
\label{decay-amplitude-0plus}
\end{eqnarray}
where exact isospin symmetry is assumed with the coupling constant
\begin{eqnarray}
&& h_{KD}^{(0+)} =
\sqrt{2}\,M_{0^+}\,g^{(0+)}_{K D}
\leftrightarrow  \frac{M_{0^+}^2-M_D^2-m_K^2}{2\,f}\,g_R\,, \qquad \quad
\nonumber\\
&& h_{\eta D_s}^{(0+)}=\sqrt{2}\,M_{0^+}\,g^{(0+)}_{\eta D_s}
\leftrightarrow \frac{M_{0^+}^2-M_{D_s}^2-m_\eta^2}{2\,\sqrt{3}\,f}\,g_R\,.
\label{match-h0plus}
\end{eqnarray}
The values of the coupling constants $g^{(0+)}_{ K D}$ and $g^{(0+)}_{\eta D_s}$ are given
in Table \ref{tab:hadronic-decay}.
Note that the tensors introduced in (\ref{def-AB-0plus}, \ref{def-barA-0plus}, \ref{def-C-0plus}, \ref{def-D-0plus})
are not antisymmetric in the indices
$\alpha \leftrightarrow \beta$ as of notational convenience. While deriving the decay parameter
$d_{0^+\to \gamma \,1^+}$ according to (\ref{def-transition-tensor-scalar}), the wave-function projects onto the re\-levant component, the application of the projection formula (\ref{result-projection-0plus}) requiring an
explicit antisymmetrization.

The contributions of the $\phi\,D^*_s$ and $K^* D$ channels to the decay parameter
as well as explicit results for the Passareno-Veltman reduction of the tensor integrals of
(\ref{decay-amplitude-0plus}) are detailed in Appendix D. Such contributions are proportional to
$g_H $.


\section{Radiative decay of axial-vector molecules}

We derive expressions for the radiative decay amplitudes of the $D^*_{s1}(2460)$-meson corresponding to the processes
$1^+ \to \gamma\,0^- $, $1^+ \to \gamma\,0^+ $ and $1^+ \to \gamma\,1^- $. We follow the same procedure as in the previous section for the $D^*_{s0}(2317)$ radiative decay, adjusting the tensor forms to the hadronic initial and final states.

\subsection{$D^*_{s1}(2460)\to \gamma\,D_s(1968)$}

The partial decay width $\Gamma_{1^+\to \gamma\,0^-}$ is expressed in terms of
the transition amplitudes
$M^{\mu, \alpha \beta }_{1^+ \to \gamma \,0^-}$
determined  by one number $d_{1^+\to \,\gamma\, 0^-}$ as
\begin{eqnarray}
&&\epsilon_\mu^\dagger(q, \lambda_q)\, M^{\mu, \alpha \beta }_{1^+ \to \,\gamma \,0^-}\,\epsilon^{\alpha \beta}(p, \lambda_p)
\nonumber\\
&& \qquad = d_{1^+ \to \,\gamma \,0^-} \,\epsilon_\mu^\dagger(q, \lambda_q)\,\Big\{ g^{\mu
\alpha}\,(\bar p\cdot q)- \bar p^\mu\,q^\alpha \Big\}\,\frac{p^\beta}{\sqrt{p^2}}\,
\epsilon^{\alpha \beta}(p, \lambda_p) \,,
\label{def-transition-tensor-axial}
\end{eqnarray}
implying
\begin{eqnarray}
&&\Gamma_{1^+\to \,\gamma \,0^-}= \frac{|d_{1^+ \to \,\gamma\,0^-}|^2}{12\pi}\,\left(\frac{M_{1^+}^2-M_{0^-}^2}{2\,M_{1^+}} \right)^3\,,
\nonumber\\
&& d_{1^+ \to \,\gamma\,0^-} = \frac{1}{(p \cdot q)\,M_{1^+}}\,\Bigg\{
g_{\mu \alpha}- \frac{p_\mu\,q_\alpha}{(p \cdot q)} \Bigg\}\,p_\beta\,M^{\mu, \alpha \beta } \,,
\label{result:width-1plus:a}
\end{eqnarray}
with $\bar p^2=(p-q)^2 =M_{0^-}^2$ and $p^2=M_{1^+}^2$. Like in (\ref{result-projection-0plus})  we exploit
the fact that the decay amplitude is antisymmetric in $\alpha \leftrightarrow \beta$.

In this case again we have to consider the 5 classes of contributions depicted in Fig.  \ref{fig:generic-decay-a}.
Any diagram is proportional to one of the three coupling constants
$\tilde g_R, \tilde g_H$ or $\hat g_H$. In the first case, the thin and thick lines
attached to the resonance vertex stand for the propagation of Goldstone bosons and vector D-mesons. In the
second case, the role of thin and thick lines is interchanged in the sense that the thin lines correspond to light
vector mesons, whereas the thick lines describe heavy pseudoscalar mesons. In the third case both lines represent
vector mesons. The outgoing line is always a pseudoscalar $D_s$-meson.

We discuss the various
contributions, starting with the diagrams proportional to the coupling constant $\tilde g_R$.
There are two types of gauge-invariant combinations proportional to $e\,g_P/f $. We
form two corresponding transverse tensors $A^{\mu,\alpha \beta}(\bar p,p)$ and $B^{\mu,\alpha \beta }(\bar p,p)$. They receive contributions from classes 1) + 2) + 4) and 2) + 3) + 4) of Fig.  \ref{fig:generic-decay-a}
and describe
the processes where the photon couples to the charge of the light and heavy meson respectively,
\begin{eqnarray}
&&A^{\mu, \alpha \beta  }_{ab}(\bar p,p)=+i\,\int
\frac{d^4l}{(2\pi)^4}\,S_a(l)\,\Bigg\{ (l+\bar p)_{\bar \beta}\,S^{ \bar \beta\mu ,\beta}_{b}(p+l)\,p^\alpha
\nonumber\\
&& \qquad  \quad
 + \bar p_{\bar \beta }\, S^{\bar \beta  \beta }_{b}(\bar p+l)\,g^{\mu \alpha}
+S_a(l+q)\,\bar p_{\bar \beta }\,S^{\bar \beta  \beta}_{b}(p+l)\,(q+2\,l)^\mu\,p^\alpha
\Bigg\} \,,
\nonumber\\ \nonumber\\
&& B^{\mu, \alpha \beta  }_{ab}(\bar p,p)=+i\,\int
\frac{d^4l}{(2\pi)^4}\,S_a (l)\,\Bigg\{
\bar p^{\bar \alpha}\,S_{b,\bar \alpha \bar \beta}(\bar p+l)\,S^{\mu \bar \beta,\beta}_{b}(p+l)\,p^\alpha
\nonumber\\
&& \qquad \quad
+ \bar p_{\bar \alpha } \,S_{b}^{ \bar \alpha, \mu \bar \beta}(\bar p+l)\,
S^{\;\;\;\;\beta}_{b,\bar \beta}(p+l)\, p^\alpha\, -
l_{\bar \beta }\,S^{  \mu \bar \beta,\beta}_{b}(p+l)\,p^\alpha
\nonumber\\
&& \qquad \quad+ \bar p_{\bar \beta }\,S^{ \bar \beta,\mu \beta}_{b}(\bar p+l)\,p^\alpha
+ \bar p_{\bar \beta }\, S^{ \bar \beta \beta}_{b}(\bar p+l)\,g^{\mu \alpha}
\Bigg\}\,.
\label{def-AB-1plus}
\end{eqnarray}
The notations are the same as in (\ref{def-Sa}, \ref{def-Sb}).
The
tensors (\ref{def-AB-1plus}) are constructed with an effective molecule
vertex of the generic form $ ( \partial_\alpha D^{\alpha \tau}) \,\Phi \,( \partial^\beta \,\bar R_{\beta \tau}) $
as implied by the projector technique of Section 2.2. The latter is equivalent
to the term proportional to $\tilde g_R$ in (\ref{def-gR}) (see
the discussion following (\ref{eq:75})).

The contributions driven by interactions involving the electromagnetic field
strength tensor $F_{\mu \nu}$ are encoded into the transverse tensor $\bar A^{\mu, \alpha \beta, }(\bar p,p)$,
\begin{eqnarray}
&&\bar A^{\mu, \alpha \beta  }_{ab}(\bar p,p)=+i\,\int
\frac{d^4l}{(2\pi)^4}\,S_a(l)\,\Bigg\{ (l\cdot q)\,g^{\mu}_{\;\; \sigma} -l^\mu\,q_\sigma
\Bigg\}\,\bar p_{\bar \beta}\,S^{ \bar \beta \sigma  ,\beta}_{b}(p+l) \,p^\alpha\,.
\label{def-barA-1plus}
\end{eqnarray}
Additional tensors
$C^{\mu,\alpha \beta }(\bar p,p)$ and $\bar C^{\mu,\alpha \beta}(\bar p,p)$ describe the processes where the
Goldstone boson, which emits the photon, is converted into a light vector meson. They
are proportional to the parameter combinations $e_A\,\tilde g_T/f^2$ and $e_A\,g_E/f^2$ respectively.
We introduce
\begin{eqnarray}
&& C^{\mu, \alpha \beta  }_{abc}(\bar p,p)=+2\,i\,\int
\frac{d^4l}{(2\pi)^4}\,S_a (l)\, \bar p^{\bar \beta}\,S_{b}^{\sigma
\tau, \tilde \alpha}(q+l)\,S^{\bar \alpha \beta}_{c}(p+l)\,p^\alpha
\nonumber\\
&& \qquad \quad \,\times \;\epsilon_{\sigma \tau \bar \alpha \bar \beta}\,
\epsilon^{\mu  }_{\;\,\, \nu \tilde \alpha \tilde \beta}\,q^\nu \,l^{\tilde \beta}\,,
\nonumber\\ \nonumber\\
&& \bar C^{\mu, \alpha \beta  }_{abc}(\bar p,p)= -2\,i\,\int \frac{d^4l}{(2\pi)^4}\,S_a (l)\, \bar
p^{\bar \beta}\,S_{b}^{\bar \alpha \tilde \alpha}(q+l)\,S^{\sigma
\tau ,\beta}_{c}(p+l)\,p^\alpha
\nonumber\\
&& \qquad \quad \,\times \;\epsilon_{\sigma \tau \bar \alpha \bar \beta}\,
\epsilon^{\mu}_{\;\,\, \nu \tilde \alpha \tilde \beta}\,q^\nu \,l^{\tilde \beta} \,.
\label{def-C-1plus}
\end{eqnarray}
The terms proportional to $\tilde e_C, \tilde e_Q$ are described by the tensor
$D^{\mu, \alpha \beta  }_{}(\bar p,p)$. In the convention of Fig.  \ref{fig:generic-decay-a} they correspond to
contributions of class 3), which probe the anomalous magnetic moments of the vector D-mesons. We introduce
\begin{eqnarray}
&& D^{\mu, \alpha \beta  }_{ab}(\bar p,p)=+i\,\int
\frac{d^4l}{(2\pi)^4}\,S_a (l)\,\Bigg\{\, \bar p_{\bar \alpha } \,S_{b}^{ \bar \alpha \mu }(\bar p+l)\,
q_\tau \,S^{\tau\beta}_{b}(p+l)\, p^\alpha\,
\nonumber\\
&& \qquad \quad
- \bar p_{\bar \alpha}\,S_{b}^{\bar \alpha \tau}(\bar p+l)\,q_\tau\,S^{\mu \beta}_{b}(p+l)\,p^\alpha
\Bigg\} \,.
\label{def-D-1plus}
\end{eqnarray}
Each of the tensor integrals introduced in
(\ref{def-AB-1plus}-\ref{def-D-1plus})
is gauge-invariant separately.

Collecting all terms for the $K D^*$ and $\eta \,D_s^*$ channels,
we arrive at the decay amplitude
\begin{eqnarray}
&& M^{\mu, \alpha \beta }_{1^+ \to \gamma \,0^-}=
-\frac{e\,g_P}{f}\,h^{(1+)}_{K D^*}\, \Big\{ A_{KD^*}^{\mu,  \alpha \beta}(\bar p, p)
+B_{KD^*}^{\mu, \alpha \beta }(\bar p, p) \Big\}
\nonumber\\
&& \qquad \;\;\; -\frac{e_P}{f\,m_V^2}\,h^{(1+)}_{K D^*}\,  \bar A_{KD^*}^{\mu,  \alpha \beta}(\bar p, p)
-\frac{2}{\sqrt{3}}\,\frac{e\,g_P}{f}\,h^{(1+)}_{\eta D^*_s}\, B_{\eta D_s^*}^{ \mu, \alpha \beta}(\bar p, p)
\nonumber\\
&& \qquad \;\;\;
+\frac{e_A}{48\,f^2\,m_V}\,h^{(1+)}_{K D^*}\, \Big\{
\tilde g_T\,C_{K K^*D^*}^{\mu, \alpha \beta }(\bar p, p)
+g_E\, \bar C_{K K^*D^*}^{\mu, \alpha \beta }(\bar p, p) \Big\}
\nonumber\\
&& \qquad  \;\;\;
+\frac{e_A}{\sqrt{3}\,12\,f^2\,m_V}\,h^{(1+)}_{\eta D^*_s}\,
\Big\{ \tilde g_T\,C_{\eta  \phi D_s^*}^{\mu, \alpha \beta }(\bar p, p)
+ g_E\,\bar C_{\eta  \phi D_s^*}^{\mu, \alpha \beta }(\bar p, p) \Big\}
\nonumber\\
&& \qquad  \;\;\;
-\frac{g_P}{f\,M^2_V}\,\Big\{\frac{2}{\sqrt{3}}\, \frac{3\,\tilde e_C+\tilde e_Q -3\,e}{3}\,h^{(1+)}_{\eta D^*_s}\,
D_{\eta  D_s^*}^{\mu, \alpha \beta}(\bar p, p)
\nonumber\\
&& \qquad   \qquad \qquad \quad +\frac{6\,\tilde e_C-\tilde e_Q-3\,e}{3}\,h^{(1+)}_{K D^*}\,
D_{K  D^* }^{\mu, \alpha \beta}(\bar p, p)\Big\}  \,,
\label{decay-amplitude-1plus-a1}
\end{eqnarray}
where
\begin{eqnarray}
&& h_{KD^*}^{(1+)}=\sqrt{2}\,\frac{g^{(1+)}_{K D^*}}{M_{D^*}}
\leftrightarrow  \frac{M_{1^+}^2-M_{D^*}^2-m_K^2}{2\,f\,M_{1^+}^2}\,\tilde g_R\,, \qquad \quad
\nonumber\\
&& h_{\eta D^*_s}^{(1+)}=\sqrt{2}\,\frac{g^{(1+)}_{\eta D^*_s}}{M_{D_s^*}}
\leftrightarrow  \frac{M_{1^+}^2-M_{D^*_s}^2-m_\eta^2}{2\,\sqrt{3}\,f\,M_{1^+}^2}\,\tilde g_R\,.
\label{match-h1plus}
\end{eqnarray}
The values of the coupling constants $g^{(1+)}_{K D}$ and $g^{(1+)}_{\eta D_s}$ are given
in Table \ref{tab:hadronic-decay}.
We reemphasize that the application of the projection formula (\ref{result:width-1plus:a}) requires an
antisymmetrization of the tensors (\ref{def-AB-1plus}-\ref{def-D-1plus}).

We turn to the contributions implied by the channels involving the $K^*$ or $\phi$ meson. Such terms
are proportional to the coupling constants $\tilde g_H$ or $\hat g_H$
introduced in (\ref{def-gH}).

While the terms proportional to $\hat g_H$ are deferred to Appendix E, we
detail those proportional to $\tilde g_H$ here. The former encode the physics of the
$K^* D^*$ and $\phi\, D^*$ channels. Appendix E provides in addition explicit results for the
Passareno-Veltman reduction of the tensor integrals (\ref{def-AB-1plus}-\ref{def-D-1plus}, \ref{def-barB-1plus}).
The terms proportional to $\tilde g_H$ are evaluated like those proportional to $g_R$.
Formally the role of light and heavy intermediate lines in Fig.  \ref{fig:generic-decay-a} is interchanged but
the result involves the same tensor integrals as (\ref{def-AB-1plus}-\ref{def-D-1plus}). Additional tensors are required to describe the effects of the terms proportional to
$e_V \neq 0$ or $e_E \neq 0$ introduced in (\ref{def-eV-eT-eE}). The latter give rise to
contributions of class 2) in Fig.  \ref{fig:generic-decay-a}. They are analogous to the contributions proportional to
$e_P$ in (\ref{decay-amplitude-1plus-a1}). We form the
gauge-invariant tensors $\bar B^{\mu, \alpha \beta  }_{}(\bar p,p)$ and
$\tilde B^{\mu, \alpha \beta  }_{}(\bar p,p)$,
\begin{eqnarray}
&& \bar B^{\mu, \alpha \beta  }_{ab}(\bar p,p)= -i\,\int
\frac{d^4l}{(2\pi)^4}\,S_a (l)\,
S^{ \tau \beta}_{b}(p+l)\,p^\alpha \,
\nonumber\\
&& \qquad \qquad \qquad \times \Big(
(l \cdot q)\,g^{\mu}_{\;\;\tau}-  l^\mu\,q_\tau
-(\bar p \cdot q)\,g^{\mu}_{\;\;\tau}+  \bar p^\mu\,q_\tau\Big)\,,
\nonumber\\
&& \tilde B^{\mu, \alpha \beta  }_{ab}(\bar p,p)=-2\,i\,\int
\frac{d^4l}{(2\pi)^4}\,S_a (l)\,
q_{\tau }\,S^{ \tau \mu,\beta}_{b}(p+l)\,p^\alpha \,
(\bar p \cdot l)\,.
\label{def-barB-1plus}
\end{eqnarray}
We collect the contributions of the $K^* D$ and $\phi\, D_s$ channels to the decay amplitude
\begin{eqnarray}
&& M^{\mu, \alpha \beta }_{1^+\to \gamma 0^-}=
-\frac{e\,g_V\,\tilde g_H}{f^2}\, \Big\{ A_{D K^*}^{\mu,  \alpha \beta}(\bar p, p)
+B_{DK^*}^{\mu, \alpha \beta }(\bar p, p) +A_{ D_s \phi }^{ \mu, \alpha \beta}(\bar p, p)\Big\}
\nonumber\\
&& \quad \;\;\;
-\,\frac{e_V\,\tilde g_H}{2\,f^2\,m^2_V}\, \bar B_{DK^*}^{\mu, \alpha \beta }(\bar p, p)
+ \frac{e_E\,\tilde g_H}{3\,f^2\,m^2_V}\, \tilde B_{DK^*}^{\mu, \alpha \beta }(\bar p, p)
+ \frac{2\,e_E\,\tilde g_H}{3\,f^2\,m^2_V}\, \tilde B_{D_s\phi}^{\mu, \alpha \beta }(\bar p, p)
\nonumber\\
&& \quad \;\;\;
+\,\frac{(e_C+e_Q/6)\,\tilde g_H}{4\,f^2\,M^2_V}\,\Big\{ g_E\,C_{D D^*K^*}^{\mu, \alpha \beta }(\bar p, p)
+\tilde g_T\, \bar C_{D D^*K^*}^{\mu, \alpha \beta }(\bar p, p) \Big\}
\nonumber\\
&& \quad \;\;\;
+\,\frac{(e_C-e_Q/3)\,\tilde g_H}{8\,f^2\,M^2_V}\,\Big\{ g_E\,C_{D_s D^*_s\phi}^{\mu, \alpha \beta }(\bar p, p)
+\tilde g_T\, \bar C_{D_s D_s^* \phi}^{\mu, \alpha \beta }(\bar p, p) \Big\}\,,
\label{decay-amplitude-1plus-a2}
\end{eqnarray}
in terms of the tensor integrals of
(\ref{def-AB-1plus}-\ref{def-C-1plus}, \ref{def-barB-1plus}).
There is no term involving the tensor (\ref{def-D-1plus}) since we
neglect the effect of the anomalous magnetic moment of the light vector mesons.


\subsection{$D^*_{s1}(2460)\to \gamma\,D^*_{s0}(2317)$}

The partial decay width $\Gamma_{1^+\to \gamma\,0^+}$ is expressed  in terms of the transition amplitudes
$M^{\mu, \alpha \beta }_{1^+ \to \gamma \,0^+}$
determined  by the number $d_{1^+\to \,\gamma\, 0^+}$ as
\begin{eqnarray}
&&\epsilon_\mu^\dagger(q, \lambda_q)\, M^{\mu, \alpha \beta }_{1^+ \to \,\gamma \,0^+}\,\epsilon^{\alpha \beta}(p, \lambda_p)
\nonumber\\
&& \qquad = d_{1^+ \to \,\gamma \,0^+} \,\epsilon_\mu^\dagger(q, \lambda_q)\;
q_{\tau }\,\epsilon^{\mu \tau}_{\quad \sigma \alpha}\,p^\sigma \,
\,\frac{p^\beta}{\sqrt{p^2}}\,
\epsilon^{\alpha \beta}(p, \lambda_p) \,,
\label{def-transition-tensor-axial-2}
\end{eqnarray}
implying
\begin{eqnarray}
&&\Gamma_{1^+\to \,\gamma \,0^+}= \frac{|d_{1^+ \to \,\gamma\,0^+}|^2}{12\pi}\,\left(\frac{M_{1^+}^2-M_{0^+}^2}{2\,M_{1^+}} \right)^3\,,
\nonumber\\
&& d_{1^+ \to \,\gamma\,0^+} = \frac{-1}{(p \cdot q)^2\,M_{1^+}}\,
q^{\tau }\,\epsilon_{\mu \tau  \sigma \alpha}\,p^\sigma \,p_\beta\,M^{\mu, \alpha \beta } \,,
\label{result:width-1plus:b}
\end{eqnarray}
with $\bar p^2=(p-q)^2 =M_{0^+}^2$ and $p^2=M_{1^+}^2$.

\begin{figure}[b]
\begin{center}
\includegraphics[width=11cm,clip=true]{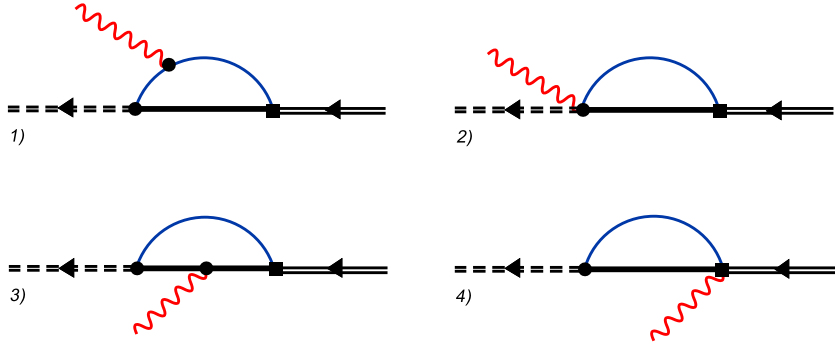}
\end{center}
\caption{Diagrams contributing to the process $1^+ \to \gamma \,0^+$.  }
\label{fig:final-state-0plus}
\end{figure}

There are 4 classes of contributions. They are depicted in Fig.  \ref{fig:final-state-0plus}.
Like in Fig. ~\ref{fig:generic-decay-a} the solid lines stand for the propagation of
pseudoscalar or vector mesons. The thick lines are used for the heavy mesons, the thin ones for the light mesons.
The solid and dashed double lines represent the molecule of the initial and final state respectively.
Any diagram is proportional to one of the four products of coupling constants
$g_R\, \tilde g_R$, $g_R\,\tilde g_H$, $g_H \,\tilde g_R $ or $g_H\,\tilde g_H$. We discuss the four possibilities
case by case.

In the first two cases (terms proportional to $g_R\, \tilde g_R$ or $g_R\,\tilde g_H$),
the thin and thick lines attached to the final molecule vertex represent
the pro\-pagation of pseudoscalar light and heavy mesons. There is only one generic tensor
$A^{\mu,\alpha \beta}_{+,abc}(\bar p,p)$ describing these processes.
They are of class 1) or 3) in the convention of Fig.  \ref{fig:final-state-0plus}.
The contributions probe anomalous electromagnetic vertices proportional to
$e_A$ or $e_C$ and $e_Q$. We define
\begin{eqnarray}
&&A^{\mu,\alpha \beta}_{+,abc}(\bar p,p) = -2\,i\,q_\tau\,\epsilon^{\mu \tau}_{\quad \sigma \rho}\,\int \frac{d^4 l}{(2\pi)^4}\,S_a(l)\,
S_b(l + \bar p)\,(l+\bar p)^\rho\,S^{ \sigma \beta}_c(l+p)\,p^\alpha \,.\nonumber\\
\label{def-A-1plus:0plus}
\end{eqnarray}
For the terms proportional to $g_H \,\tilde g_R$, there is
a class 1) contribution only, with thick lines describing the propagation of vector mesons. The thin line
changes from a pseudoscalar to a vector line at the photon vertex. The contributions are proportional
to the anomalous coupling constant $e_A$. We introduce the corresponding gauge-invariant tensor
\begin{eqnarray}
&&D^{\mu,\alpha \beta}_{+,abc}(\bar p,p) = -2\,i\,q_\nu\,\epsilon^{\mu \nu}_{\quad  \sigma \tau}\,g_{\bar \kappa \kappa}
\int \frac{d^4 l}{(2\pi)^4}\,l^\tau\,p^\alpha\,S_a(l)\,S_{b}^{ \bar \kappa \sigma  }(l+q)\,
S^{ \kappa \beta}_{c}(p+l)\,. \nonumber\\
\label{def-D-1plus:0plus}
\end{eqnarray}
For the contributions proportional to $g_H\,\tilde g_H$,
there are two types of gauge-invariant combinations. We
form two transverse tensors $B_+^{\mu,\alpha \beta}(\bar p,p)$ and
$C_+^{\mu,\alpha \beta }(\bar p,p)$
receiving contributions from classes  2) + 3) + 4) and 1) + 2) + 4)
of Fig.~\ref{fig:final-state-0plus} respectively. The tensors $B^{\mu,\alpha \beta}_{+}(\bar p,p)$ and
$C^{\mu,\alpha \beta }_{+}(\bar p,p)$ describe
the processes where the photon couples to the charge of the heavy and light mesons. We introduce the gauge-invariant
tensors
\begin{eqnarray}
&&B^{\mu,\alpha \beta}_{+,ab}(\bar p,p) = -\,i\,\epsilon_{\tilde \alpha \tilde \beta \sigma}^{\quad\;\,\beta}\,
g_{\bar \rho \rho} \,
\int \frac{d^4 l}{(2\pi)^4}\,S^{\bar \rho \sigma}_a(l)\,\Bigg\{
g^{\mu \alpha}\,S^{\rho ,\tilde \alpha \tilde \beta}_b(l+\bar p)
\nonumber\\
&& \qquad \qquad+\, p^\alpha\,S^{\mu \rho, \tilde \alpha \tilde \beta}_b(l+p)
+ g_{\bar \kappa \kappa}\,p^\alpha\,\Big\{S_{b}^{\rho \bar \kappa  }(\bar p+l)\,S^{\mu \kappa, \tilde \alpha \tilde \beta}_{b}(p+l)
\nonumber\\
&& \qquad \qquad \qquad +\, S_{b}^{ \bar \rho ,\mu \bar \kappa }(\bar p+l)\,
S^{ \kappa ,\tilde \alpha \tilde \beta}_{b}(p+l) \Big\}\Bigg\} \,,
\nonumber\\
&&C^{\mu,\alpha \beta}_{+,ab}(\bar p,p) = +\,i\,\epsilon_{\tilde \alpha \tilde \beta \tau}^{\quad \;\, \beta}\,
g_{\bar \rho \rho} \,
\int \frac{d^4 l}{(2\pi)^4}\,S^{\bar \rho ,\tilde \alpha \tilde \beta}_a(l)\,\Bigg\{
p^\alpha S^{\rho , \mu \tau}_b(l+\bar p)
\nonumber\\
&& \qquad \qquad
+\, g^{\mu \alpha}\,S^{\rho \tau}_b(l+\bar p)+ p^\alpha\,S^{\mu \rho, \tau}_b(l+p)
\nonumber\\
&& \qquad \qquad
+\, g_{\bar \kappa \kappa}\,p^\alpha\,\Big\{S_{b}^{\rho \bar \kappa  }(\bar p+l)\,S^{\mu \kappa, \tau }_{b}(p+l)
+ S_{b}^{ \bar \rho ,\mu \bar \kappa }(\bar p+l)\,S^{ \kappa \tau}_{b}(p+l)\Big\} \Bigg\} \,.\nonumber\\
\label{def-BC-1plus:0plus}
\end{eqnarray}
The tensor integrals introduced in
(\ref{def-A-1plus:0plus}-\ref{def-BC-1plus:0plus})
are gauge-invariant separately.
Collecting all terms we arrive at the decay amplitude
\begin{eqnarray}
&&M^{\mu, \alpha \beta}_{1^+\to \gamma 0^+} =
\frac{e_C+e_Q/6}{2\,M_V^2}\,
h_{KD}^{(0^+)}\,h_{KD^*}^{(1^+)}\, A^{\mu , \alpha \beta}_{+,K D D^*}(\bar p,p)
\nonumber\\
&& \quad  +\,\frac{e_C-e_Q/3}{2\,M_V^2}\,
h_{\eta D_s}^{(0^+)}\,h_{\eta D_s^*}^{(1^+)}\, A^{\mu , \alpha \beta}_{+,\eta D_s D_s^*}(\bar p,p)
\nonumber\\
&& \quad  +\,\frac{e_A\,}{6\,\sqrt{3}\,f^2\,m_V}\,\Big\{\tilde g_H\,h_{\eta D_s}^{(0^+)}\,
 A^{\mu , \alpha \beta}_{+,D_s \eta \phi}(\bar p,p)
 +g_H\,h_{\eta D_s}^{(1^+)}\,
 D^{\mu , \alpha \beta}_{+,\phi \eta D_s^*}(\bar p,p) \Big\}
 \nonumber\\
&& \quad +\,\frac{e_A\,}{24\,f^2\,m_V}\,\Big\{ \tilde g_H\,h_{K D}^{(0^+)}\,
 A^{\mu , \alpha \beta}_{+,D K K^*}(\bar p,p)
+g_H\,h_{K D}^{(1^+)}\,
 D^{\mu , \alpha \beta}_{+,K^* K D}(\bar p,p) \Big\}
\nonumber\\
&& \quad  +\,e\,\frac{\hat g_H\,g_H}{4\,f^2} \, \Big\{
B^{\mu , \alpha \beta}_{+,\phi D_s^* }(\bar p,p) +B^{\mu , \alpha \beta}_{+,K^* D^* }(\bar p,p)
+C^{\mu , \alpha \beta}_{+, D^* K^* }(\bar p,p) \Big\}
\nonumber\\
&& \quad + \frac{g_H\,\tilde g_H}{2\,f^2}\,\Big\{
\frac{e_C+e_Q/6}{2\,M_V^2}\,
D^{\mu , \alpha \beta}_{+,D D^* K^*}(\bar p,p)
+\frac{e_C-e_Q/3}{2\,M_V^2}\,
D^{\mu , \alpha \beta}_{+,D_s D_s^* \phi}(\bar p,p)  \Big\}\,.
\label{decay-amplitude-1plus-b}
\end{eqnarray}
where the coupling constants $h_{KD}$ and $h_{\eta D_s}$ are specified in (\ref{match-h0plus}, \ref{match-h1plus})
in terms of the values given in Table \ref{tab:hadronic-decay}.
Appendix F provides  the results of a Passareno-Veltman reduction
of the tensor integrals (\ref{def-A-1plus:0plus}-\ref{def-BC-1plus:0plus}).


\subsection{$D^*_{s1}(2460)\to \gamma\,D^*_{s}(2112)$}

This radiative decay mode is described by a rank-five
transition tensor $M^{\bar \alpha \bar \beta, \mu, \alpha \beta }_{1^+ \to \gamma \,1^-}$.  The decay
amplitude has a slightly more complicated structure than those displayed in
(\ref{def-transition-tensor-scalar}, \ref{def-transition-tensor-axial-2}).
It is characterized by two scalar decay parameters as the decay may go
via an s-wave or a d-wave transition. Separating these two transitions is an original and interesting feature of our work leading to angular distributions.
We write
\begin{eqnarray}
&& \epsilon_\mu^\dagger(q,\lambda_q)\,\epsilon^\dagger_{\bar \alpha \bar \beta}(\bar p, \lambda_{\bar p})\,
M^{\bar \alpha \bar \beta, \mu, \alpha \beta}_{1^+ \to \gamma \,1^-} \,\epsilon_{\alpha \beta}(p, \lambda_p)
 =  \epsilon_\mu^\dagger(q,\lambda_q)\, \frac{-i\,\bar p^{\bar \beta}\,\epsilon^\dagger_{\bar \alpha \bar \beta }(\bar p,
\lambda_{\bar p}) }{\sqrt{\bar p^2}} \,
\nonumber\\
&&   \times \,q_\tau \,\epsilon^{ \mu \tau \bar \alpha \sigma}\,\Bigg\{
d^{(1)}_{1^+\to \gamma \,1^-} \,\frac{q^\alpha\,p_\sigma}{(q \cdot p)}
+d^{(2)}_{1^+\to \gamma \,1^-} \left(g^\alpha_{\;\, \sigma}-\frac{q^\alpha\,p_\sigma}{(q \cdot p)}\right)
 \Bigg\}\,\frac{p^\beta\,\epsilon_{\alpha \beta}(p,\lambda_p) }{\sqrt{p^2}} \,.
\nonumber\\
\label{def-transition-amplitude-axial-b}
\end{eqnarray}
To verify that there are indeed only two independent decay parameters requires using the identity
\begin{eqnarray}
&& g_{\sigma\tau}\epsilon_{\alpha\beta\gamma\delta}=
g_{\alpha\tau}\epsilon_{\sigma\beta\gamma\delta}
+g_{\beta\tau}\epsilon_{\alpha\sigma\gamma\delta}
+g_{\gamma\tau}\epsilon_{\alpha\beta\sigma\delta}
+g_{\delta\tau}\epsilon_{\alpha\beta\gamma\sigma} \,.
\label{epsilon-property}
\end{eqnarray}
We obtain for the decay width
\begin{eqnarray}
&&\Gamma_{1^+ \to \gamma\,1^-} =\frac{1}{12\,\pi}\,
\Bigg\{
\Bigg|\frac{ d^{(1)}_{1^+\to \gamma \,1^-} }{M_{1^+} }\Bigg|^2
+\Bigg|\frac{ d^{(2)}_{1^+\to \gamma \,1^-} }{M_{1^-} }\Bigg|^2
 \Bigg\} \left(\frac{M_{1^+}^2-M_{1^-}^2}{2\,M_{1^+}} \right)^3\,,
\label{result:width-1plus:c}
\end{eqnarray}
with $\bar p^2 =M_{1^-}^2$ and $p^2=M_{1^+}^2$. As indicated by (\ref{epsilon-property}), it is quite cumbersome to
extract the two decay parameters from a given amplitude. Fortunately, this task can be streamlined considerably
by the projection identities,
\begin{eqnarray}
&& d^{(1)}_{1^+\to \gamma \,1^-} =
\frac{+i\,16\,M_{1^+}}{M_{1^-}\,(M_{1^+}^2-M_{1^-}^2)^3}\,p^{\sigma}\,q^{\tau}\,
\epsilon_{\sigma  \tau  \bar \alpha \mu }\,\bar p_{\bar \beta}\,
M^{\bar \alpha \bar \beta, \mu, \alpha \beta}_{1^+ \to \gamma\,1^-}\,q_{\alpha}\,p_\beta \,,
\nonumber\\
&& d^{(2)}_{1^+\to \gamma \,1^-} = \frac{+i\,16\,M_{1^-}}{M_{1^+}\,(M_{1^+}^2-M_{1^-}^2)^3}\,p^{\sigma}\,q^{\tau}\,
\epsilon_{\sigma  \tau\mu \alpha }\,q_{\bar \alpha}\,\bar p_{\bar \beta}\,
M^{\bar \alpha \bar \beta, \mu, \alpha \beta}_{1^+ \to \gamma\,1^-} \,p_\beta\,,
\nonumber\\
\label{result-projection-1plus}
\end{eqnarray}
resulting from the antisymmetry and transversality of the decay amplitude.

We consider the 5 classes of contributions depicted in Fig.  \ref{fig:generic-decay-a}.
Any diagram is proportional to one of the three coupling constants
$\tilde g_R, \tilde g_H$ or $\hat g_H$. In the first case the thin and thick lines
attached to the resonance vertex stand for the propagation of Goldstone bosons and vector D-mesons. In the
second case the thin lines correspond to light
vector mesons whereas the thick lines describe heavy pseudoscalar mesons. In the third case both lines stand for
vector mesons. The outgoing line is always a vector $D^*_s$-meson.

We discuss the various
contributions starting with diagrams proportional to the coupling constant $\tilde g_R$.
There are two types of gauge-invariant combinations proportional to $e\,\tilde g_P/f $. We
form the two corresponding transverse tensors $A_+^{\bar \alpha \bar \beta ,\mu,\alpha \beta}(\bar p,p)$ and
$B_+^{\bar \alpha \bar \beta ,\mu,\alpha \beta }(\bar p,p)$,
 receiving contributions from classes 1) + 2) + 4) + 5) and 2) + 3) + 4) + 5) of Fig.  \ref{fig:generic-decay-a}
respectively. The tensors $A^{\bar \alpha \bar \beta ,\mu,\alpha \beta}_{+}(\bar p,p)$ and
$B^{\bar \alpha \bar \beta ,\mu,\alpha \beta }_{+}(\bar p,p)$ describe
the processes where the photon couples to the charge of the light and heavy mesons,
\begin{eqnarray}
&& A^{\bar \alpha \bar \beta ,\mu, \alpha \beta}_{\pm, ab}(\bar p, p)=- i\, \int \frac{d^4l}{(2\pi)^4}\,S_a(l)\,
\Bigg\{ \epsilon^{\bar \alpha \bar \beta}_{\;\;\;\;\; \sigma \tau} \,\Big[
-g^{\mu \sigma}\,S_{b}^{\tau \beta }(p+l)\, p^\alpha
\nonumber\\
&& \qquad
+ \,l^\sigma\, S^{ \tau \beta}_{b}(\bar p+l)\,g^{\mu \alpha } \,
+(l+q)^\sigma\,S_a(l+q)\,S^{ \tau \beta}_{b}(p+l)\,(q+2\,l)^\mu\,p^\alpha
\Big]
\nonumber\\
&& \qquad
+\,\epsilon^{\tilde \alpha \tilde \beta}_{\;\;\;\;\; \sigma \tau}\,
  \bar p^{\bar \alpha} \,S^{\mu \bar \beta}_{\bar p,\;\;\; \tilde \alpha \tilde \beta }(p)
  \,l^\sigma \,S_{b}^{\tau \beta }(p+l)\, p^\alpha
\mp  \epsilon_{ \sigma \tau}^{\;\;\;\;\; \rho \bar \beta  }\,\Big[
-g^{\mu}_{\;\, \rho}\,\bar p^{\bar \alpha}\, S_{b}^{\sigma \tau,\beta }(p+l)\,p^\alpha
\nonumber\\
&& \qquad
+\,l_\rho\,g^{\mu \bar \alpha }\,S_{b}^{ \sigma \tau,\beta }(p+l)\,  p^\alpha
+  \bar p^{\bar \alpha}\,l_\rho\,S^{\sigma \tau,\beta }_{b}(\bar p+l)\,
 \,g^{\mu \alpha }
\nonumber\\
&& \qquad
+\,\bar p^{\bar \alpha }\,(l+q)_\rho\,
S_a(l+q)\,S^{\sigma \tau, \beta }_{b}(p+l)\,(q+2\,l)^\mu\,
p^\alpha
\Big] \Bigg\} \,,
\nonumber\\
&& B^{\bar \alpha \bar \beta ,\mu, \alpha \beta}_{\pm, ab}  (\bar p,p)=-i\, \int
\frac{d^4l}{(2\pi)^4}\,S_a (l)\,\Bigg\{ \epsilon^{\bar \alpha \bar \beta}_{\;\;\;\;\;  \sigma \tau}\,\Big[
+l_\sigma\,S_{b}^{ \mu \tau ,\beta}(p+l)\,p^\alpha
\nonumber\\
&& \qquad
+ \,l^\sigma \,S^{ \tau \beta}_{b}(\bar p+l)\, g^{\mu \alpha }
+ l^\sigma\,S_{b}^{  \tau,\mu \beta }(\bar p+l)\, p^\alpha\,
\nonumber\\
&& \qquad
+ \,l^\sigma\,g_{\bar \kappa \kappa}\,\Big\{
S_{b}^{ \tau ,\mu \bar \kappa}(\bar p+l)\,
S^{\kappa \beta }_{b}(p+l) + S_{b}^{ \tau \bar \kappa}(\bar p+l)\,S^{\mu \kappa ,\beta }_{b}(p+l)\Big\}\,
p^\alpha \Big]
\nonumber\\
&& \qquad
+\,   \epsilon^{\tilde \alpha \tilde \beta}_{\;\;\;\;\; \sigma \tau}\,
\bar p^{\bar \alpha} \,S^{\mu \bar \beta}_{\bar p,\;\;\; \tilde \alpha \tilde \beta }(p)
  \,l^\sigma \,S_{b}^{\tau \beta }(p+l)\, p^\alpha
 \mp  \epsilon_{ \sigma \tau}^{\;\;\;\;  \rho \bar \beta}\,\Big[
+l_\rho\,g^{\mu\bar \alpha }\,S_{b}^{ \sigma \tau ,\beta}(p+l)\, p^\alpha\,
\nonumber\\
&& \qquad
+\, l_\rho\,\bar p^{\bar \alpha} \,S^{\sigma \tau ,\beta}_{b}(\bar p+l)\,
g^{\mu \alpha } +l_\rho\,\bar p^{\bar \alpha }\,
S_{b}^{ \sigma \tau ,\mu \beta}(\bar p+l)\, p^\alpha\,
\nonumber\\
&& \qquad
+\,l_\rho\,\bar p^{\bar \alpha}\,g_{\bar \kappa \kappa}\,
\Big\{ S_{b}^{\sigma \tau,\bar \kappa  }(\bar p+l)\,S^{\mu \kappa, \beta }_{b}(p+l)
+ S_{b}^{ \sigma \tau ,\mu \bar \kappa }(\bar p+l)\,S^{ \kappa \beta}_{b}(p+l)\Big\}\,p^\alpha
\Big]\Bigg\}\,.
\label{def-AB-1plus:b}
\nonumber\\
\end{eqnarray}
In the tensors $A^{\bar \alpha \bar \beta ,\mu, \alpha \beta}_{\pm, ab}$ and
$B^{\bar \alpha \bar \beta ,\mu, \alpha \beta}_{\pm ,ab}$
there are contributions where the photon couples to the final vector particle. In the convention of
Fig.  \ref{fig:generic-decay-a}, these are diagrams of class 5). They
give rise to terms proportional to $S^{\bar \alpha \bar \beta,\alpha \beta}_{\bar p} (p)$, with
$m_{\bar p}^2 =\bar p^2$. Such contributions are not at odds with parity conservation since the
tensor field carries spin one quanta with both parities.  Analogous terms where the photon couples to the
initial vector meson do not arise due to parity conservation. Since the resonance field couples always
with $(\partial_\tau R^{\tau \alpha })$, only the positive parity component is accessible.

The contributions induced by interactions involving the electromagnetic field
strength tensor $F_{\mu \nu}$ are proportional to the anomalous coupling strengths
$e_A$, $e_Q$, $e_C$, $\tilde e_Q$ and $\tilde e_C$  of (\ref{def-eA}, \ref{def-eC-eQ}, \ref{def-tilde-eC-eQ})
or the parameter $\tilde e_P$ of (\ref{def-eP}).
The terms proportional to $\tilde e_P/(f\,m_V^2)$ probe the class 2) of Fig.  \ref{fig:generic-decay-a} only.
Their effect is encoded into the transverse tensor $\bar A_+^{\bar \alpha \bar \beta, \mu, \alpha \beta, }(\bar p,p)$.
The tensor $\bar B_{ab}^{\bar \alpha \bar \beta ,\mu, \alpha \beta}(\bar p, p)$
is associated with contributions probing the anomalous magnetic moment of the $D$ mesons, which are
proportional to $\tilde e_C$ and $\tilde e_Q$,
\begin{eqnarray}
&& \bar A^{\bar \alpha \bar \beta ,\mu, \alpha \beta}_{\pm ,ab}(\bar p, p)=- i\, \int \frac{d^4l}{(2\pi)^4}\,S_a(l)\,
\Bigg\{ \epsilon^{\bar \alpha \bar \beta}_{\;\;\;\;\; \sigma \tau} \,
\Big(l^\mu\,q^\sigma -(l\cdot q)\,g^{\mu \sigma} \Big)\,S_{b}^{\tau \beta }(p+l)
\nonumber\\
&& \qquad
\mp \, \epsilon_{ \sigma \tau}^{\;\;\;\;\; \rho \bar \beta  }\,
\Big(l^\mu\,q_\rho-(l \cdot q)\,g^{\mu}_{\;\, \rho}\Big)\,\bar p^{\bar \alpha}\, S_{b}^{\sigma \tau,\beta }(p+l)
\Bigg\}\, p^\alpha \,,
\nonumber\\ \label{def-barA-1plus:b}\\
&&  \bar B_{ab}^{\bar \alpha \bar \beta ,\mu, \alpha \beta}(\bar p, p)
=+i\, \int
\frac{d^4l}{(2\pi)^4}\,S_a (l)\,\Bigg\{ \epsilon^{\bar \alpha \bar \beta}_{\;\;\;\;\;  \sigma \tau}\,\Big[
l^\sigma\,q_\kappa\Big\{
S_{b}^{ \tau \kappa}(\bar p+l)\,
S^{\mu \beta }_{b}(p+l)
\nonumber\\
&& \qquad -\, S_{b}^{ \tau \mu}(\bar p+l)\,S^{\kappa \beta }_{b}(p+l)\Big\}\,
p^\alpha \Big]
 - \epsilon_{ \sigma \tau}^{\;\;\;\;  \rho \bar \beta}\,\Big[
l_\rho\,q_\kappa\,\bar p^{\bar \alpha}\,
\Big\{ S_{b}^{\sigma \tau,\kappa  }(\bar p+l)\,S^{\mu  \beta }_{b}(p+l)
\nonumber\\
&& \qquad  -\, S_{b}^{ \sigma \tau ,\mu  }(\bar p+l)\,S^{ \kappa \beta}_{b}(p+l)\Big\}\,p^\alpha
\Big]
\Bigg\}\,.
\end{eqnarray}
The additional tensors
$C_+^{\bar \alpha \bar \beta ,\mu,\alpha \beta }(\bar p,p)$ and
$\bar C^{\bar \alpha \bar \beta ,\mu,\alpha \beta}(\bar p,p)$ describe the processes where the
Goldstone boson, which emits the photon, is converted into a light vector meson as included in class 1).
They are proportional to the parameter combinations $e_A\,\tilde g_V/f^2$ and $e_A\,g_T/f^2$ respectively.
We define
\begin{eqnarray}
&& C_{\pm,abc}^{\bar \alpha \bar \beta ,\mu, \alpha \beta}(\bar p, p)
=+2\,i\, \int
\frac{d^4l}{(2\pi)^4}\,S_a (l)\,q_\nu\,\epsilon^{\mu \nu }_{\quad  \sigma \tau}\,l^\tau\,
\Bigg\{S_b^{\bar \alpha \sigma}(l+q)\,S_c^{\bar \beta \beta}(l+p)\,
p^\alpha\,
\nonumber\\
&& \qquad \pm \,\bar p^{\bar \alpha }\,g_{\bar \kappa \kappa }\,S_b^{\bar \kappa \sigma}(l+q)\,S_c^{\kappa \bar \beta, \beta}(l+p)\,
p^\alpha\,\Bigg\} \,,
\nonumber\\ \label{def-C-1plus:b}\\
&& \bar C_{abc}^{\bar \alpha \bar \beta ,\mu, \alpha \beta}(\bar p, p)
=+2\,i\, \int
\frac{d^4l}{(2\pi)^4}\,S_a (l)\,q_\nu\,\epsilon^{\mu \nu}_{\quad \sigma \tau}\,l^\tau\,
 \bar p^{\bar \alpha }\,
 \nonumber\\
&& \qquad  \times \,g_{\bar \kappa \kappa }\,S_b^{ \bar \kappa \bar \beta, \sigma}(l+q)\,S_c^{\kappa \beta}(l+p)\,
p^\alpha\, \,.
\end{eqnarray}
Anomalous processes analogous to those described by the tensors (\ref{def-C-1plus:b}) are driven by $e_Q$ and
$ e_C$. In this case the photon is emitted by a pseudoscalar D-meson, which is converted
into a vector D-meson. The contribution is included in class 3) of Fig.  \ref{fig:generic-decay-a}.
We introduce the corresponding loop tensor
\begin{eqnarray}
&& D_{abc}^{\bar \alpha \bar \beta ,\mu, \alpha \beta}(\bar p, p)
=+2\,i\, \int \frac{d^4l}{(2\pi)^4}\,S_a (l)\,
S_b(l+\bar p)\,
\nonumber\\
&& \qquad  \times l^{\bar \alpha} \,\bar p^{\bar \beta}\,q_\tau\,
\epsilon^{\mu \tau }_{\quad \rho \sigma}\,(l+\bar p)^{\sigma}\,S^{\rho\beta}_c(l+p)\,p^{\alpha} \,.
\label{def-D-1plus:b}
\end{eqnarray}
The tensor integrals introduced in
(\ref{def-AB-1plus:b}, \ref{def-barA-1plus:b}, \ref{def-C-1plus:b}, \ref{def-D-1plus:b})
are again gauge-invariant separately.

Summing all terms, we get the decay amplitude
\begin{eqnarray}
&& -i\,M^{\bar \alpha \bar \beta, \mu, \alpha \beta }_{1^+ \to \gamma
\,1^-}=
\frac{e\,\tilde g_P}{4\,f}\,h^{(1+)}_{K D^*}\, \Big\{ A_{+,KD^*}^{\bar \alpha \bar \beta,\mu,  \alpha \beta}(\bar p, p)
+B_{+,KD^*}^{\bar \alpha \bar \beta,\mu, \alpha \beta }(\bar p, p) \Big\}
\nonumber\\
&&  \;\;\;
+\frac{\tilde e_P}{4\,f\,m_V^2}\,h^{(1+)}_{K D^*}\,  \bar A_{+,KD^*}^{\bar \alpha \bar \beta,\mu,  \alpha \beta}(\bar p, p)
+\,\frac{2}{\sqrt{3}}\,\frac{e\,\tilde g_P}{4\,f}\,h^{(1+)}_{\eta D^*_s}\, B_{+,\eta D_s^*}^{ \bar \alpha \bar \beta,\mu, \alpha \beta}(\bar p, p)
\nonumber\\
&&   \;\;\;
+\,\frac{\tilde g_P}{4\,f\,M^2_V}\,\Big\{\frac{2}{\sqrt{3}}\, \frac{3\,\tilde e_C+\tilde e_Q -3\,e}{3}\,h^{(1+)}_{\eta D^*_s}\,
\bar B_{\eta  D_s^*}^{\bar \alpha \bar \beta,\mu, \alpha \beta}(\bar p, p)
\nonumber\\
&&  \qquad  \qquad \qquad \quad
+\,\frac{6\,\tilde e_C-\tilde e_Q-3\,e}{3}\,h^{(1+)}_{K D^*}\,
\bar B_{K  D^* }^{\bar \alpha \bar \beta,\mu, \alpha \beta}(\bar p, p)\Big\}
\nonumber\\
&&  \;\;\;
-\,\frac{e_A}{24\,f^2\,m_V}\,h^{(1+)}_{K D^*}\,\Big\{
\tilde g_V\, C_{+,K K^*D^*}^{\bar \alpha \bar \beta,\mu, \alpha \beta }(\bar p, p)
+ g_T\, \bar C_{K K^*D^*}^{\bar \alpha \bar \beta,\mu, \alpha \beta }(\bar p, p) \Big\}
\nonumber\\
&&   \;\;\;
-\,\frac{e_A}{\sqrt{3}\,6\,f^2\,m_V}\,h^{(1+)}_{\eta D^*_s}\,
\Big\{ \tilde g_V\,C_{+,\eta  \phi D_s^*}^{\bar \alpha \bar \beta,\mu, \alpha \beta }(\bar p, p)
+ g_T\,\bar C_{\eta  \phi D_s^*}^{\bar \alpha \bar \beta,\mu, \alpha \beta }(\bar p, p) \Big\}
\nonumber\\
&& \;\;\;
-\,\frac{g_P}{f\,M^2_V}\,\Big\{\frac{2}{\sqrt{3}}\, \frac{3\,e_C-e_Q }{6}\,h^{(1+)}_{\eta D^*_s}\,
D_{\eta  D_s D_s^*}^{\bar \alpha \bar \beta,\mu, \alpha \beta}(\bar p, p)
\nonumber\\
&&  \qquad \qquad \quad
+\,\frac{6\,e_C+e_Q}{6}\,h^{(1+)}_{K D^*}\,
D_{K D D^* }^{\bar \alpha \bar \beta,\mu, \alpha \beta}(\bar p, p)\Big\}\,.
\label{decay-amplitude-1plus-c1}
\end{eqnarray}

We turn to the contributions induced by the coupling constants $\tilde g_H$ and $\hat g_H$
introduced in (\ref{def-gH}). The terms proportional to $\hat g_H$ are deferred to Appendix G. We
discuss those proportional to $\tilde g_H$. The former are expected to be more relevant than the latter due
to phase space. Appendix G provides in addition explicit results for the Passareno-Veltman reduction
of the tensor integrals (\ref{def-AB-1plus:b}-\ref{def-D-1plus:b}, \ref{def-ABCD-1plus:cextra}).

The evaluation of terms proportional to $\tilde g_H$ is similar to those proportional to $g_R$.
Formally the role of light and heavy intermediate lines in Fig.  \ref{fig:generic-decay-a} is interchanged so that the tensor integrals formed in (\ref{def-AB-1plus:b}-\ref{def-D-1plus:b})
will occur again. Additional tensors are required to describe the implications of
$\tilde e_T \neq 0$  as introduced in (\ref{def-eV-eT-eE}). The latter give rise to
contributions of class 2) in Fig.  \ref{fig:generic-decay-a}. They are analogous to the contributions proportional to
$\tilde e_P$ in (\ref{decay-amplitude-1plus-b}). We form the
gauge-invariant tensor $\tilde B^{\mu, \alpha \beta  }_{}(\bar p,p)$
\begin{eqnarray}
&& \tilde B^{\bar \alpha \bar \beta ,\mu, \alpha \beta}_{ab}  (\bar p,p)=+\,i\, \int
\frac{d^4l}{(2\pi)^4}\,S_a (l)\, \bar p^{\bar \alpha}\,l_\rho\,
\epsilon^{\rho \bar \beta}_{\;\;\;\;\;  \bar \sigma \tau}\,
S_{b}^{ \sigma \tau ,\beta}(p+l)\,
\nonumber\\
&& \qquad \qquad \qquad \qquad \times \Big( g^{\mu }_{\;\;\sigma}\,q^{\bar \sigma}
-  g^{\mu \bar \sigma}\,q^\sigma \Big)\,p^\alpha \,.
\label{def-ABCD-1plus:cextra}
\end{eqnarray}

The contributions proportional to $\tilde g_H$ are
\begin{eqnarray}
&& -i\,M^{\bar \alpha \bar \beta, \mu, \alpha \beta }_{1^+ \to \gamma\,1^-}=
-\frac{e\,(g_E+\tilde g_T)\,\tilde g_H}{8\,f^2}\, \Big\{ A_{+,DK^*}^{\bar \alpha \bar \beta,\mu,  \alpha \beta}(\bar p, p)
+B_{+,DK^*}^{\bar \alpha \bar \beta,\mu, \alpha \beta }(\bar p, p) \Big\}
\nonumber\\
&& \quad \;\;\;
-\,\frac{e\,(g_E-\tilde g_T)\,\tilde g_H}{8\,f^2}\, \Big\{ A_{-,DK^*}^{\bar \alpha \bar \beta,\mu,  \alpha \beta}(\bar p, p)
+B_{-,DK^*}^{\bar \alpha \bar \beta,\mu, \alpha \beta }(\bar p, p) \Big\}
\nonumber\\
&& \quad \;\;\;
-\frac{\tilde e_T\,\tilde g_H}{4\,f^2\,m_V^2}\,\tilde B_{DK^*}^{\bar \alpha \bar \beta,\mu, \alpha \beta }(\bar p, p)
\nonumber\\
&& \quad \;\;\;
-\,\frac{e\,\tilde g_H}{8\,f^2}\, \Big\{(g_E+\tilde g_T)\,
A_{+,D_s \phi }^{ \bar \alpha \bar \beta,\mu, \alpha \beta}(\bar p, p) +
(g_E-\tilde g_T)\,A_{-,D_s \phi }^{ \bar \alpha \bar \beta,\mu, \alpha \beta}(\bar p, p) \Big\}
\nonumber\\
&& \quad \;\;\;
+\,\frac{(e_C+e_Q/6)\,\tilde g_H}{4\,f^2\,M^2_V}\, \Big\{ (\tilde g_V+g_T)\,C_{+,D D^*K^*}^{\bar \alpha \bar \beta,\mu, \alpha \beta }(\bar p, p)
+2\,\tilde g_V\, \bar C_{D D^*K^*}^{\bar \alpha \bar \beta,\mu, \alpha \beta }(\bar p, p)
\nonumber\\
&&  \;\;\; \qquad \qquad \qquad +\,(\tilde g_V-g_T)\,C_{-,D D^*K^*}^{\bar \alpha \bar \beta,\mu, \alpha \beta }(\bar p, p) \Big\}
\nonumber\\
&& \quad \;\;\;
+ \frac{(e_C-e_Q/3)\,\tilde g_H}{8\,f^2\,M^2_V}\, \Big\{ (\tilde g_V+g_T)\,C_{+,D_s D_s^*\phi}^{\bar \alpha \bar \beta,\mu, \alpha \beta }(\bar p, p)
+2\,\tilde g_V\, \bar C_{D_s D_s^*\phi}^{\bar \alpha \bar \beta,\mu, \alpha \beta }(\bar p, p)
\nonumber\\
&& \;\;\; \qquad \qquad \qquad +(\tilde g_V-g_T)\,  C_{-,D_s D_s^*\phi}^{\bar \alpha \bar \beta,\mu, \alpha \beta }(\bar p, p) \Big\}
\nonumber\\
&& \quad \;\;\;  +\,\frac{e_A\,g_P\,\tilde g_H}{12\,f^3\,m_V}\,
D_{D K K^* }^{\bar \alpha \bar \beta,\mu, \alpha \beta}(\bar p, p)
+\frac{e_A\,g_P\,\tilde g_H}{9\,f^3\,m_V}\,
D_{D_s \eta \phi  }^{\bar \alpha \bar \beta,\mu, \alpha \beta}(\bar p, p ) \,.
\label{decay-amplitude-1plus-c2}
\end{eqnarray}


\section{Numerical results}

We confront the results of the previous sections with the experimental data on
the radiative and strong decays of the scalar and axial-vector $D_s$-mesons reviewed in the introduction.

Our prediction of
$140$ keV for the isospin-violating strong width
$D_{s0}^*(2317)^\pm \, \rightarrow \, D_{s}(1968)^\pm\, \pi^0$
is compatible with the empirical bound $\Gamma < 3.8$ MeV  \cite{Aubert2} but this comparison does not provide any significant constraint on the underlying coupled-channel dynamics.
The present upper limit
on the ratio of the radiative to pionic decay width \cite{PDG:2006}
\begin{eqnarray}
\frac{\Gamma \,[D_{s0}^*(2317)^\pm \, \rightarrow \,  D^*_{s}(2112)^\pm\, \gamma]}
{\Gamma\,[D_{s0}^*(2317)^\pm \, \rightarrow \, D_{s}(1968)^\pm\, \pi^0]}
<\, 0.059
\label{Dstar0widthtoDss6}
\end{eqnarray}
implies for the decay constant $d_{0^+ \to \gamma\, 1^-}$
defined in (\ref{result:width-0plus}) the inequality,
\begin{eqnarray}
\Big| d_{0^+ \to \gamma\, 1^-} \Big|\, < \,0.117 \,{\rm GeV}^{-1}\,,
\label{d-0plus-exp}
\end{eqnarray}
using our predicted value for the $D_{s0}^*\rightarrow \, D_{s}\, \pi^0$
 decay width (140 keV).

The total width of the D$_{s1}^*$(2460)$^\pm$ meson is less than 3.5 MeV
\cite{Aubert2}. Our prediction of $140$ keV for the isospin-violating strong decay width
$D_{s1}^*(2460) \, \rightarrow \, D_{s}^*(2112)\, \pi^0$
is compatible with that upper bound but this is again not very significant. The constraints on the radiative decays of the D$_{s1}^*$(2460)$^\pm$ to the D$_{s}$(1968)$^\pm$,
the D$_{s}^*$(2112)$^\pm$  and the D$_{s0}^*$(2317)$^\pm$ given in (\ref{Dstar1widthtoDsfirst}), (\ref{Dstar1widthtoDstarfirst}) and (\ref{Dstar1widthtoDstar0first})
imply the following relations for
the decay constants $d_{1^+ \to \gamma\, 0^-}$,
 $d_{1^+ \to \gamma\, 0^+}$ and $d^{(1,2)}_{1^+ \to \gamma\, 1^-} $ introduced in
 (\ref{result:width-1plus:a}), (\ref{result:width-1plus:b}) and (\ref{result:width-1plus:c}),
\begin{eqnarray}
&& \Big| d_{1^+ \to \,\gamma\, 0^-} \Big|\, = \,0.138\,^{+0.012}_{-0.014} \,{\rm GeV}^{-1}\,, \qquad \quad
 \Big| d_{1^+ \to \,\gamma\, 0^+} \Big|\, < \,0.665 \,{\rm GeV}^{-1} \,,
\nonumber\\ \nonumber\\
&& \qquad \quad \quad \Big| d^{(1)}_{1^+ \to \, \gamma\, 1^-} \Big|^2
+ (1.164 )^2\,\Big| d^{(2)}_{1^+ \to \gamma\, 1^-} \Big|^2 < \Big(0.39\Big)^2\, .
\label{d-1plus-exp}
\end{eqnarray}

\begin{table}[b]
\begin{center}
\begin{tabular}{|c||l l l||}
\hline
 &$10\,\times \,d_{0^+ \to \gamma\, 1^-}$[GeV$^{-1}$]& &\\
\hline
\hline
$ K D$ &$+2.371\,g_P +3.036\,e_P $ &
$-2.252\,(e_C+e_Q/6)\,\tilde g_P$ &
$-( 6.095\,\tilde g_T+0.950\,g_E)\,e_A$        \\
\hline
$\eta D_s$  &$+6.822\,g_P             $ &
$-1.018\,(e_C-e_Q/3)\,\tilde g_P$ &
$-( 14.93\,\tilde g_T+3.575\,g_E)\,e_A$            \\
\hline
\hline
 &$10\,\times \,d_{1^+ \to \gamma\, 0^-}$[GeV$^{-1}$]& &\\
\hline
\hline
$ K D$ &$+1.641\,g_P +2.920\,e_P $ &
$+2.272\,(\tilde e_C-\tilde e_Q/6)\,g_P$ &
$+( 0.500\,\tilde g_T+0.501\,g_E)\,e_A$        \\
\hline
$\eta D_s$  &$+3.445\,g_P             $ &
$+1.222\,(\tilde e_C+\tilde e_Q/3)\,g_P$ &
$+(3.149\,\tilde g_T+0.883\,g_E)\,e_A$            \\
\hline
\hline
 &$10\,\times \,d_{1^+ \to \gamma\, 0^+}$[GeV$^{-1}$]& &\\
\hline
\hline
$ K D$ & &
$+2.104\,(e_C+e_Q/6)$ &        \\
\hline
$\eta D_s$  & &
$+0.891\,(e_C-e_Q/3)$ &
           \\
\hline
\end{tabular}
\caption{Contributions to the decay constants $d_{0^+ \to \gamma\, 1^-}$ and $d_{1^+ \to \gamma\, 0^+}$ implied by
(\ref{decay-amplitude-0plus}), (\ref{decay-amplitude-1plus-a1}) and (\ref{decay-amplitude-1plus-b}).
We use the coupling constants of Table \ref{tab:hadronic-decay}, $f=90$ MeV, $m_V=776$ MeV and $M_V=2000$ MeV.}
\label{tab:decay-parameters:1}
\end{center}
\end{table}

It is interesting to confront the empirical constraints (\ref{d-0plus-exp}, \ref{d-1plus-exp}) to
the predictions of heavy-quark symmetry. To leading order, the heavy-quark symmetry implies the
relations
\begin{eqnarray}
&& M_c\,d_{0^+ \to \,\gamma\, 1^-} = M_c\,d_{1^+ \to \,\gamma\, 0^-} =
d^{(1)}_{1^+ \to \,\gamma\, 1^-} =  d^{(2)}_{1^+ \to \,\gamma\, 1^-}  \equiv  d\,,
\label{hqs-decay-parameters}
\end{eqnarray}
where $M_c \simeq M_V$ is a typical mass of a charmed meson.
The result (\ref{hqs-decay-parameters})
can be derived from  \cite{Mehen-Springer-2004} by a matching of corresponding decay amplitudes.
In the heavy-quark mass limit the parameter $d$ scales linearly with the charm quark mass. The value given in
(\ref{d-1plus-exp}) for $\Big| d_{1^+ \to \,\gamma\, 0^-} \Big|$ suggests the range  $ 0.25 < |d| < 0.30  $ for $M_c =M_V= 2000$ MeV. This value is barely
compatible with the bound (\ref{d-0plus-exp}) and requires that $\Big|d_{0^+ \to \gamma\,1^-}\Big|$ be close to that upper bound.

\begin{table}[b]
\begin{center}
\begin{tabular}{|c||l l l||}
\hline
\hline
 &$10\,\times \,d^{(1)}_{1^+ \to \gamma\, 1^-} $& &\\
\hline
\hline
$ K D$ &$+5.147\,\tilde g_P +7.347\,\tilde e_P $ &
$+0.221\,(e_C + e_Q/6)\,g_P$ &
$+(8.544\,g_T- 1.458\,\tilde g_V)\,e_A$        \\
&  & $+3.214\,(\tilde e_C - \tilde e_Q/6)\,\tilde g_P$&  \\
\hline
$\eta D_s$  &$+11.92\,\tilde g_P             $ &
$-0.227\,(e_C- e_Q/3)\,g_P$ &
$+( 26.85\,g_T- 5.914\,\tilde g_V)\,e_A$    \\
&& $+1.756\,(\tilde e_C+ \tilde e_Q/3)\,\tilde g_P$  &        \\
\hline
\hline
 &$10\,\times\,d^{(2)}_{1^+ \to \gamma\, 1^-} $& &\\
\hline
\hline
$  K D$ &$+2.799\,\tilde g_P +6.305\,\tilde e_P $ &
$+2.153\,(e_C + e_Q/6)\,g_P$ &
$-(8.802\,g_T- 1.335\,\tilde g_V)\,e_A$        \\
&& $+0.0650\,(\tilde e_C - \tilde e_Q/6)\,\tilde g_P$& \\
\hline
$\eta D_s$  &$+8.852\,\tilde g_P             $ &
$+0.489\,(e_C-e_Q/3)\,g_P$ &
$-( 21.62\,g_T- 3.469\,\tilde g_V)\,e_A$            \\
&& $+0.080\,(\tilde e_C+\tilde e_Q/3)\,\tilde g_P$& \\
\hline
\end{tabular}
\vglue 0.3 true cm
\caption{Contributions to the decay parameters $d^{(1)}_{1^+ \to \gamma\, 1^-}$ and $d^{(2)}_{1^+ \to \gamma\, 1^-}$ implied by
(\ref{decay-amplitude-1plus-c1}).
We use the same coupling constants as in Table \ref{tab:decay-parameters:1}}
\label{tab:decay-parameters:2}
\end{center}
\end{table}

We discuss the electromagnetic decay parameters obtained using the expressions derived in Sections 4 and 5. The contributions
from the $K D$ and $\eta D_s$ channels are shown in Table \ref{tab:decay-parameters:1} for radiative transitions between scalar and vector states and in Table \ref{tab:decay-parameters:2} for the radiative decay of the $1^+$ to the $1^-$ state. The first column displays contributions involving the 4-point vertices (\ref{gP-gauging}) and (\ref{def-eP}). The second column shows the effect of anomalous processes and the third column contributions induced by the light vector mesons as intermediate states.

We consider the first column of these tables and try to reproduce the constraints
(\ref{d-0plus-exp}) and (\ref{d-1plus-exp}) using these terms only and the empirical value $g_P \simeq 0.57$.
This exercise leads immediately to the conclusion that $g_P$ and $e_P$ must have opposite signs.
Adding constructively the contributions from the $K D$ and $\eta \,D_s$ channels would produce
decay parameters that are much too large.
We recall that $e_P$ parameterizes gauge-invariant 4-point vertices describing the process where
a D-meson emits a photon and a charged Goldstone boson simultaneously. Such interactions must be taken into account
in effective field theories. The parameter $e_P$ can be varied to achieve
consistency with (\ref{d-0plus-exp}) and (\ref{d-1plus-exp}). To fit the $1^+ \to \gamma \,0^-$ decay, the range of
values for $e_P$ is limited to the intervals
$-0.57 < e_P < -0.49$ and $-1.50  < e_P <-1.42 $.  The constraints on the process $1^+ \to \gamma \,1^-$ suggest
$ -1.54 < \tilde e_P < -0.84$ for $g_P=\tilde g_P =0.57 $. From the $0^+$ decay we deduce
$-2.08 < e_P <-1.37$. With $e_P =\tilde e_P \simeq -1.50$
we arrive at an acceptable scenario, given the assumption of Section 3 that the chiral
power assigned to the $e_P$ vertex is promoted from order $Q^3_\chi$ to order $Q^2_\chi$.
The corresponding decay parameters are collected in
the first column of Table \ref{tab:result-decay-constant-A}. We observe that this scenario, though compatible
with the constraints
(\ref{d-0plus-exp}) and (\ref{d-1plus-exp}), is characterized by decay parameters in significant disagreement
with the heavy-quark symmetry relation (\ref{hqs-decay-parameters}).

A value of $e_P$ of the order of $-1.5$ has many consequences.
As mentioned in the discussion following (\ref{eP-hqs}), this coupling is unnaturally large and points to the effectiveness of the theory. It also induces very important cancellations in the $KD$ channel and between the $KD$ and the $\eta D_s$ channels.  The $D^*_{s0}(2317)\to \gamma\,D^*_{s}(2112)$
transition is actually dominated by the contribution of the $\eta D_s$ channel. The corresponding radiative width is 2.85 keV
and the calculated value for the ratio (\ref{Dstar0widthtoDstarfirst}) is 0.02. The comparison with other
coupled-channel calculations is not easy as the mechanism driving the $D^*_{s0}(2317)\to \gamma\,D^*_{s}(2112)$ transition is not simple. Our radiative width is larger than the value of 0.49 keV obtained in \cite{Gamermann3} but we note that similar destructive effects between channels are observed in both approaches. Our result is quantitatively comparable
 to the radiative width of $\sim$1 keV found in \cite{Gutsche1}
in a molecular picture, despite the fact that the $\eta D_s$ channel is not taken into account in \cite{Gutsche1} and important in our work as well as in \cite{Gamermann3}. This particular example illustrates the need for a consistent
description of the spectroscopic properties of the $D_s$-mesons guided by general principles such as strong interaction symmetries. For the decays $D^*_{s1}(2460)\to \gamma\,D_{s}(1968)$,
$D^*_{s1}(2460)\to \gamma\,D_{s0}(2317)$ and $D^*_{s1}(2460)\to \gamma\,D^*_{s}(2112)$, we obtain partial widths of 50 keV, 0 and 18 keV respectively (assuming always a strong width of 140 MeV).
We stress that performing a formal
expansion of the full expressions for the decay parameter in the inverse charm-quark mass leads to results that are
compatible with the expectation from the heavy-quark symmetry. With phenomenological charm quark masses, the decay parameters obtained for the $K D$ channel
distort the pattern implied by the relation (\ref{hqs-decay-parameters}) obtained in the limit of infinitely heavy charm
quarks. Such breaking pattern is even larger in the $\eta D_s$ channel. We note also from the numbers provided in
Tables \ref{tab:decay-parameters:1} and \ref{tab:decay-parameters:2} that the contribution proportional to $g_P$ is larger in the $\eta D_s$ channel than in the $K D$ channel. This is because there is only one contribution in the $\eta D_s$
channel while there are two contributions of opposite sign in the $K D$ channel corresponding to the graphs where the photon couples to the $K^+$ or to the
$D^+$.

\begin{table}[t]
\begin{center}
\begin{tabular}{|c||c c cc||}
\hline
& I)& II) &III)&IV)\\
\hline
$d_{0^+ \to \gamma\, 1^-}$[GeV$^{-1}$]  &$ +0.069$&  $+0.035$ &$-0.003$ & $+0.073$\\
\hline
$d_{1^+ \to \gamma\, 0^-}$[GeV$^{-1}$]  &$ -0.148 $& $-0.130$ &$-0.120$&$-0.139$\\
\hline
$d_{1^+ \to \gamma\, 0^+}$[GeV$^{-1}$]  &$ 0 $&  $+0.055$ &$+0.055$& $+0.055$\\
\hline
$d^{(1)}_{1^+ \to \gamma\, 1^-}$  &$-0.129$& $-0.097$ &$-0.105$& $-0.090$\\
\hline
$d^{(2)}_{1^+ \to \gamma\, 1^-}$  &$-0.282$& $-0.251$ &$-0.251$& $-0.251$\\
\hline
\end{tabular}
\caption{Decay constants that are implied by  $e_P=\tilde e_P= -1.50 $  and $g_H=\tilde g_H =\hat g_H=0$.
The values used for $e_A$, $e_C=\tilde e_C$ and $e_Q=\tilde e_Q$ are discussed in the text. The four
scenarios are characterized by
I) $e_A=0$ and $e_Q=e_C =0$,
II) $e_A=0$,
III)
$e_A>0$ and
IV) $e_A <0$.
We use  $f=90$ MeV, $m_V=776$ MeV, $M_V=2000$ MeV.
}
\label{tab:result-decay-constant-A}
\end{center}
\end{table}

We display the effect of the anomalous contributions in the second column of Tables \ref{tab:decay-parameters:1} and \ref{tab:decay-parameters:2}.
Without light vector mesons as explicit degrees of freedom, these terms
determine entirely the $1^+ \to \gamma\,0^+$ process. According to Appendix A, we should take $e_C \simeq  0.13$ and
adjust the value of $e_Q$ to the channel under consideration ($e_Q \simeq  0.91$ for the
$K D$ channel and $e_Q \simeq 0.52$ for the $\eta D_s$ channel). Given
the coupling constants of Table \ref{tab:hadronic-decay}, we predict
$d_{1^+ \to \gamma\,0^+ } = 0.055$ GeV$^{-1}$, a result compatible with the bound of (\ref{d-1plus-exp}).

We discuss now the effects induced by the presence of light vector mesons when they do not couple directly to the scalar and axial-vector
molecules ($g_H=\tilde g_H = \hat g_H=0$) but influence their radiative decays as intermediate states through the anomalous
vertex introduced in (\ref{def-eA}). This contribution is proportional to $e_A$
and given in the third column of Tables \ref{tab:decay-parameters:1} and \ref{tab:decay-parameters:2}.
The parameter $e_A$ exhibits significant flavour SU(3) breaking as indicated in (\ref{value-eA}). To take
it into account we use
\begin{eqnarray}
&&  e_A^{(\eta \,D_s)}= e^{(\phi \to \gamma\,\eta )}_A  \simeq \pm 0.053\,,
\nonumber\\
&&e_A^{(K \,D)} = 2\,e^{(K^{*0} \to \gamma\,K^0)}_A - e^{(K^{*+} \to \gamma\,K^+)}_A \simeq \pm 0.148 \,,
\label{value-eA-result}
\end{eqnarray}
in the $\eta D_s$ and $K D$ channels respectively. The phase of the
parameter $e_A$ is not determined by experiment. We consider both signs and
provide the numerical values of the electromagnetic decay parameters in the third and fourth columns of Table \ref{tab:result-decay-constant-A} for $e_A>$0 and $e_A<$0.
The phases and the size of the parameters $g_T,\, g_E,\, \tilde g_T$ and $\tilde g_V$
were estimated in Section 3 from the assumption of universally coupled light vector mesons,
together with the ansatz of a combined heavy-quark and flavour SU(4) symmetry. We used the values $\tilde g_V \simeq 0.71$,
$g_E =  g_P  $ and $g_T =\tilde g_T =0.5\, m_V\,g_P/M_V $ with $g_P=0.57$. We caution that these numbers were
derived with significant approximations.
As can be seen in Table \ref{tab:result-decay-constant-A},
the sign of $e_A$ matters mostly for the radiative decay width of the $D^*_{s0}(2317)$: this width is extremely small for $e_A>$0 (5.4 eV) and much larger for $e_A<$0 (3.2 keV). An accurate measurement of the $D^*_{s1}(2460)\to \gamma\,D^*_{s}(2112)$ decay width would therefore be most helpful in determining whether the Lagrangian (\ref{def-eA}) leads to constructive or destructive interference effects in that quantity.
We note that the
$K^+ \,D^0$ and $K^0\,D^+$ channels contribute to the decay amplitudes with opposite signs,
the contribution of the $K^0\,D^+$ channel being twice as large as the contribution of the $K^+\, D^0$ channel.

We enlarge the discussion of the role played by light vector mesons by assuming that the $K^*D^*$ and $\phi\, D_s^*$ channels can couple directly to the scalar and
axial-vector molecules and the $K^*D$ and $\phi\, D_s$ channels to the axial-vector molecules through the interactions defined in (\ref{def-gH}). In the heavy-quark mass limit, the coupling constants are equal, i.e. $\tilde g_H=g_H= \hat g_H\neq 0$ .

We consider first the transition $1^+ \to \gamma \,0^+$ between molecules. The corresponding decay parameter reads
\begin{eqnarray}
&& 10\,\times \,d_{1^+\to \gamma\,0^+} [{\rm GeV}^{-1}]= \hat g_H\,g_H\,\Big[ 1.187 - 0.377 \Big]
\nonumber\\
&& \qquad -\,  \tilde g_H\,g_H\,\Big[5.363\, (e_C+e_Q/6)\,+5.607\,(e_C-e_Q/3)\,\Big]
\nonumber\\
&& \qquad +\, e_A^{(K D)} \,\Big[0.429\, g_H + 9.244\,\tilde g_H \Big]+
e_A^{(\eta D_s)} \,\Big[ 1.704\,g_H+ 13.62\,\tilde g_H \Big]\,,
\label{1plus-0plus-VV}
\end{eqnarray}
where we separate the contributions from channels involving the $D$ and $D_s$ mesons. In the first term, the numbers $1.187$ and $-0.377$ represent the contributions from the $\phi\, D_s^*$ and the $K^* D^*$ channels respectively. Similarly the contributions proportional to
$e_C+e_Q/6 $ and $e_C-e_Q/3 $ reflect the $K^* D \to K^* D^*$
and $\phi\, D_s \to \phi\,D_s^*$ transitions. With the values of $e_A$, $e_Q$ and $e_C$ discussed above,
we may use (\ref{1plus-0plus-VV}) and (\ref{d-1plus-exp}) to constrain the parameters $g_H, \tilde g_H$ and $\hat g_H$. In the heavy-quark limit
($ g_H=\tilde g_H= \hat g_H$), we derive the conditions
\begin{eqnarray}
-10.5 < g_H < 2.18\,, \qquad \qquad -2.18 < g_H < 10.5 \,,
\label{result-gH}
\end{eqnarray}
for the positive and negative values of $e_A$ respectively. Clearly the result (\ref{result-gH}) does not
provide a strong constraint on the size of the parameter $g_H$.

We turn to the radiative transitions of the scalar and axial-vector molecules to $0^-$ and $1^-$ ground states.
We display in Tables \ref{tab:decay-parameters:3} and \ref{tab:decay-parameters:4} the
contributions of the channels involving light
vector mesons to these decays. The terms shown in Table \ref{tab:decay-parameters:3} correspond to the expressions
(\ref{decay-amplitude-0plus-AppendixD}) and (\ref{decay-amplitude-1plus-a2}) (obtained in Appendix D
and in Section 5) for the $0^+ \rightarrow \gamma 1^-$ transition
and (\ref{decay-amplitude-1plus-AppendixE}) derived in Appendix D for the $1^+ \rightarrow \gamma 0^-$ transition. The terms quoted in Table \ref{tab:decay-parameters:4} correspond to the expressions
(\ref{decay-amplitude-1plus-c2}) and  (\ref{decay-amplitude-1plus-AppendixG}) (obtained in Section 5
and in Appendix G) for the $1^+ \rightarrow \gamma 1^-$ transition. They involve the additional parameters
$e_T, \tilde e_T, e_V, \tilde e_V$ and $e_E,\tilde e_E$ introduced in (\ref{def-eV-eT-eE}).
Clearly, it is not possible to determine these numbers
by the constraints (\ref{d-0plus-exp}) and (\ref{d-1plus-exp}) only.

\begin{table}[t]
\begin{center}
\begin{tabular}{|c||l l ||}
\hline
 &$10\,\times \,d_{0^+ \to \gamma\, 1^-}$[GeV$^{-1}$]& \\
\hline
\hline
$ K^* D^*$ &$+10.66\,g_T +2.976\,\tilde g_V $ &
$+1.758\,e_T+2.343\,\tilde e_E -2.765\,\tilde e_V $ \\
$\sim g_H:$ & $-27.78\,e_A\,\tilde g_P$  &
$+(e_C+e_Q/6)\,( 3.408\,\tilde g_T +0.611\,g_E)$ \\
&&$+(\tilde e_C-\tilde e_Q/6)\,(2.283\,\tilde g_T +0.653\,\tilde g_V)$\\
\hline
$\phi \;D^*_s$  &$+29.69\,g_T +8.369\,\tilde g_V $ &
$+3.091\,\tilde e_E  $ \\
$\sim g_H:$ & $-60.24\,e_A\,\tilde g_P$  &
$+(e_C-e_Q/3)\,(1.383 \,\tilde g_T +0.322\,g_E)$\\
&&$+(\tilde e_C+\tilde e_Q/3)\,(1.278\,\tilde g_T +0.592\,\tilde g_V)$\\
\hline
 &$10\,\times \,d_{1^+ \to \gamma\, 0^-}$[GeV$^{-1}$]& \\
\hline
\hline
$ K^* D$ &$+0.852\,g_V $ &
$+6.902\, e_E -6.127\,e_V $ \\
$\sim \tilde g_H:$ &   &
$+(e_C+e_Q/6)\,(6.777\,\tilde g_T +1.337\,g_E)$ \\
\hline
$\phi \;D_s$  &$-2.902\,g_V $ &
$+6.462\,e_E  $ \\
$\sim \tilde g_H:$ &   &
$+(e_C-e_Q/3)\,(3.548\,\tilde g_T +0.847\,g_E)$ \\
\hline
$ K^* D^*$ &$+9.798\,\tilde g_T +1.603\,g_E $ &
$+3.338\,\tilde e_T$ \\
$\sim \hat g_H:$ & $-9.070\,e_A\,g_P$  &
$+1.568\,(e_C+e_Q/6)\,g_V $ \\
&&$-(\tilde e_C-\tilde e_Q/6)\,(2.705\,\tilde g_T+0.547\,g_E) $\\
\hline
$\phi \;D^*_s$  &$+16.92\,\tilde g_T +3.807\,g_E $ &
 \\
$\sim \hat g_H:$ & $-15.44\,e_A\, g_P$  &
$+1.072\,(e_C-e_Q/3)\,g_V$ \\
&&$-(\tilde e_C+\tilde e_Q/3)\,(1.478\,\tilde g_T+0.359\,g_E) $\\
\hline
\end{tabular}
\caption{Contributions to the decay parameters $d_{0^+ \to \gamma\, 1^-}$ and $d_{1^+ \to \gamma\, 0^-}$ implied by the light vector meson couplings as given by
(\ref{decay-amplitude-0plus-AppendixD}, \ref{decay-amplitude-1plus-a2}, \ref{decay-amplitude-1plus-AppendixE}).
The various terms have to be multiplied by either $g_H, \tilde g_H$ or $\hat g_H$ as indicated in the first column.}
\label{tab:decay-parameters:3}
\end{center}
\end{table}

\begin{table}[t]
\begin{center}
\begin{tabular}{|c||l l ||}
\hline
 &$d^{(1)}_{1^+ \to \gamma\, 1^-}$& \\
\hline
\hline
$ K^* D$ &$+1.024\,\tilde g_T+0.152\,g_E $ &
$+0.618\, \tilde e_T  $ \\
$\sim \tilde g_H:$ & $+2.049\,e_A\,g_P$  &
$-(e_C+e_Q/6)\,(2.930\,g_T +0.483\,\tilde g_V)$ \\
\hline
$\phi \;D_s$  &$+2.674\,\tilde g_T +0.597\,g_E$ &
 \\
$\sim \tilde g_H:$ & $+5.714\,e_A\,g_P$  &
$-(e_C-e_Q/3)\,(1.505\,g_T +0.147\,\tilde g_V)$ \\
\hline
$ K^* D^*$ &$+1.065\,g_T +0.569\,\tilde g_V $ &
$+0.528\,e_T- 0.979\,\tilde e_V +1.069\,\tilde e_E$\\
$\sim \hat g_H:$ & $-5.914\,e_A\,\tilde g_P$  &
$-(e_C+e_Q/6)\,(0.209\,\tilde g_T +0.037\,g_E) $ \\
&& $+(\tilde e_C-\tilde e_Q/6)\,(0.439\,g_T +0.002\,\tilde g_V) $\\
\hline
$\phi \;D^*_s$  &$+2.597\, g_T +0.746\,\tilde g_V $ &
$+1.328\,\tilde e_E$ \\
$\sim \hat g_H:$ & $-10.99\,e_A\, \tilde g_P$  &
$-(e_C-e_Q/3)\,(0.047\,\tilde g_T+0.011\,g_E)$ \\
&& $+(\tilde e_C+\tilde e_Q/3)\,(0.260\,g_T +0.058\,\tilde g_V) $\\
\hline
 &$d^{(2)}_{1^+ \to \gamma\, 1^-}$& \\
\hline
\hline
$ K^* D$ &$+0.683\,\tilde g_T+0.091\,g_E $ &
$+0.521\, \tilde e_T  $ \\
$\sim \tilde g_H:$ & $-6.998\,e_A\,g_P$  &
$+(e_C+e_Q/6)\,(1.567\,g_T -0.202\,\tilde g_V)$ \\
\hline
$\phi \;D_s$  &$+2.276\,\tilde g_T +0.504\,g_E$ &
 \\
$\sim \tilde g_H:$ & $-13.99\,e_A\,g_P$  &
$+(e_C-e_Q/3)\,(0.793\,g_T -0.190\,\tilde g_V)$ \\
\hline
$ K^* D^*$ &$+1.512\,g_T +0.446\,\tilde g_V $ &
$+0.669\,e_T- 0.907\,\tilde e_V+0.881\,\tilde e_E$ \\
$\sim \hat g_H:$ & $+1.435\,e_A\,\tilde g_P$  &
$+(e_C+e_Q/6)\,(0.247\,\tilde g_T +0.044\,g_E) $ \\
&& $-(\tilde e_C-\tilde e_Q/6)\,(0.886\,g_T -0.037\,\tilde g_V) $\\
\hline
$\phi \;D^*_s$  &$+2.820\, g_T +0.587\,\tilde g_V $ &
$+1.041\,\tilde e_E $ \\
$\sim \hat g_H:$ & $+2.858\,e_A\, \tilde g_P$  &
$+(e_C-e_Q/3)\,(0.157\,\tilde g_T+0.037\,g_E)$ \\
&& $-(\tilde e_C+\tilde e_Q/3)\,(0.443\,g_T -0.014\,\tilde g_V) $\\
\hline
\end{tabular}
\caption{Contributions to the decay parameters $d^{(1)}_{1^+ \to \gamma\, 1^-}$ and $d^{(2)}_{1^+ \to \gamma\, 1^-}$ implied by the light vector meson couplings as given by
(\ref{decay-amplitude-1plus-c2}, \ref{decay-amplitude-1plus-AppendixG}).
The various terms have to be multiplied by either $\tilde g_H$ or $\hat g_H$ as indicated in the first column.}
\label{tab:decay-parameters:4}
\end{center}
\end{table}

To achieve a qualitative understanding of these terms,
we assume that all parameters are correlated as dictated by the heavy-quark mass limit (i.e.
$e_P= \tilde e_P, e_V = \tilde e_V$,...). In this limit, there remain five unknown parameters $g_H$, $e_P$, $e_V$, $e_T$ and
$e_E$. One may expect to learn little from the constraints (\ref{d-0plus-exp}) and (\ref{d-1plus-exp}) only. However, this
is not quite so. For specific values of $g_H$ and $e_P$, one can always adjust the values for $e_V$, $e_T$ and $e_E$ so as
to reproduce a given value of the decay parameter $d_{0^+ \to \gamma\,1^-}$
and the partial decay widths of the process $1^+\to \gamma\,0^-$ and $1^+ \to \gamma\,1^-$ derived in (\ref{d-0plus-exp}) and (\ref{d-1plus-exp}).
We assume as before $\tilde g_V \simeq 0.71$, $g_E =  g_P $ and $g_T =\tilde g_T = 0.5\,m_V\,g_P/M_V $ with
$g_P=0.57$.
The requirement that the
parameters $e_V, e_T$ and $e_E$  be real, as implied by charge conjugation symmetry, defines stringent
conditions on the ranges of values allowed for
$g_H$ and $e_P$. Note that the latter are determined by quadratic equations, the solution of which
involves a square root. The constants $e_V$, $e_T$ and $e_E$ be real numbers only if the argument of that square root is positive. For a given value of $e_P$, we derive the condition  that the parameter $g_H$
has to be confined in a small interval
\begin{eqnarray}
g_H^{crit,-} < g_H <  g^{crit,+}_H \,.
\label{def-gHcrit}
\end{eqnarray}

We discriminate between the scenarios implied by positive or negative values of the parameter $e_A$, for which we find respectively
\begin{eqnarray}
&& g_H^{crit,\pm } \simeq  -0.390+1.332\,d_{0^+ \to \gamma\,1^-} [{\rm GeV}^{-1}] -0.242\,e_P\pm 0.131\,,
\nonumber\\
&& g_H^{crit,\pm } \simeq  -0.285+0.751\,d_{0^+ \to \gamma\,1^-} [{\rm GeV}^{-1}] -0.137\,e_P\pm 0.074\,.
\label{value-gHcrit}
\end{eqnarray}
We checked the stability of this result against
reasonable variations of the parameters $g_V,g_T$ and $g_E$. In (\ref{value-gHcrit}) we allow the decay
parameter $d_{0^+ \to \gamma\,1^-}$ to take any
value. The result (\ref{value-gHcrit}) is very significant: for any reasonable
range of $e_P$ it requires the coupling constant $g_H$  to be quite small, typically $|g_H|< 0.5$. This
justifies in retrospect the coupled-channel computation of Section 2, which assumed that the light vector
mesons are not relevant for the formation of the scalar and axial-vector $D_s$ molecules.

\begin{table}[t]
\begin{center}
\begin{tabular}{|c||c c ||}
\hline
& I)& II) \\
\hline
$d_{0^+ \to \gamma\, 1^-}$[GeV$^{-1}$]  &$ +0.057$&  $+0.104$ \\
\hline
$d_{1^+ \to \gamma\, 0^-}$[GeV$^{-1}$]  &$ -0.139 $& $-0.140$ \\
\hline
$d_{1^+ \to \gamma\, 0^+}$[GeV$^{-1}$]  &$ -0.043 $&  $+0.091$ \\
\hline
$d^{(1)}_{1^+ \to \gamma\, 1^-}$  &$+0.303$& $+0.278$ \\
\hline
$d^{(2)}_{1^+ \to \gamma\, 1^-}$  &$+0.202$& $+0.071$ \\
\hline
\end{tabular}
\caption{Decay constants obtained using
$e_P=e_V=e_T=e_E=0 $ and $\tilde e_P=\tilde e_V=\tilde e_T=\tilde e_E=0 $.
In set I) we use $e_A>0$ with $g_H=\tilde g_H = \hat g_H =-0.46$.
Set II) assumes $e_A <0$ and $g_H=\tilde g_H = \hat g_H =-0.25$.
We assume  $f=90$ MeV, $m_V=776$ MeV, $M_V=2000$ MeV.
}
\label{tab:result-decay-constant-B}
\end{center}
\end{table}

The light vector mesons may nevertheless change the radiative decay para\-meters significantly
because of cancellations between terms.
 Our main result is that the introduction of light vector mesons leads to a very consistent picture. We obtain values for all decay parameters that are compatible with
the empirical constraints using vanishing values for all gauge-invariant counter terms $e_P=e_V=e_T=e_E=0$ and
$\tilde e_P=\tilde e_V=\tilde e_T=\tilde e_E=0$.
In such a scenario there is one free parameter only,
$g_H=\tilde g_H = \hat g_H$, which can be dialed to recover all empirical constraints. The results for the
positive and negative $e_A$ scenarios are collected in Table \ref{tab:result-decay-constant-B}. It is interesting to
observe that we predict a negative decay constant for the $1^+ \to \gamma\,0^-$ decay contradicting the
naive expectation of heavy-quark symmetry. We emphasize that this follows even though performing a formal
expansion of the full expressions for the decay parameter in the inverse charm-quark mass leads to results
compatible with the expectation from the heavy-quark symmetry. Such an expansion assumes for instance
$m_\phi \ll M_c$, which is not realized too well in nature. Our results provide a physical justification
for promoting the counter terms proportional to $e_P$ and $\tilde e_P$ to chiral order $Q^2_\chi$ rather
than the expected power $Q^3_\chi$. Once the light vector mesons are introduced as important physical degrees of freedom,
the naturalness assumption for the residual size of $e_P$ and $\tilde e_P$ appears justified.

We emphasize that it does not appear possible to predict precise values for the decay parameters.
The results of Table \ref{tab:result-decay-constant-B} should be viewed as possible and natu\-ral scenarios.
We provide therefore the radiative widths associated with the decay constants of Table \ref{tab:result-decay-constant-B}, assuming the strong width of the D$_{s0}^*$(2317)$^\pm$ and D$_{s1}^*$(2460)$^{\pm}$ to be 140 keV, as a mere indication of their expected range. For the scenarios labeled I and II, we find $\Gamma_{0^+ \to \gamma\,1^-}$
= 1.94 and 6.47 keV, $\Gamma_{1^+ \to \gamma\,0^-}$=44.50 and 45.14 keV, $\Gamma_{1^+ \to \gamma\,0^+}$=0.13 and 0.59 keV and $\Gamma_{1^+ \to \gamma\,1^-}$=21.8 and 12.47 keV respectively.
Precise unquenched lattice QCD simulations for the hadronic coupling constants of the Goldstone bosons and
light vector mesons to the D-mesons would be very helpful.


\section{Summary}

Based on the chiral Lagrangian properties of scalar and axial-vector meson molecules with open-charm
content were studied. Chiral correction terms were incorporated systematically in the coupled-channel dynamics,
where we relied on constraints from large-$N_c$ QCD and the heavy-quark symmetry. We focused on the
$D^*_{s0}(2317)$ and $D^*_{s1}(2460)$ states and computed
their isospin-violating hadronic and electromagnetic decay widths. In order to
establish manifestly gauge-invariant results for the electromagnetic decay parameters
the spin-one particles
were represented in terms of antisymmetric tensor fields rather than by the more conventional
vector fields. The role of explicit light vector mesons in the radiative decays was investigated.

The main findings of this work for the strong and radiative decay widths of the $D^*_{s0}(2317)$ and $D^*_{s1}(2460)$ states can be summarized as follows.

The hadronic isospin-violating decay widths of both the  $D^*_{s0}(2317)$ and $D^*_{s1}(2460)$ states are
predicted to be 140 keV. Chiral corrections to order $Q_\chi^2$ lead to a significant enhancement of these widths. We point out the importance of treating consistently the $\eta \, \pi $ mixing and the isospin-mixing effects in the $KD$ system. The observed upper limits on these strong widths of 3.8 and 3.5 MeV respectively do not provide any useful constraint on the theory yet.

The radiative decay parameters of the $D^*_{s0}(2317) \to \gamma\,D_s^*$ and
$D^*_{s1}(2460) \to \gamma\,D_s, \gamma\,D_s^*$ and $D^*_{s1}(2460) \to \gamma\,D^*_{s0}(2317) $ were computed
in the hadrogenesis conjecture. They involve only one-loop diagrams. We find that both the $\eta D_s$ and $\eta \,D_s^*$ channels contribute significantly and that large cancellations occur between the different channels.
The results are compatible with all empirical constraints if one gauge-invariant
contact term is considered to be more important than expected from a naive naturalness assumption. The decay parameters
are subject to a cancellation mechanism, which makes a  prediction of their precise values unreliable at the moment.

 When considering light vector mesons as explicit degrees of freedom coupled to the scalar and axial-vector molecules, radiative decay parameters that are compatible
with the empirical constraints, can be obtained without envoking subleading contact terms. We understand this effect as an indication of the dynamical role of light vector mesons in the electromagnetic transition processes.
We predict that the $D^*_{s0}(2317)$ and $D^*_{s1}(2460)$ have small coupling strength to the
$K^*D^*$ and $\phi\, D_s^*$ channels despite the decisive role played by these degrees of freedom in the radiative decay processes.

Finally we point out that the invariant $\eta \,D^*$ invariant mass distribution shows a
signal of a member of an exotic axial-vector sextet at a mass of 2568 MeV and with a width of 18 MeV. While that state
decouples from the $\pi D^*$ spectrum, its heavy-quark partner defines a narrow dip at a mass of 2410 MeV and with a width of 2 MeV
in the  $\pi\, D$ mass distribution.

{\bfseries{Acknowledgments}}

M.F.M.L. acknowledges useful discussions with J. Hofmann, E.E. Kolomeitsev, St. Leupold, M.A. Nowak and
A. Semke. We acknowledge the support of the European Community-Research Infrastructure Activity under
the FP6 "Structuring the European Research Area" programme
(Hadron Physics, contract number RII3-CT-2004-506078).


\section*{Appendix A}

The modulus of the coupling constant $g_P$ appearing in the Lagrangian (\ref{def-gP}) can be determined from the measured strong D$^*$-meson decay widths \cite{PDG:2006}
\begin{eqnarray}
&& \Gamma_{D^{*+}\to D^0\,\pi^+} = (65.0 \pm 14.9) \,{\rm keV} \,,
\nonumber\\
&& \Gamma_{D^{*+}\to D^+\,\pi^0} = (29.5 \pm 6.8)\,{\rm keV} \,.
\label{Dstar-decays-empirical}
\end{eqnarray}
Using
\begin{eqnarray}
&& \Gamma_{D^{*+}\to D^0\,\pi^+} = \frac{|g_P|^2}{12\,\pi}\,\frac{q_{cm}^3}{f^2}
\,,\qquad \qquad q_{cm}= 39.60 \,{\rm MeV} \,,
\nonumber\\
&& \Gamma_{D^{*+}\to D^+\,\pi^0} = \frac{|g_P|^2}{24\,\pi}\,\frac{q_{cm}^3}{f^2}
\,,\qquad \qquad q_{cm}= 38.30 \,{\rm MeV} \,,
\label{determination-gP}
\end{eqnarray}
leads to
\begin{eqnarray}
|g_P| = 0.57 \pm 0.07 \,.
\label{value-gP-Appendix}
\end{eqnarray}

The modulus of the coupling constants
$e_Q$ and $e_C$ introduced in the Lagrangian (\ref{def-eC-eQ}) can be determined from the observed branching ratios for the radiative decays $D^{*+}\to D^+\,\gamma $, $D^{*0}\to D^0\,\gamma $ and
$D^{*+}_s\to D^+_s\,\gamma $ \cite{PDG:2006}. The total widths of the $D^{*0}$ and
of the $D^{*+}_s$ are not known. They are the sums of the strong and radiative widths. The strong widths associated with $\pi^0$ emission,
\begin{eqnarray}
&& \Gamma_{D^{*0}\to D^0\,\pi^0} = \frac{|g_P|^2}{24\,\pi}\,\frac{q_{cm}^3}{f^2} \simeq 42 \,{\rm keV}
\,,\qquad \qquad q_{cm} = 43.12 \,{\rm MeV} \,,
\nonumber\\
&& \Gamma_{D^{*+}_s\to D^+_s\,\pi^0} = \frac{|\epsilon\,g_P|^2}{18\,\pi}\,\frac{q_{cm}^3}{f^2}\,
\simeq \epsilon^2\,77 \,{\rm keV}
\,,\qquad\;  q_{cm}= 47.87 \,{\rm MeV} \,,
\label{determination-gP-width}
\end{eqnarray}
 are calculated using the value of $|g_P|$ derived above. $\epsilon$ is the $\pi_0 \eta$ mixing angle (\ref{eps-def}).
From the data on the $D^{*+}\rightarrow D^+ \gamma$ partial width and the known branching ratios of the $D^{*0}$ and $D^{*+}_s$ into the $\pi^0 D^0$ and $\pi^0 D^+_s$
channels \cite{PDG:2006}, we obtain
\begin{eqnarray}
&& \Gamma_{D^{*+}\to D^+\,\gamma} = \,(1.5 \pm 0.8 )\,{\rm keV} \,,
\nonumber\\
&& \Gamma_{D^{*0}\to D^0\,\gamma} = (\Gamma_{D^{*0}\to D^0\,\gamma}+42\,{\rm keV})\times (0.381 \pm 0.029 )
\nonumber\\
&& \qquad \qquad \,= \,(26.0 \pm 3) \,{\rm keV} \,,
\nonumber\\
&& \Gamma_{D^{*+}_s\to D^+_s\,\gamma} = (\Gamma_{D^{*+}_s\to D^+_s\,\gamma}+\epsilon^2\,77\,{\rm keV})\times (0.942 \pm 0.007 )
\nonumber\\
&& \qquad \qquad \,= \epsilon^2\,(1328 \pm 180) \,{\rm keV} \,.
\label{D-gamma-decay}
\end{eqnarray}
From the Lagrangian (\ref{def-eC-eQ}), we derive
\begin{eqnarray}
&& \Gamma_{D^{*+} \to D^+\,\gamma} =  \frac{M_{D^{*+}}^2}{4\,\pi}\,\frac{(e_Q-3\,e_C)^2}{9\,M_V^4} \,q^3_{cm} \,,\qquad
  q_{cm}= 135.78 \,{\rm MeV} \,,
\nonumber\\ \nonumber\\
&& \Gamma_{D^{*0} \to D^0\,\gamma} = \frac{M_{D^{*0}}^2{4\,\pi}}\,\frac{(2\,e_Q+3\,e_C)^2}{9\,M_V^4} \,q^3_{cm} \,,
 \qquad  q_{cm}= 137.16 \,{\rm MeV} \,,
\nonumber\\ \nonumber\\
&& \Gamma_{D^{*+}_s \to D^+_s\,\gamma}
= \frac{M_{D^{*+}_s}^2}{4\,\pi}\,\frac{(e_Q-3\,e_C)^2}{9\,M_V^4} \,q^3_{cm} \,,
\;\, \qquad  q_{cm}= 138.91 \,{\rm MeV} \,.
\label{}
\end{eqnarray}
From the first two constraints of  (\ref{D-gamma-decay}), we get
\begin{eqnarray}
&& |e_Q-3\,e_C| \simeq 0.537^{+0.128}_{-0.170} \,,\qquad
 |2\,e_Q+3\,e_C| \simeq 2.201^{+0.124}_{-0.131}\,,
\nonumber\\ \nonumber\\
i.e. && \qquad e_Q  =0.91 \pm 0.10  \,, \qquad e_C =0.13 \pm 0.05\,,
\label{values-A-eQ-eC}
\end{eqnarray}
where we assume $M_V=2000$ MeV. Using (\ref{values-A-eQ-eC}) and taking $\epsilon\simeq 0.01 $ in the third constraint overestimates
the $D^{*+}_s\to D^+_s \gamma$ decay rate by an order of magnitude ($\sim$1.9 keV rather than  $\sim$ 0.13 keV).
This may be a signal of SU(3) breaking effects in $e_Q$ or suggest the need for an increased
value of $\epsilon $. For $\epsilon =0.01$ and $e_C = 0.13 $, an effective $e_Q \simeq 0.52$ reproduces the
empirical width.

The modulus of the coupling constant $e_A$ appearing in (\ref{def-eA}) is determined from the radiative decay of the
light vector mesons, $K^{*0} \to K^0 \gamma,
K^{*\pm} \to K^\pm \gamma $ and $\phi\to \eta\gamma $.
From the Lagrangian (\ref{def-eA}), we obtain
\begin{eqnarray}
&&\Gamma_{K^{*\pm} \to K^\pm \gamma}  = \frac{|e_A|^2}{24\times 144\,\pi}\,\frac{m^5_{K^{*\pm}}}{m^2_V\,f^2} \,
 \left(1- \frac{m_{K^\pm}^2}{m_{K^{*\pm}}^2} \right)^3\,,
\nonumber \\
&&\Gamma_{K^{*0} \,\to K^0\, \gamma}  =  \frac{|e_A|^2}{24\times \;\;36\,\pi}\,\frac{m^5_{K^{*0}}}{m^2_V\,f^2}  \, \left(1- \frac{m_{K^0}^2}{m_{K^{*0}}^2} \right)^3\,,
\nonumber \\
&& \Gamma_{\;\phi \;\to \;\eta \;\gamma} \;\, =  \frac{|e_A|^2}{24\times \;\;54\,\pi}\,\frac{m^5_{\phi}}{m^2_V\,f^2}  \,
 \left(1- \;\frac{m_{\eta}^2}{m_{\phi}^2} \;\right)^3\,,
\label{vec-to-gam-decays}
\end{eqnarray}
where we will assume $f=90$ MeV and $m_V = 776$ MeV.
The experimental decay widths \cite{PDG:2006} imply conflicting values for
$e_A$. We find
\begin{eqnarray}
&& \Gamma_{K^{*0}\, \to \,K^0 \gamma}\, =\, (116 \, \pm \, 10) \, {\rm keV} \,, \qquad \to \qquad |e_A| =0.119 \pm 0.006 \,,
\nonumber\\
&& \Gamma_{K^{*\pm} \to K^\pm \gamma}\, =\, (50 \, \pm \, 5) \,{\rm  keV}\,, \qquad \quad \to \qquad |e_A| = 0.090 \pm 0.004\,,
\nonumber\\
&&\Gamma_{\phi \;\to\; \eta\;\gamma\;}\, =\, (55.38 \, \pm \, 1.68) \,{\rm  keV} \,, \quad\, \to \qquad |e_A| = 0.053 \pm 0.001 \,.
\label{widthetagamma}
\end{eqnarray}


\section*{Appendix B}

To work out the implications of the heavy-quark symmetry of QCD, it is useful to introduce auxiliary
fields, $P_\pm(x)$ and $P^\mu_\pm(x)$ such that
\begin{eqnarray}
&& D(x) \;\;\,\,\!= e^{-i\,(v\cdot x) \,M_c}\,P_{+}(x) +e^{+i\,(v\cdot x) \,M_c}\,P_{-}(x)\,,
\nonumber\\
&& D^{\mu \nu}(x)=
i\,e^{-i\,(v\cdot x) \,M_c}\,\Big\{v^\mu\,P^\nu_{+}(x)-v^\nu \,P^\mu_{+}(x)
+ \frac{i}{M_c}\,\Big( \partial^\mu P^\nu_+-\partial^\nu P^\mu_+\Big)\Big\}
\nonumber\\
&& \qquad \quad \;\,+\,i\,e^{+i\,(v\cdot x) \,M_c}\,\Big\{v^\mu\,P^\nu_{-}(x)-v^\nu \,P^\mu_{-}(x)
- \frac{i}{M_c}\,\Big( \partial^\mu P^\nu_--\partial^\nu P^\mu_-\Big)\Big\}\,,
\label{non-relativistic-expansion}
\end{eqnarray}
with a 4-velocity normalized by $v^2=1$. The mass $M_c$ is an ave\-raged value of the pseudoscalar and
vector charmed meson ground state masses. The fields $P_\pm(x)$ and $P_\pm^\mu(x)$ are therefore varying slowly
in space and time and their derivatives
are small compared to $M_c\,v_\alpha\,P$. In the limit $M_c \to \infty$, these derivatives can be neglected.

We rewrite the interaction terms introduced in (\ref{def-gP}, \ref{def-eP}, \ref{def-gV-gT}) in  terms of the
auxiliary fields. Restricting the Lagrangian density to the 'plus' components and dropping that index for simplicity, we have
\begin{eqnarray}
{\mathcal L}&=& i\,\frac{g_P\,M_c}{f}\,\Big\{P_{\mu }\,(\partial^\mu
\Phi )\,\bar P  -  P \,(\partial^\mu
\Phi)\,\bar P_{\mu } \Big\}
+ i\,\frac{\tilde g_P\,M_c}{f}\,v_\mu\,\epsilon^{\mu \nu \alpha \beta}\,
P_{\nu}\,(\partial_\alpha \Phi)\,
 \bar P_{ \beta}
\nonumber\\
&-&\frac{g_V\,M_c}{f}\, v_\mu\, P \, (\partial_\alpha V^{\alpha \mu}) \,\bar P
+\frac{\tilde g_V\,M_c}{f}\, v_\mu\,P_{  \nu} \,(\partial_\alpha V^{\alpha \mu})\,\bar P^\nu
\nonumber\\
&+& i\,\frac{\tilde g_T\,M_c^2}{4\,f}\,v_\alpha \,\epsilon^{\mu \nu \alpha \beta}\,\Big\{
P_{ \beta}\,V_{\mu \nu } \, \bar P - P\,V_{\mu \nu } \,\bar P_{\beta}\Big\}\,
+ i\,\frac{g_T\,M_c^2}{2\,f}\,P_{\mu} \,V^{\mu \nu} \, \bar P_{\nu}
\nonumber\\
&+&\frac{e_P\,M_c}{f}\,F^{\mu \nu}\,\Big\{P_{\mu }\,\big[(\partial_\nu
\Phi ),\,Q\big]\,\bar P  -  P \,\big[(\partial_\nu
\Phi ),\,Q\big]\,\bar P_{\mu } \Big\}
\nonumber\\
&+& \frac{\tilde e_P\,M_c}{f}\,F_{\mu \nu}\,v_\tau\,\epsilon^{\tau \alpha \mu \beta}\,
P_{\alpha}\,\big[(\partial_\nu \Phi),\,Q\big]\,
 \bar P_{ \beta}
\nonumber\\
&+& \cdots \,,
\label{rewrite-def-gP-gV-gT}
\end{eqnarray}
where the ellipses stand for additional terms involving derivatives of the soft fields $P=P_+$ and $P^\mu=P_+^\mu$ and the 'minus' components. In deriving (\ref{rewrite-def-gP-gV-gT})
we used the equation of motion of the
auxiliary field $P^\mu_\pm(x)$,
\begin{eqnarray}
&&\mp i\,M_c\,v_\mu \, P^\mu_\pm + \partial_\mu \, P^\mu_\pm =0\,,
\label{rewrite-eom-tensor}
\end{eqnarray}
obtained from the equation of motion of the field $D^{\mu \nu}$,
\begin{eqnarray}
&& \partial^\mu \partial_\alpha D^{\alpha \nu}-\partial^\nu \partial_\alpha D^{\alpha \mu} +M_c^2\,D^{\mu \nu} =0\,.
\label{eomtensor}
\end{eqnarray}

The QCD action depends linearly on the charm-quark mass. The effective Lagrangian (\ref{rewrite-def-gP-gV-gT})
should therefore be proportional to $M_c$.
The coupling constants $g_{P,V}$ and $\tilde g_{P,V}$ must then approach a finite value in the limit of infinite
charm-quark mass.  The same
argument requires the coupling constants $g_T$ and $\tilde g_T$ to scale with $1/M_c$. The fact that there is
no term proportional to $g_E$ displayed in (\ref{rewrite-def-gP-gV-gT}) signals that the corresponding
interaction term is at least subleading in the heavy-quark mass expansion. Note that this observation is
compatible with a possible finite and non-zero value of the asymptotic value of $g_E$ in the heavy-quark mass limit.

In the limit of infinite quark mass, the fields $P_\pm$ and $P_\pm^\mu$ may be combined into a
multiplet involving the $1^-$ and $0^-$ fields. Therefore the properties of pseudoscalar and
vector states should be closely related. We follow the formalism developed in
\cite{Wise92,YCCLLY92,BD92,Jenkins94,Casalbuoni} and introduce the multiplet field $H$,
\begin{eqnarray}
&& H = \frac{1}{2}\,\Big( 1 + \vslash \Big)\,\Big(\gamma_\mu\,P^\mu_+  +i\, \gamma_5\,P_+ \Big)
\nonumber\\
&& \bar H = \gamma_0\,H^\dagger\,\gamma_0 =
\Big(P^\dagger_{+,\mu} \,\gamma^\mu +P^\dagger_+ \,i\, \gamma_5 \Big)\,\frac{1}{2}\,\Big( 1 + \vslash \Big) \,,
\nonumber\\
&& P^\mu_+ \,v_\mu =0 \, , \qquad  v^2=1\,,
\label{phase-convention}
\end{eqnarray}
in terms of which the interaction Lagrangian can be recast. Let S denote a transformation of the
heavy-quark spin symmetry group SU(2)$_v$ whose elements are
characterized by the 4-vector $\theta^\alpha$ with $\theta \cdot v=0$.
According to \cite{Georgi:1990}
under such a transformation the field H transforms as
\begin{eqnarray}
&&H \to  e^{-i\,S_\alpha\,\theta^\alpha}\,H
\,, \qquad \quad
\bar H \to  \gamma_0\,(e^{-i\,S_\alpha\,\theta^\alpha}\,H)^\dagger\,\gamma_0 =
\bar H\,e^{+i\,S_\alpha\,\theta^\alpha}\,,
\nonumber\\
&& S^\alpha =\frac{1}{2}\,\gamma_5\, [\vslash, \gamma^\alpha] \,,
\qquad \;\;S^\dagger_\alpha \, \gamma_0=\gamma_0\,S_\alpha \,, \qquad \;\;
[\vslash, S_\alpha ]_- = 0 \,.
\label{spin-rotation}
\end{eqnarray}
Under a Lorentz transformation characterized by the antisymmetric tensor $\omega_{\mu \nu}$,
the spinor part of the fields transforms as
\begin{eqnarray}
&& H \to e^{i\,S_{\mu \nu}\,\omega^{\mu \nu}}\,H\,e^{-i\,S_{\mu \nu}\,\omega^{\mu \nu}}\,, \qquad \quad
\bar H \to e^{i\,S_{\mu \nu}\,\omega^{\mu \nu}}\,\bar H\,e^{-i\,S_{\mu \nu}\,\omega^{\mu \nu}}\,,
\nonumber\\
&& S_{\mu \nu} = \frac{i}{4}\,[\gamma_\mu, \gamma_\nu]\,.
\end{eqnarray}
Only the combinations where Dirac matrices are to the right of the field $H$ or to the left of the field
$\bar H$ are invariant under the spin group $SU_v(2)$. It is straightforward to construct those terms that can be matched
to the structures given in (\ref{rewrite-def-gP-gV-gT}). We introduce
\begin{eqnarray}
&&{\mathcal L} = -\,\frac{f_P}{2}\,{\tr } \Big( H  \,(\partial^\mu \Phi)\,\gamma_5\,\gamma_\mu \,\bar H \Big)
 +i\,\frac{\bar f_P}{2}\,F^{\mu \nu} \,{\tr } \Big(H\,\gamma_5\,\gamma_\mu\,\Big[(\partial_\nu \,\Phi),\,Q\Big]\,\bar H \Big)
\nonumber\\
&& \qquad
 - \,\frac{f_V}{2}\,{\tr } \,\Big( H  \,(\partial_\alpha V^{\alpha \mu})\,\gamma_\mu\,\bar H \Big)
 + i\,\frac{\bar f_V}{2}\,F^{\mu \nu} \,{\tr } \,\Big( H \,\gamma_\mu \,\Big[(\partial^\alpha V_{\alpha \nu}),\,Q\Big]\,\bar H \Big)
\nonumber\\
&& \qquad +\,\frac{f_T}{8}\,{\tr } \,\Big(H \,V^{\mu \nu}\,\sigma_{\mu \nu} \,\bar H \Big)
-i\,\frac{\bar f_T}{8}\,{\tr } \,F^{\mu}_{\;\; \nu}\,\Big(H \,\big[V^{\nu \alpha},\,Q\big]\,\sigma_{\mu \alpha} \,\bar H \Big)
\,,
\nonumber\\
&& \qquad +\, \frac{\bar f_E}{4}\,F^{\mu \nu}\,{\tr } \,\Big(H \,\{V^{\mu \nu},\,Q\} \,\bar H \Big),
\label{spin-symmetric}
\end{eqnarray}
where we note that the field $H$ is a three-dimensional row in flavour space, each of its components
consisting of a 4-dimensional Dirac matrix. We have
\begin{eqnarray}
{\tr } \gamma_5\,\gamma_\mu \,\gamma_\nu \,\gamma_\alpha \,\gamma_\beta =
-4\,i\,\epsilon_{\mu \nu \alpha \beta }\,,
\label{convention-epsilon}
\end{eqnarray}
in our convention. Matching the expressions (\ref{rewrite-def-gP-gV-gT}, \ref{spin-symmetric})
we obtain
\begin{eqnarray}
&& \frac{g_P\,M_c}{f} = \frac{\tilde g_P\,M_c}{f} =f_P\,, \qquad \quad
\frac{e_P\,M_c}{f\,m_V^2} = \frac{\tilde e_P\,M_c}{f\,m_V^2} =\bar f_P \,,
\nonumber\\
&& \frac{g_V\,M_c}{f} = \frac{\tilde g_V\,M_c}{f} =f_V \,, \qquad \quad
\frac{e_V\,M_c}{f\,m_V^2} = \frac{\tilde e_V\,M_c}{f\,m_V^2} =\bar f_V \,,
\nonumber\\
&& \frac{g_T\,M^2_c}{f} = \frac{\tilde g_T\,M^2_c}{f} =f_T\,, \qquad \quad
\frac{e_T\,M^2_c}{f} = \frac{\tilde e_T\,M^2_c}{f} =\bar f_T\,,
\nonumber\\
&&\frac{e_E\,M^2_c}{f} = \frac{\tilde e_E\,M^2_c}{f} =\bar f_E \,.
\label{match-g-tildeg}
\end{eqnarray}

We turn to the terms (\ref{def-eC-eQ}, \ref{def-tilde-eC-eQ}). Applying the ansatz (\ref{non-relativistic-expansion})
we derive
\begin{eqnarray}
&& {\mathcal L}_{e.m.} =
\frac{i}{2} \,F^{\mu \nu} \,v^\alpha \,\epsilon_{\mu \nu \alpha \beta}\,\Big\{
P^{ \beta} \,\Big(e_C+e_Q\,Q\Big)\,\bar P
-P\,\Big(e_C+e_Q\,Q\Big)\,\bar P^{\beta}  \Big\}
\nonumber\\
&& \qquad \;\,+\, i \,F^{\mu \nu} \,P_{ \mu} \,\Big( -\tilde e_C+\tilde e_Q\,Q\Big)\,
\bar P_{\nu } + \cdots \,.
\label{rewrite-def-eQ-eC}
\end{eqnarray}
where we identify $M_c = M_V$ for convenience. Like in (\ref{rewrite-def-gP-gV-gT}),  we drop in (\ref{rewrite-def-eQ-eC})
additional irrelevant terms. The interaction terms displayed in (\ref{rewrite-def-eQ-eC}) are readi\-ly
reproduced using the heavy-quark multiplet field $H$. We follow \cite{Hai-Yang-Cheng-1993,Stewart-1998} and write
\begin{eqnarray}
&& {\mathcal L}_{e.m.}=
+\frac{e_Q}{4}\,F_{\mu \nu} \,{\tr } H\,Q\,\sigma_{\mu \nu}\,\bar H
+\frac{e_C}{4}\,F_{\mu \nu} \,{\tr } \sigma_{\mu \nu}\,H\,\bar H \,,
\label{expect-eQ-eC}
\end{eqnarray}
where the term proportional to $e_Q$ respects the heavy-quark symmetry, but the term proportional to $e_C$ breaks it.
We expect $|e_Q| \gg |e_C|$ since the parameter $e_C$ should vanish in the heavy-quark limit.
The requirement that (\ref{rewrite-def-eQ-eC}) and (\ref{expect-eQ-eC}) agree yields the desired  relations
\begin{eqnarray}
\tilde e_C = e_C \,, \qquad \qquad  \tilde e_Q = e_Q \,.
\label{result-tilde eQ-eC}
\end{eqnarray}
Note that the empirical values (\ref{values-A-eQ-eC}) confirm the expectation $|e_Q| \gg |e_C|$.

Finally we provide an analysis of the effective resonance interaction terms (\ref{def-gR}, \ref{def-gH}).
Assuming a decomposition  analogous to (\ref{non-relativistic-expansion}) for the resonance fields $R$ and $R_{\mu \nu}$,
we introduce a resonance multiplet field
\begin{eqnarray}
&& S = \frac{1}{2}\,\Big( 1 + \vslash \Big)\,\Big(\gamma_5\,\gamma_\mu\, R^\mu_+  +R_+ \Big) \,,
\nonumber\\
&& \bar S = \gamma_0\,S^\dagger\,\gamma_0 =
\Big(R^\dagger_{+,\mu} \,\gamma_5\,\gamma_\mu +R^\dagger_+ \Big)\,\frac{1}{2}\,\Big( 1 + \vslash \Big)\,, \qquad
v^\mu \,R^\mu_+ =0\,,
\label{phase-convention-resonance}
\end{eqnarray}
where the field $S$ transforms like the field $H$ under a spin rotation (see \ref{spin-rotation}).
At leading order in a heavy-quark mass expansion, the interaction terms (\ref{def-gR}, \ref{def-gH})  are described by
\begin{eqnarray}
&&{\mathcal L} = \frac{f_H}{4}\,{\tr }  \Big( S \, \gamma_\mu \,( \partial_\tau V^{\tau \mu})\,\bar H\
+H \,\gamma_\mu \,( \partial_\tau V^{\tau \mu})\,\bar S \Big)
\nonumber\\
&& \quad + \,\frac{f_R}{4}\,{\tr }  \Big( S \, \gamma_5\,\gamma_\mu\,(\partial^\mu \Phi)\,\bar H
+H \,\gamma_5\,\gamma_\mu \,(\partial^\mu \Phi)\,\bar S \Big)\,,
\label{spin-symmetric2}
\end{eqnarray}
which implies the identifications
\begin{eqnarray}
&& \frac{g_R\,M_c}{f} = \frac{\tilde g_R\,M_c}{f}= f_R \,, \qquad
 \frac{g_H\,M^2_c}{f} = \frac{\tilde g_H\,M^2_c}{f}= \frac{\hat g_H\,M^2_c}{f} =f_H\,.
\label{match-g-tildeg:b}
\end{eqnarray}


\section*{Appendix C}

All tensor integrals will be decomposed into scalar objects of the form
\begin{eqnarray}
&& I_{ab}= -i\,\int
\frac{d^4l}{(2\pi)^4}\,S_a (l)\,S^{}_{b}(l+p)\,, \qquad
 \bar I_{ab}= -i\,\int
\frac{d^4l}{(2\pi)^4}\,S_a (l)\,S^{}_{b}(l+\bar p)\,,
\nonumber\\
&& J_{abc}= +i\,\int
\frac{d^4l}{(2\pi)^4}\,S_a (l)\,S^{}_{b}(l+q)\,S^{}_{c}(l+p)\,,
\nonumber\\
&& \bar J_{abc}= +i\,\int
\frac{d^4l}{(2\pi)^4}\,S_a (l)\,S^{}_{b}(l+\bar p)\,S^{}_{c}(l+p)\,.
\label{def-master-loops}
\end{eqnarray}
We introduce the two integrals $J_{abc}$ and $\bar J_{abc}$ even though $J_{abc}=\bar J_{cba}$ as this redundancy will be used for a consistency check of the numerical simulation.
The integrals $I_{ab}$ and $\bar I_{ab}$ are ultraviolet divergent and read (see(\ref{i-def}))
\begin{eqnarray}
&& I_{ab}(s)=\frac{1}{16\,\pi^2}
\left( \frac{p_{ab}}{\sqrt{s}}\,
\left( \ln \left(1-\frac{s-2\,p_{ab}\,\sqrt{s}}{m_b^2+m_a^2} \right)
-\ln \left(1-\frac{s+2\,p_{ab}\sqrt{s}}{m_b^2+m_a^2} \right)\right)
\right.
\nonumber\\
&&\qquad \qquad + \left.
\left(\frac{1}{2}\,\frac{m_b^2+m_a^2}{m_b^2-m_a^2}
-\frac{m_b^2-m_a^2}{2\,s}
\right)
\,\ln \left( \frac{m_b^2}{m_a^2}\right) +1 \right)+I_{ab}(0) \,,
\nonumber\\
&& p^2_{ab} =\frac{s-\left(m_a+m_b \right)^2}{4\,s}
\left(s-\left(m_a-m_b \right)^2\right) \,,
\label{i-def-appendix}
\end{eqnarray}
where the logarithmic divergence sits in the subtraction term
$I_{ab}(0)$. The difference $I_{ab}(s)-I_{ab}(0)$ is finite. According to the discussion of
Section 3.3, we define the renormalized expressions
\begin{eqnarray}
I_{ab} \to  I_{ab}(p^2)-I_{ab}(\mu^2_M)\,, \qquad \bar I_{ab}\to
I_{ab}(\bar p^2)- I_{ab}(\mu^2_M) \,,
\end{eqnarray}
with the matching scale $\mu_M$.
The integrals $J_{abc}$ and $\bar J_{abc}$ are finite. They may be evaluated in terms of
their dispersion-integral representations
\begin{eqnarray}
J_{abc}&=&
\int_{(m_a+m_c )^2}^{\infty }
\frac{d\, s}{\pi } \, \frac{\rho^{(+)}_{abc}(s)}{s-p^2-i\,\epsilon } \,, \qquad
\bar J_{abc}=
\int_{(m_a+m_c )^2}^{\infty }
\frac{d\, s}{\pi } \, \frac{\rho^{(-)}_{abc}(s)}{s-p^2-i\,\epsilon }\,,
\label{dispersion-integral-representation}
\end{eqnarray}
with the spectral densities
\begin{eqnarray}
&& \rho^{(\pm)}_{abc}(s) = \frac{\sqrt{p_{ac}^2}}{8\,\pi\,\sqrt{s}}\,
\frac{ \log \left(
-\mu_{\pm,abc}^2 +\sqrt{p_{ac}^2\,k^2} \,\right)
- \log \left( -\mu_{\pm,abc}^2 -\sqrt{p_{ac}^2\,k^2}\,\right) } {2\,\sqrt{p_{ac}^2\,k^2}\,}\,,
\nonumber\\
&&  \mu_{\pm,abc}^2 = \left(\frac{m_c^2-m_a^2\mp \bar p^2 }{2\,\sqrt{s}}\right)^2  - p_{ac}^2- {\textstyle{1\over 4}}\,k^2 - m_b^2\,,
\nonumber\\
&& k^2 = \left(\frac{\bar p^2}{\sqrt{s}}\right)^2 +s-2\,\bar p^2\,, \qquad
\sqrt{s^{\phantom{2}}_{\phantom{a}}}= \sqrt{m_a^2+p_{ac}^2}+\sqrt{m_c^2+p_{ac}^2}\,,
\label{result-spectral-densities}
\end{eqnarray}
We wrote the logarithm so as to ensure a smooth spectral density.
It is important to impose the dispersion-integral representation  in terms of the proper variables, i.e.
keeping $q^2=0 $ and $\bar p^2= (p-q)^2$ fixed. This implies that the spectral densities have an implicit dependence on
$\bar p^2 =p^2-2\,p \cdot q $.
The alternative representation of the integrals $J_{abc}$ and $\bar J_{abc}$ derived through an application of Feynman's parametrization is also useful. We have
\begin{eqnarray}
&& J_{abc} = \int \frac{\Theta[z_2-z^2_2]\,\Theta[z_1-z^2_1]\,\Theta[z_1-z_2]d z_1\,d z_2/(16 \pi^2)}{
(m_c^2-z_1\,p^2)\,(1-z_1) + z_1\,m_a^2+2\,(1-z_1)\,z_2\,(p \cdot q)+z_2\,\mu^2_{ba}}\,,
\nonumber\\
&& \bar J_{abc} = \int  \frac{\Theta[z_2-z^2_2]\,\Theta[z_1-z^2_1]\,\Theta[1-z_1-z_2]\,d z_1\,d z_2/(16 \pi^2)}{
(m_c^2-z_1\,p^2)\,(1-z_1) +z_1\,m_a^2 +2\,z_1\,z_2\,(p \cdot q)\,+z_2\,\mu^2_{bc}}\,,
\label{integral-check}
\end{eqnarray}
where we used $ q^2=0 $ and $\mu^2_{bc} = m_b^2-m_c^2$. Depending
on the values of the various mass parameters either
(\ref{dispersion-integral-representation}) or
(\ref{integral-check}) may be more economical in a numerical
simulation. If $p^2 > (m_a+m_c)^2$ for instance, the
representation (\ref{dispersion-integral-representation}) is advantageous, the integrals being complex.
We performed numerical checks using (\ref{dispersion-integral-representation}) or (\ref{integral-check}) and verify that
the identity $J_{abc}=\bar J_{cba}$ is indeed satisfied.


\section*{Appendix D}

We provide explicit expressions for the contributions of the $\phi\,D_s^*$ and $K^*D^*$ channels to the decay amplitude
(\ref{def-transition-tensor-scalar}). The latter are linear in the resonance
coupling constant $g_H$ introduced in (\ref{def-gH}). We obtain
\begin{eqnarray}
&& i\,M^{\alpha \beta, \mu }_{0^+ \to \gamma \,1^-} =
\frac{e\,g_H}{2\,f^2}\, \Big\{
g_T\,E_{+,\phi D_s^*}^{ \alpha \beta, \mu }(\bar p, p)+ \tilde g_V \,\Big(E_{\phi D_s^*}^{ \alpha \beta, \mu }(\bar p, p)
+E_{-,\phi D_s^*}^{ \alpha \beta, \mu }(\bar p, p) \Big)
 \Big\}
\nonumber\\
&& \qquad  \;\;\;
+\,\frac{g_H}{4\,f^2\,M_V^2}\,\Big[\tilde e_C-e +\frac{\tilde e_Q}{3}\Big]\,  \Big\{
g_T\,\hat E_{+,\phi D_s^*}^{ \alpha \beta, \mu }(\bar p, p)
\nonumber\\
&& \qquad \qquad \qquad  \;\;\;+ \tilde g_V \,\Big(\hat E_{\phi D_s^*}^{ \alpha \beta, \mu }(\bar p, p)
+\hat E_{-,\phi D_s^*}^{ \alpha \beta, \mu }(\bar p, p) \Big)
 \Big\}
\nonumber\\
&& \qquad  \;\;\;
+\,\frac{e\,g_H }{2\,f^2}\, \Big\{ g_T\,E_{+, K^*D^*}^{ \alpha \beta, \mu }(\bar p, p) +
 \tilde g_V  \,\Big( E_{K^*D^*}^{ \alpha \beta, \mu }(\bar p, p)+E_{-,K^*D^*}^{ \alpha \beta, \mu }(\bar p, p) \Big) \Big\}
\nonumber\\
&& \qquad  \;\;\;
+\,\frac{g_H }{2\,f^2\,M_V^2}\,\Big[\tilde e_C-\frac{e}{2} -\frac{\tilde e_Q}{6}\Big]\,  \Big\{ g_T\,\hat E_{+, K^*D^*}^{ \alpha \beta, \mu }(\bar p, p)
\nonumber\\
&& \qquad \qquad \qquad  \;\;\;+
 \tilde g_V  \,\Big( \hat E_{K^*D^*}^{ \alpha \beta, \mu }(\bar p, p)+\hat E_{-,K^*D^*}^{ \alpha \beta, \mu }(\bar p, p) \Big) \Big\}
\nonumber\\
&& \qquad  \;\;\; -\,
\frac{e\,g_H }{2\,f^2}\, \Big\{ g_T\,E_{-, D^*K^*}^{ \alpha \beta, \mu }(\bar p, p) +
 \tilde g_V  \,\Big( E_{D^*K^*}^{ \alpha \beta, \mu }(\bar p, p)+E_{+,D^*K^*}^{ \alpha \beta, \mu }(\bar p, p) \Big) \Big\}
\nonumber\\
&& \qquad  \;\;\; -\,
\frac{g_H }{2\,f^2\,m_V^2}\, \Big\{ e_T\,\bar E_{-, D^*K^*}^{ \alpha \beta, \mu }(\bar p, p) +
 \tilde e_V  \,\Big( \bar E_{D^*K^*}^{ \alpha \beta, \mu }(\bar p, p)+\bar E_{+,D^*K^*}^{ \alpha \beta, \mu }(\bar p, p) \Big) \Big\}
\nonumber\\
&& \qquad  \;\;\; +\,
\frac{\tilde e_E\,g_H }{3\,f^2\,m_V^2}\, \Big\{ \tilde E_{ D^*K^*}^{ \alpha \beta, \mu }(\bar p, p) +
2\,\tilde E_{D_s^* \phi}^{ \alpha \beta, \mu }(\bar p, p) \Big\}
\nonumber\\
&& \qquad  \;\;\;
+\frac{(e_C+e_Q/6)\,g_H}{4\,f^2\,M_V^2}\, \Big\{ \tilde g_T \,\bar F_{K^*D D^*}^{ \alpha \beta, \mu }(\bar p, p)
+ g_E\, F_{K^*DD^*}^{ \alpha \beta, \mu }(\bar p, p) \Big\}
\nonumber\\
&& \qquad  \;\;\;
+\,\frac{(e_C-e_Q/3)\,g_H}{8\,f^2\,M_V^2}\, \Big\{ \tilde g_T \,\bar F_{\phi D_s D_s^*}^{ \alpha \beta, \mu }(\bar p, p)
+ g_E\, F_{\phi D_s D_s^*}^{ \alpha \beta, \mu }(\bar p, p) \Big\}
\nonumber\\
&& \qquad  \;\;\;
-\,\frac{e_A\,\tilde g_P\,g_H}{48\,f^3\,m_V} \Big\{ F_{D^*K K^*}^{ \alpha \beta, \mu }(\bar p, p)
+ \bar F_{D^*K K^*}^{ \alpha \beta, \mu }(\bar p, p) \Big\}
\nonumber\\
&& \qquad  \;\;\;
-\,\frac{e_A\,\tilde g_P\,g_H}{36\,f^3\,m_V}\, \Big\{ F_{D_s^*\,\eta\; \phi\;}^{ \alpha \beta, \mu }\,(\bar p, p)
+ \bar F_{D_s^*\,\eta\; \phi\;}^{ \alpha \beta, \mu }\,(\bar p, p) \Big\} \,,
\label{decay-amplitude-0plus-AppendixD}
\end{eqnarray}
with
\begin{eqnarray}
&& E^{\alpha \beta, \mu}_{ab}(\bar p, p) =- i\,\int
\frac{d^4l}{(2\pi)^4}\,g_{\bar \rho \rho}\,S^{ \alpha \bar \rho}_a(l)\, \Bigg\{
S_{b}^{\beta ,\mu \rho  }(\bar p+l) +S_{b}^{\mu \beta ,\rho  }(p+l)
\nonumber\\
&& \qquad  \quad +
g_{\bar \kappa \kappa}\,\Big\{ S_{b}^{\beta \bar \kappa  }(\bar p+l)\,S^{\mu \kappa, \rho }_{b}(p+l)
+ S_{b}^{ \beta ,\mu \bar \kappa }(\bar p+l)\,S^{ \kappa \rho}_{b}(p+l)\Big\} \Bigg\}
\,,
\nonumber\\
&& E^{\alpha \beta, \mu}_{-,ab}(\bar p, p) =- i\,\int
\frac{d^4l}{(2\pi)^4}\,S_{a, \sigma  \rho} (l)\, \Bigg\{
\bar p^\alpha\,S_{b}^{\sigma \beta,\mu \rho }(\bar p+l)
 +g^{\mu \alpha }\,S_{b}^{\sigma \beta,\rho  }(p+l)
\nonumber\\
&& \qquad  \quad  + \bar p^\alpha\,g_{\bar \kappa \kappa}\,
\Big\{ S_{b}^{\sigma \beta,\bar \kappa  }(\bar p+l)\,S^{\mu \kappa, \rho }_{b}(p+l)
+ S_{b}^{ \sigma \beta ,\mu \bar \kappa }(\bar p+l)\,S^{ \kappa \rho}_{b}(p+l)\Big\} \Bigg\}\,,
\nonumber\\
&& E^{\alpha \beta, \mu}_{+,ab}(\bar p, p) = -\,  i\,\int
\frac{d^4l}{(2\pi)^4}\,g_{\bar \rho \rho}\,g_{\bar \sigma \sigma}\,S^{ \bar \sigma \beta  ,\bar \rho}_a(l)\,\Bigg\{
\bar p^\alpha\, S_{b}^{\sigma ,\mu \rho }(\bar p+l)
\nonumber\\
&& \qquad  \quad +\bar p^\alpha\, S_{b}^{\mu \sigma ,\rho }(p+l)
+g^{\alpha \mu}\, S_{b}^{\sigma \rho  }( p+l)
\nonumber\\
&& \qquad  \quad + \bar p^\alpha\,g_{\bar \kappa \kappa}\,
\Big\{ S_{b}^{\sigma \bar \kappa  }(\bar p+l)\,S^{\mu \kappa, \rho }_{b}(p+l)
+ S_{b}^{ \sigma ,\mu \bar \kappa }(\bar p+l)\,S^{ \kappa \rho}_{b}(p+l)\Big\} \Bigg\}\,,
\nonumber\\
&& \hat E^{\alpha \beta, \mu}_{ab}(\bar p, p) =- i\,\int
\frac{d^4l}{(2\pi)^4}\,g_{\bar \rho \rho}\,S^{ \alpha \bar \rho}_a(l)\,
\nonumber\\
&& \qquad  \quad \Bigg\{
q_\kappa\,\Big\{ S_{b}^{\beta \kappa  }(\bar p+l)\,S^{\mu  \rho }_{b}(p+l)
- S_{b}^{ \beta \mu  }(\bar p+l)\,S^{ \kappa \rho}_{b}(p+l)\Big\} \Bigg\}
\,,
\nonumber\\
&& \hat E^{\alpha \beta, \mu}_{-,ab}(\bar p, p) =- i\,\int
\frac{d^4l}{(2\pi)^4}\,S_{a, \sigma  \rho} (l)\,
\nonumber\\
&& \qquad  \quad  \Bigg\{\bar p^\alpha\,q_\kappa\,
\Big\{ S_{b}^{\sigma \beta,\kappa  }(\bar p+l)\,S^{\mu  \rho }_{b}(p+l)
- S_{b}^{ \sigma \beta ,\mu  }(\bar p+l)\,S^{ \kappa \rho}_{b}(p+l)\Big\} \Bigg\}\,,
\nonumber\\
&& \hat E^{\alpha \beta, \mu}_{+,ab}(\bar p, p) = -\,  i\,\int
\frac{d^4l}{(2\pi)^4}\,g_{\bar \rho \rho}\,g_{\bar \sigma \sigma}\,S^{ \bar \sigma \beta  ,\bar \rho}_a(l)\,
\nonumber\\
&& \qquad  \quad \Bigg\{ \bar p^\alpha\,q_\kappa\,
\Big\{ S_{b}^{\sigma \kappa  }(\bar p+l)\,S^{\mu  \rho }_{b}(p+l)
- S_{b}^{ \sigma \mu  }(\bar p+l)\,S^{ \kappa \rho}_{b}(p+l)\Big\} \Bigg\}\,,
\nonumber\\
&& \bar E^{\alpha \beta, \mu}_{ab}(\bar p, p) =- i\,\int
\frac{d^4l}{(2\pi)^4}\,g_{\bar \rho \rho}\,S^{ \alpha \bar \rho}_a(l)\,
S_{b}^{\tau \rho  }(p+l)\,\Big( g^{\mu }_{\;\;\tau}\,q^\beta - g^{\mu \beta}\,q_\tau \Big)
\,,
\nonumber\\
&& \tilde E^{\alpha \beta, \mu}_{ab}(\bar p, p) =+\,2\,i\,\int
\frac{d^4l}{(2\pi)^4}\,g_{\bar \rho \rho}\,S_{a}^{\beta \bar \rho} (l)\, \bar p^\alpha\,S_{b}^{ \tau \mu,\rho  }(p+l)
\,q_\tau \,,
\nonumber\\
&& \bar E^{\alpha \beta, \mu}_{+,ab}(\bar p, p) = -\, i\,\int
\frac{d^4l}{(2\pi)^4}\,g_{\bar \rho \rho}\,S^{ \sigma \beta  ,\bar \rho}_a(l)\,
\bar p^\alpha\, S_{b}^{\tau \rho }(p+l)\,\Big( g^{\mu}_{\;\;\tau}\,q_\sigma - g^{\mu}_{\;\;\sigma}\,q_\tau \Big)
\,,
\nonumber\\
&& \bar E^{\alpha \beta, \mu}_{-,ab}(\bar p, p) =- \,i\,\int
\frac{d^4l}{(2\pi)^4}\,g_{\bar \rho \rho}\,S_{a}^{\sigma  \bar \rho} (l)\, \bar p^\alpha\,S_{b}^{ \beta \tau,\rho  }(p+l)
\,\Big( g^{\mu \tau}\,q_\sigma - g^{\mu \sigma}\,q_\tau \Big)\,
\nonumber\\
&& \bar F^{\alpha \beta, \mu}_{abc}(\bar p, p) = -\,2\, i\,\int
\frac{d^4l}{(2\pi)^4}\,S^{ \bar \tau \rho}_a(l)\,q_\nu\, \epsilon^{\mu \nu }_{\quad \sigma \tau}
\,\epsilon^{\alpha \beta}_{\quad  \bar \sigma \bar \tau}
\nonumber\\
&& \qquad  \quad \times  (\bar p+l)^\tau\,(\bar p+l)^{\bar \sigma}\,
 S_{b}(\bar p+l)\,S^{\sigma }_{c,\rho}(p+l)\,,
\nonumber\\
&& F^{\alpha \beta, \mu}_{abc}(\bar p, p) =-\,2\, i\,\int
\frac{d^4l}{(2\pi)^4}\,S^{\bar \mu \bar \nu,\rho}_a(l)\,q_\nu\, \epsilon^{\mu \nu}_{ \quad \sigma \tau}
\,\epsilon_{\bar \mu \bar \nu}^{\quad \bar \sigma \beta}
\nonumber\\
&& \qquad  \quad  \times  (\bar p+l)^\tau\,(\bar p+l)_{\bar \sigma}\,\bar p^{\alpha}\,
S_{b}(\bar p+l)\,S^{\sigma }_{c, \rho}(p+l)\,.
\label{def-EF-0plus-d}
\end{eqnarray}
The evaluation of the decay constant $d_{0^+\to \gamma 1^-}$ defined in (\ref{result-projection-0plus}) is
done by contracting the gauge-invariant tensors
(\ref{def-AB-0plus}, \ref{def-barA-0plus}, \ref{def-C-0plus}, \ref{def-D-0plus}) and (\ref{def-EF-0plus-d})
with the antisymmetric tensor
\begin{eqnarray}
&& \qquad P^{(1-)}_{\alpha \beta, \mu} =-\frac{1}{2}\,
\Bigg( \Big\{g_{\mu \alpha}- \frac{p^\mu\,q^\alpha}{p\cdot q}\Big\}\,\bar p_\beta
- \Big\{g_{\mu \beta }- \frac{p^\mu\,q^\beta}{p\cdot q}\Big\}\,\bar p_\alpha \Bigg) \,.
\label{}
\end{eqnarray}
We derive the required contractions in terms of the master loop integrals $I_{ab}, \bar I_{ab}, J_{abc}$ and
$\bar J_{abc}$ introduced in (\ref{def-master-loops}). Following the
arguments of Section 3.3, reduced tadpole integrals are dropped systematically. Using the notation $p^2=M_i^2$ and
$(p-q)^2=M_f^2$, we have
\begin{eqnarray}
&& 8\,(p \cdot q)\,M_i^2\,P^{(1-)}_{\alpha \beta, \mu}\,A^{\alpha \beta, \mu }_{ab} =
\Big[M_i^6-\left(m_a^2-m_b^2+2\, M_f^2\right) M_i^4
\nonumber\\
&&  \qquad \,-\,M_f^2 \left(2 \,m_a^2-2 \,m_b^2+M_f^2\right) M_i^2+\left(m_a^2-m_b^2\right)
   M_f^4 \Big]\,I_{ab}
\nonumber\\
&&  \,+\,2\, M_f^2 \left(m_a^2-m_b^2+M_f^2\right) M_i^2\,\bar I_{ab}
  + 4\, m_a^2\, M_f^2\, M_i^2 \left(M_i^2-M_f^2\right) J_{aab}\,,
\nonumber\\ \nonumber\\
&& 48\,(p \cdot q)\,M_i^4\,P^{(1-)}_{\alpha \beta, \mu}\,\bar A^{\alpha \beta, \mu }_{ab} =
-\left(M_f^2-M_i^2\right)^2 \Big[-\left(3 \left(m_a^2+m_b^2\right)-M_f^2\right) M_i^4
\nonumber\\
&&  \qquad \,+\,3\, M_i^6+\left(m_a^2+m_b^2\right)
   M_f^2\, M_i^2-2 \left(m_a^2-m_b^2\right)^2 M_f^2 \Big]\,I_{ab}\,,
\nonumber\\ \nonumber\\
&& 8\,(p \cdot q)\,M_i^2\,P^{(1-)}_{\alpha \beta, \mu}\,B^{\alpha \beta, \mu }_{ab} =
\Big[\left(M_f^4-2 \,M_i^2 M_f^2-M_i^4\right) m_a^2
\nonumber\\
&& \qquad \,+\,M_i^2 \left(M_f^4+2\, M_i^2 M_f^2-M_i^4\right)+m_b^2 \left(-M_f^4+2\, M_i^2
   M_f^2+M_i^4\right) \Big]\,I_{ab}
\nonumber\\
&&  \,  -\,2\, M_f^2 \left(-m_a^2+m_b^2+M_f^2\right) M_i^2\,\bar I_{ab}
  + 4 \,m_b^2\, M_f^2\, M_i^2 \left(M_f^2-M_i^2\right)\bar J_{abb}\,,
\nonumber\\ \nonumber\\
&& 8\,(p \cdot q)\,M_i^2\,P^{(1-)}_{\alpha \beta, \mu}\,C^{\alpha \beta, \mu }_{abc} =
\Big[2 \left(m_a^2-m_c^2\right) M_f^6+6\, M_i^4 M_f^4
\nonumber\\
&&  \qquad \,-2\, \left(3\, m_a^2-2 \,m_b^2-m_c^2+M_f^2\right)
M_i^2 \,M_f^4-4 \,M_i^6 \,M_f^2\Big]\,I_{ac}
\nonumber\\
&& \, +\,2 \,M_f^2
   M_i^2 \,\Big[\left(2 \,m_a^2-3 m_b^2+m_c^2-M_f^2\right) M_f^2+\left(m_b^2-m_c^2+M_f^2\right) M_i^2\Big]\, \bar I_{bc}
\nonumber\\
&&  \,-\,4\, M_f^2\, M_i^2\,
   \Big[-m_b^2\, M_f^4+\left(m_a^2-m_b^2\right)^2 M_f^2+2\, m_b^2 \,M_i^2 \,M_f^2-m_b^2\, M_i^4\Big]\, J_{abc} \,,
\nonumber\\ \nonumber\\
&& 12\,(p \cdot q)\,M_i^4\,P^{(1-)}_{\alpha \beta, \mu}\,\bar C^{\alpha \beta, \mu }_{abc} =
-\Big[
\left( 4\,m_c^2-2\,m_a^2\right) \left(M_f^3-M_f M_i^2\right)^2 m_a^2
\nonumber\\
&&  \qquad \,+\Big(\left(M_f^2-M_i^2\right)^2 \left(M_f^2+3 \,M_i^2\right)-3\, m_b^2
   \left(M_f^4-3\, M_f^2\, M_i^2\right)\Big)\, M_i^2\, m_a^2
\nonumber\\
&&  \qquad \,-\left(m_c^2-M_i^2\right) \left(M_f^2-M_i^2\right)^2
   \left(-3\, M_i^4+M_f^2\, M_i^2+2\, m_c^2\, M_f^2\right)
\nonumber\\
&&  \qquad\, +\,3\, m_b^2\, M_i^2
   \left(M_i^2-M_f^2\right) \left(2\, M_i^4-M_f^2\, M_i^2-m_c^2\, M_f^2\right)
   \nonumber\\
&&  \qquad\,  -\,6 \,m_b^4\, M_f^2\, M_i^4\Big]\,I_{ac}
\nonumber\\
&&  \,-\,
   3\, m_b^2\, M_i^4 \, \Big[M_f^2 \left(-2\, m_a^2+3\,
   m_b^2-m_c^2+M_f^2\right)
     \nonumber\\
&&  \qquad\,-\,\left(m_b^2-m_c^2+M_f^2\right) M_i^2\Big]\,\bar I_{bc}
\nonumber\\
&&  \,-\,   6 \,m_b^2\, M_i^4 \,\Big[-m_b^2\,
   M_f^4+\left(m_a^2-m_b^2\right)^2 M_f^2+2\, m_b^2 \,M_i^2 \,M_f^2-m_b^2\, M_i^4\Big]\, J_{abc} \,,
\nonumber\\ \nonumber\\
&& 12\,(p \cdot q)\,M_i^4\,P^{(1-)}_{\alpha \beta, \mu}\,D^{\alpha \beta, \mu }_{abc} =
\Big[ -6 \left(m_b^2-m_c^2\right) \left(m_b^2+M_i^2\right) M_i^6
\nonumber\\
&&  \qquad \,-\,3 \left(M_i^2-M_f^2\right)
   \Big(\left(m_a^2-5\, m_b^2+3\, m_c^2\right) M_i^2
   \nonumber\\
&&  \qquad \,+\,m_b^2 \left(m_a^2-2\, m_b^2+m_c^2\right)\Big)
   M_i^4
\nonumber\\
&&  \qquad \,+\,2\left(M_i^2-M_f^2\right)^3 \left(m_a^4-2 \left(m_c^2+M_i^2\right) m_a^2+m_c^4-2 \,M_i^4+m_c^2 M_i^2\right)
\nonumber\\
&&  \qquad \,-\,
   \left(M_i^3-M_f^2\, M_i\right)^2 \Big(2\, m_a^4-\left(3\, m_b^2+4\, m_c^2+10\, M_i^2\right) m_a^2
\nonumber\\
&&  \qquad \,+\, 2 \left(m_c^2-M_i^2\right)^2 +3\, m_b^2
   \left(m_c^2+M_i^2\right)\Big)+3 \left(M_i^2-M_f^2\right) M_i^8\Big]\, I_{ac}
\nonumber\\
&&  \,+\,3 \left(m_b^2+M_f^2\right) M_i^4 \,\Big[M_f^2 \left(-m_a^2+3\, m_b^2-2\,
   m_c^2+M_f^2\right)
   \nonumber\\
&&  \qquad \,-\,\left(-m_a^2+m_b^2+M_f^2\right) M_i^2\Big]\,\bar I_{ab}
\nonumber\\
&&  \,+\,6 \left(m_b^2+M_f^2\right) M_i^4\, \Big[M_f^2 \,m_b^4-\Big(2\,
   m_c^2 \,M_f^2+\left(M_f^2-M_i^2\right)^2\Big)\, m_b^2
   \nonumber\\
&&  \qquad \,+\,m_c^4\, M_f^2\Big]\, \bar J_{abc}\,,
\nonumber\\
&& 96\,M_f^2\,M_i^4\,(p \cdot q) \,P^{(1-)}_{\alpha \beta, \mu}\,E^{\alpha \beta, \mu }_{ab} =
-3 \,M_f^2 \,M_i^2 \,\Big[-4 \,\Big(\left(m_a^2-M_i^2\right)^2-m_b^4\Big)\, M_i^4
\nonumber\\
&&  \qquad\,  +\,\left(-m_a^2+m_b^2+M_i^2\right)
   \left(M_i^2-M_f^2\right) \left(-m_a^2-3\, m_b^2+7 \,M_i^2\right) M_i^2
   \nonumber\\
&&  \qquad \,+\,\left(M_f^2-M_i^2\right)^2 \left(-3\, M_i^4-2 \left(m_a^2+3\,
   m_b^2\right) M_i^2+\left(m_a^2-m_b^2\right)^2\right)\Big]\, I_{ab}
\nonumber\\
&&  \,+\,\Big[-\, \Big(5\, M_f^4-\left(7\, m_a^2+m_b^2\right) M_f^2+2
   \left(m_a^2-m_b^2\right)^2\Big)\, M_i^8  -17\,M_f^4 \,m_a^4\,M_i^4
   \nonumber\\
&&  \qquad\,-\,M_f^2 \left(-7 \,m_a^4+2 \left(m_b^2+M_f^2\right) m_a^2+5\, m_b^4+5 \,M_f^4-22\, m_b^2\,
   M_f^2\right) M_i^6
   \nonumber\\
&&  \qquad\,-\,M_f^4 \left(\left(2 \,m_b^2-19\, M_f^2\right) m_a^2-19\, m_b^4+2\, M_f^4
    +23 \,m_b^2\, M_f^2\right) M_i^4\,\Big]\bar I_{ab}
\nonumber\\
&&  \,-\,6\,
   m_b^2\, M_f^2\, M_i^4 \left(M_f^2-M_i^2\right) \Big[-M_i^4+\left(-5 \,m_a^2+m_b^2+M_f^2\right) M_i^2
\nonumber\\
&&  \qquad \, +\,M_f^2 \left(9\, m_a^2+3\, m_b^2-4\,
   M_f^2\right)\Big]\,\bar J_{abb} \,,
\nonumber\\ \nonumber\\
&& 96\,M_f^2\,M_i^4\,(p \cdot q) \,P^{(1-)}_{\alpha \beta, \mu}\,E^{\alpha \beta, \mu }_{+,ab} =
6 \,M_f^2\, M_i^2 \,\Big[ M_i^8-2 \left(m_a^2-2 m_b^2\right) M_i^6
\nonumber\\
&& \qquad\,+\,\left(m_a^4+4\, m_b^2\, m_a^2-5\, m_b^4-4 \,m_b^2\, M_f^2\right)
   M_i^4
   \nonumber\\
&& \qquad\,+\,M_f^2 \,\Big(2\, m_a^4+\left(8\, m_b^2-3\, M_f^2\right) m_a^2
 -10 \,m_b^4+M_f^4-7\, m_b^2 \,M_f^2\Big)\, M_i^2
   \nonumber\\
&&  \qquad\,+\,\left(m_a^2-m_b^2\right)
   M_f^4 \left(-m_a^2-5 m_b^2+M_f^2\right) \Big]\,I_{ab}
\nonumber\\
&&   \,+\,   6\, M_f^4 \,M_i^4\left(-m_a^2+m_b^2+M_f^2\right)  \left(2 \,m_a^2+10 \,m_b^2+M_f^2-3\,
   M_i^2\right) \bar I_{ab}
   \nonumber\\
&& \,-\,12\, m_b^2\, M_f^4 \,M_i^4\, \left(2\, m_a^2+10\, m_b^2+M_f^2-3 M_i^2\right) \left(M_f^2-M_i^2\right) \bar J_{abb}\,,
\nonumber\\ \nonumber\\
&& 96\,M_f^2\,M_i^4\,m_b^2\,(p \cdot q) \,P^{(1-)}_{\alpha \beta, \mu}\,E^{\alpha \beta, \mu }_{-,ab} =
3\, M_f^2\, M_i^2
\nonumber\\
&&  \qquad \Big[\,-\,4\, m_b^2 \left(5 \,m_a^2+m_b^2-M_i^2\right) \left(-m_a^2+m_b^2+M_i^2\right)
   M_i^4
\nonumber\\
&&  \qquad \,+\left(M_f^2-M_i^2\right)^3 \left(m_a^4-m_b^4+M_i^4-2 \left(m_a^2-2\, m_b^2\right) M_i^2\right)
\nonumber\\
&&  \qquad \, +\,\left(M_i^2-M_f^2\right) \Big( m_a^4+36 \,m_b^2\, m_a^2+11\, m_b^4+M_i^4
\nonumber\\
&&  \qquad \,-\, 2 \left(m_a^2+10\, m_b^2\right) M_i^2\Big)\,   M_i^4
\nonumber\\
&&  \qquad \,+\,2\, m_b^2
   \left(M_f^2-M_i^2\right)^2 \Big(-5 \,m_a^4+4 \,m_b^2 m_a^2+m_b^4+9 \,M_i^4
\nonumber\\
&&  \qquad \, -\,2\left(5\, m_a^2+2\, m_b^2\right) M_i^2 \Big)
   \Big] \,I_{ab}
\nonumber\\
&& \,+ \, M_f^2\,
   M_i^4 \,\Big[ 2\, M_f^8-\left(m_a^2+25 \,m_b^2\right) M_f^6+\left(-m_a^4+56 \,m_b^2 \,m_a^2+17\, m_b^4\right) M_f^4
\nonumber\\
&& \qquad\,-\,\left(-5 \,m_a^4+\left(4\,
   M_f^2-8\, m_b^2\right) m_a^2+13\, m_b^4+M_f^4-2 \,m_b^2\, M_f^2\right) M_i^2\, M_f^2
   \nonumber\\
&& \qquad \,+\,12\, m_b^2 \left(-5 \,m_a^4+4 \,m_b^2 \,m_a^2+m_b^4\right)
   M_f^2
   \nonumber\\
&& \qquad \,-\,\Big(M_f^4-\left(5\, m_a^2+11 \,m_b^2\right) M_f^2
 +4 \left(m_a^2-m_b^2\right)^2\Big) \,M_i^4\Big]\,\bar I_{ab}
\nonumber\\
&& \,+\,   6\, m_b^2 \,M_f^4\, M_i^4
   \left(M_i^2-M_f^2\right) \Big[4 \left(5 m_a^2+m_b^2-M_i^2\right) m_b^2-2\,\left(M_f^2-M_i^2\right)^2
   \nonumber\\
&& \qquad +\,
   \left(M_i^2-M_f^2\right) \left(-3 \,m_a^2+m_b^2+3\, M_i^2\right)\Big]\, \bar J_{abb}\,,
   \nonumber\\
&& 48\,M_i^4\,M_f^2 \,P^{(1-)}_{\alpha \beta, \mu}\,\hat E^{\alpha \beta, \mu }_{ab} =
M_f^2\, \Big[-3\, M_i^{10}+2 \left(3\, m_a^2+6\, m_b^2+M_f^2\right) M_i^8
\nonumber\\
&&  \qquad \,-\,\left(3\, m_a^4+2 \left(12 \,m_b^2+M_f^2\right) m_a^2+9\,
   m_b^4-M_f^4+5\, m_b^2 \,M_f^2\right) M_i^6
   \nonumber\\
&&  \qquad \,-\,2\, M_f^2 \left(m_a^4-9\, m_b^2\, m_a^2-2 m_b^4
   +2 \left(m_a^2+m_b^2\right) M_f^2\right)M_i^4
     \nonumber\\
&&  \qquad \,+\,M_f^2 \left(7\, m_b^2 \,m_a^4-8\, m_b^4 \,m_a^2-m_b^6
   +\left(5\, m_a^4-6 \,m_b^2\, m_a^2+5\, m_b^4\right) M_f^2\right) M_i^2
     \nonumber\\
&&  \qquad \,+\,2\,M_f^2\,m_a^6\,M_i^2  -2 \left(m_a^2-m_b^2\right)^2 \left(m_a^2+m_b^2\right) M_f^4\Big]\,I_{ab}
\nonumber\\
&&   \,+\,
   M_i^4\, \Big[4 \left(M_i^2-M_f^2\right) m_a^6-3
   \left(\left(m_b^2-2 M_f^2\right) M_f^2+2 \left(m_b^2+M_f^2\right) M_i^2\right) m_a^4
\nonumber\\
&&  \qquad \,+\, m_b^2\, M_f^2 \left(6\, m_b^2+25\, M_f^2-13\,
   M_i^2\right) m_a^2+
 \left(m_b^2+M_f^2\right) \Big(-2\, M_f^6
   \nonumber\\
&&  \qquad \,+\,2\, M_i^2\, M_f^4+m_b^4 \left(M_f^2+2\, M_i^2\right)
   +m_b^2 \left(4\, M_f^4-7 \,M_f^2\, M_i^2\right)\Big)\Big]\,\bar I_{ab}
\nonumber\\
&&  \, +\,   6 \,m_b^4\, M_f^2\, M_i^4 \left(3\, m_a^2+m_b^2-M_i^2\right) \left(M_i^2-M_f^2\right)\bar J_{abb} \,,
\nonumber\\ \nonumber\\
&& 24\,M_i^2\,P^{(1-)}_{\alpha \beta, \mu}\,\hat E^{\alpha \beta, \mu }_{+,ab} =
-M_f^2 \,\Big[3 \,m_b^2 \left(-m_a^2+m_b^2+M_i^2\right) M_i^2
\nonumber\\
&&  \qquad \,+\,\left(M_i^2-M_f^2\right)
\left(m_a^4+\left(m_b^2-2\, M_i^2\right)
   m_a^2-2 \,m_b^4+M_i^4+m_b^2\, M_i^2\right)\Big]\,I_{ab}
\nonumber\\
&&  \, +\,  \Big[ 3 \,\Big( m_a^4-2 \left(m_b^2+M_f^2\right) m_a^2+\left(m_b^2-M_f^2\right)^2\Big)\, M_i^4
\nonumber\\
&&  \qquad \,-\,3\, M_f^2 \left(m_a^4-\left(m_b^2+2\, M_f^2\right) m_a^2+M_f^4-3 \,m_b^2\, M_f^2\right) M_i^2 \Big]\, \bar I_{ab}
\nonumber\\
&&  \, +\, 6 \,m_b^4\, M_f^2\, M_i^2    \left(M_i^2-M_f^2\right)\bar J_{abb} \,,
\nonumber\\ \nonumber\\
&& 48\,M_i^2\,P^{(1-)}_{\alpha \beta, \mu}\,\hat E^{\alpha \beta, \mu }_{-,ab} =
\Big[-6 \left(m_a^2-2 m_b^2\right) M_i^2\, M_f^4
\nonumber\\
&&  \qquad \,+\,3 \left(m_a^4-m_b^4\right) M_f^4
+3 \left(M_f^2-4 \,m_b^2\right) M_i^4 \,M_f^2 \Big]\,I_{ab}
\nonumber\\
&&  \, +\,
M_i^2 \,\Big[\left(-5\, m_a^4+4\, m_b^2\, m_a^2+m_b^4-8\, M_f^4
+\left(13\, m_a^2+m_b^2\right) M_f^2\right) M_f^2
\nonumber\\
&&  \qquad \,+\Big(5\, M_f^4-\left(7\,
   m_a^2+m_b^2\right) M_f^2+2 \left(m_a^2-m_b^2\right)^2\Big)\, M_i^2\Big]\,\bar I_{ab}
\nonumber\\
&&  \, -\, 6\, m_b^2\, M_f^2 \,M_i^2 \left(M_f^2-M_i^2\right)
   \left(m_a^2+m_b^2-2 M_f^2+M_i^2\right) \bar J_{abb} \,,
\nonumber\\
&& 48\,M_i^4 \,P^{(1-)}_{\alpha \beta, \mu}\,\bar E^{\alpha \beta, \mu }_{ab} =
\left(M_i^2-M_f^2\right) \Big[3 \,M_i^8-\left(6 \left(m_a^2+m_b^2\right)-M_f^2\right) M_i^6
\nonumber\\
&&  \qquad \, +\, \left(3 \left(m_a^4+10\, m_b^2\,
   m_a^2+m_b^4\right)-4 \left(m_a^2+m_b^2\right) M_f^2\right) M_i^4
\nonumber\\
&&  \qquad \, +\,  \left(5\, m_a^4-6\, m_b^2 \,m_a^2+5 \,m_b^4\right) M_f^2\, M_i^2
\nonumber\\
&&  \qquad \, -\,2   \left(m_a^2-m_b^2\right)^2 \left(m_a^2+m_b^2\right) M_f^2 \Big]\,I_{ab}\,,
   \nonumber\\ \nonumber\\
&& 12\,M_i^2 \,P^{(1-)}_{\alpha \beta, \mu}\,\tilde E^{\alpha \beta, \mu }_{ab} =
M_f^2 \left(M_i^2-M_f^2\right) \Big[-5\, m_a^4+4 \left(m_b^2+M_i^2\right)
   m_a^2
   \nonumber\\
&&  \qquad \, +\, \left(m_b^2-M_i^2\right)^2\Big]\,I_{ab}\,,
\nonumber\\ \nonumber\\
&& 24\,M_i^2 \,P^{(1-)}_{\alpha \beta, \mu}\,\bar E^{\alpha \beta, \mu }_{+,ab} =
M_f^2 \left(M_i^2-M_f^2\right) \Big[m_a^4+\left(4\, m_b^2-2\, M_i^2\right) m_a^2
  \nonumber\\
&&  \qquad\,-\,5\, m_b^4+M_i^4+4 \,m_b^2\, M_i^2\Big]\,I_{ab}\,,
\nonumber\\ \nonumber\\
&& 24\,M_i^2 \,P^{(1-)}_{\alpha \beta, \mu}\,\bar E^{\alpha \beta, \mu }_{-,ab} =
-M_f^2 \left(M_i^2-M_f^2\right) \Big[-5\, m_a^4+4 \left(m_b^2+M_i^2\right)
   m_a^2
   \nonumber\\
&& \qquad\,+\,\left(m_b^2-M_i^2\right)^2\Big]\,I_{ab}\,
\nonumber\\
&& 12\,M_i^2\,M_f^2\,(p \cdot q) \,P^{(1-)}_{\alpha \beta, \mu}\,F^{\alpha \beta, \mu }_{abc} =
3\, m_a^2\, m_c^2\, M_f^2 \,\Big[2 \,M_i^6-3\, M_f^2\, M_i^4
\nonumber\\
&& \qquad \,+\,M_f^2 \left(-m_a^2-2 m_b^2+3 \,m_c^2+M_f^2\right)
   M_i^2+\left(m_a^2-m_c^2\right) M_f^4\Big]\,I_{ac}
\nonumber\\
&& \,+\, m_a^2\, M_i^2 \Big[\Big(3 \,m_c^2
   \left(3 M_f^2-M_i^2\right) M_f^2+4 \left(m_a^2+M_f^2\right) \left(M_f^2-M_i^2\right)^2\Big)\, m_b^2
\nonumber\\
&& \qquad \,-\,2\left(M_f^2-M_i^2\right)^2 m_b^4-6\, m_c^4\,
   M_f^4-2 \left(m_a^2-M_f^2\right)^2 \left(M_f^2-M_i^2\right)^2
   \nonumber\\
&& \qquad \,+\,3 \,m_c^2\, M_f^2 \left(M_f^2-m_a^2\right)
   \left(M_f^2-M_i^2\right)\Big]\,\bar I_{ab}
\nonumber\\
&& \,+\,6\, m_a^2\, m_c^2 \,M_f^2 \,M_i^2 \Big[M_f^2 \,m_b^4-\Big(2\, m_c^2
   M_f^2+\left(M_f^2-M_i^2\right)^2\Big)\, m_b^2
      \nonumber\\
&& \qquad \,+\,m_c^4\, M_f^2\Big]\,\bar J_{abc}\,,
   \nonumber\\ \nonumber\\
&& 12\,M_i^2\,M_f^2\,(p \cdot q) \,P^{(1-)}_{\alpha \beta, \mu}\,\bar F^{\alpha \beta, \mu }_{abc} =
   3\, m_c^2\, M_f^4 \,\Big[2 \,M_i^6-3\, M_f^2\, M_i^4
\nonumber\\
&& \qquad \,+\,\left(m_a^2-m_c^2\right)  M_f^4+
M_f^2 \left(-m_a^2-2\, m_b^2+3 \,m_c^2+M_f^2\right) M_i^2
\Big]\,I_{ac}
\nonumber\\
&& \,-\,M_f^2\, M_i^2 \,\Big[
-\Big(3\, m_c^2 \left(3 M_f^2-M_i^2\right) M_f^2+4
   \left(m_a^2+M_f^2\right) \left(M_f^2-M_i^2\right)^2\Big)\, m_b^2
\nonumber\\
&& \qquad \,+\,2\left(M_f^2-M_i^2\right)^2 m_b^4+6 \,m_c^4 \,M_f^4+2\left(m_a^2-M_f^2\right)^2
   \left(M_f^2-M_i^2\right)^2
   \nonumber\\
&& \qquad \,+\,3\, m_c^2\, M_f^2 \left(M_f^2-m_a^2\right) \left(M_i^2-M_f^2\right)\Big]
   \,\bar I_{ab}
\nonumber\\
&& \,+\,6\, m_c^2\, M_f^4\, M_i^2
   \Big[M_f^2 \,m_b^4-\Big(2 \,m_c^2\, M_f^2+\left(M_f^2-M_i^2\right)^2\Big)\, m_b^2
    \nonumber\\
&& \qquad \,+\,m_c^4\, M_f^2\Big]\, \bar J_{abc}\,.
\end{eqnarray}


\section*{Appendix E}

We provide the contributions of the $\phi\,D_s^*$ and $K^*D^*$ channels to the decay amplitude
$M^{\mu,\alpha \beta }_{1^+\to \gamma 0^-}$ introduced in
(\ref{def-transition-tensor-axial}). All terms are linear in the coupling constant $\hat g_H$
(see (\ref{def-gH})).
We obtain
\begin{eqnarray}
&& M^{\mu,\alpha \beta }_{1^+\to \gamma 0^-}=
\frac{e\,\tilde g_T\, \hat g_H}{8\,f^2}\, \Big\{
 F_{-,\phi D_s^*}^{ \mu, \alpha \beta }(\bar p, p)
+E_{+, D^*K^*}^{ \mu, \alpha \beta }(\bar p, p) +F_{-, K^*D^*}^{ \mu, \alpha \beta }(\bar p, p)
 \Big\}
\nonumber\\
&& \quad  \;\;\;
+\,\frac{e\,g_E\,\hat g_H }{8\,f^2}\,\Big\{
 F_{+,\phi D_s^*}^{ \mu, \alpha \beta }(\bar p, p)
+E_{-, D^*K^*}^{ \mu, \alpha \beta }(\bar p, p)+ F_{+, K^*D^*}^{ \mu, \alpha \beta }(\bar p, p)
\Big\}
\nonumber\\
&& \quad  \;\;\;
+\,\frac{\hat g_H }{16\,f^2\,M_V^2}\,\Big[ \tilde e_C-e \,+\frac{\tilde e_Q}{3}\Big]\,\Big\{
g_E\, \hat F_{+,\phi \,D_s^*}^{ \mu, \alpha \beta }(\bar p, p)
+\tilde g_T\,\hat F_{-,\phi \,D_s^*}^{ \mu, \alpha \beta }(\bar p, p)
\Big\}
\nonumber\\
&& \quad  \;\;\;
+\,\frac{\hat g_H }{8\,f^2\,M_V^2}\,\Big[ \tilde e_C-\frac{e}{2} -\frac{\tilde e_Q}{6}\Big]\,\Big\{
g_E\,\hat F_{+, K^*D^*}^{ \mu, \alpha \beta }(\bar p, p)+\tilde g_T\,\hat F_{-, K^*D^*}^{ \mu, \alpha \beta }(\bar p, p) \Big\}
\nonumber\\
&& \quad  \;\;\;
+\,\frac{\tilde e_T\, \hat g_H}{8\,f^2\,m_V^2}\,\bar E_{D^*K^*}^{ \mu, \alpha \beta }(\bar p, p)
+\frac{(2\,e_C+e_Q/3)\,\hat g_H}{4\,f^2\,M_V^2}\, g_V \,H_{K^*D D^*}^{ \mu, \alpha \beta }(\bar p, p)
\nonumber\\
&& \quad  \;\;\;
+\,\frac{(e_C-e_Q/3)\,\hat g_H}{4\,f^2\,M_V^2}\, g_V \,H_{\phi D_s D_s^*}^{ \mu, \alpha \beta }(\bar p, p)
\nonumber\\
&& \quad  \;\;\;
+\,\frac{e_A\,g_P\,\hat g_H}{24\,f^3\,m_V}\, G_{D^*K K^*}^{ \mu, \alpha \beta }(\bar p, p)
+\frac{e_A\,g_P\,\hat g_H}{18\,f^3\,m_V}\, G_{D_s^*\eta \phi}^{ \mu, \alpha \beta }(\bar p, p)\,,
\label{decay-amplitude-1plus-AppendixE}
\end{eqnarray}
with
\begin{eqnarray}
&& E^{\mu, \alpha \beta }_{+,ab}(\bar p, p) =- i\,\int
\frac{d^4l}{(2\pi)^4}\,\epsilon_{\tilde \alpha \tilde \beta  \tau}^{\quad \;\, \beta}\,
\epsilon_{\bar \alpha \bar \beta \bar \sigma \bar \tau }\,S^{\bar \tau ,\tilde \alpha \tilde \beta}_a(l)\,\Bigg\{
 \bar p^{\bar \sigma }\,p^\alpha\,  S_{b}^{\bar \alpha \bar \beta,\mu \tau  }(\bar p+l)
 \nonumber\\
&& \qquad  \quad
+\,\bar p^{\bar \sigma }\,g^{\mu \alpha}\,  S_{b}^{\bar \alpha \bar \beta, \tau  }(\bar p+l)
+ g^{\bar \sigma \mu}\,p^\alpha\,  S_{b}^{\bar \alpha \bar \beta,\tau }(p+l)
\nonumber\\
&& \qquad \quad  +\,  \bar p^{\bar \sigma }\,p^\alpha\, g_{\bar \kappa \kappa}\,
\Big\{ S_{b}^{\bar \alpha \bar \beta,\bar \kappa  }(\bar p+l)\,S^{\mu \kappa, \tau }_{b}(p+l)
 + S_{b}^{\bar \alpha \bar \beta ,\mu \bar \kappa }(\bar p+l)\,S^{ \kappa \tau}_{b}(p+l)\Big\} \Bigg\}\,,
\nonumber\\
&& E^{\mu, \alpha \beta}_{-,ab}(\bar p, p) = +\, i\,\int
\frac{d^4l}{(2\pi)^4}\,\epsilon_{\tilde \alpha \tilde \beta \tau}^{\quad \;\, \beta }\,
\epsilon_{\bar \alpha \bar \beta \bar \sigma \bar \tau }\,S^{ \bar \alpha \bar \beta ,\tilde \alpha \tilde \beta}_a(l)\,
\Bigg\{
\bar p^{\bar \sigma }\,p^\alpha\, S_{b}^{\bar \tau , \mu \tau  }(\bar p+l)
\nonumber\\
&& \qquad  \quad +\,\bar p^{\bar \sigma }\,g^{\mu \alpha}\, S_{b}^{\bar \tau \tau  }(\bar p+l)
+g^{\bar \sigma \mu}\,p^\alpha\, S_{b}^{\bar \tau \tau  }(p+l)
+\bar p^{\bar \sigma }\,p^\alpha\,  S_{b}^{\mu \bar \tau , \tau }( p+l)
\nonumber\\
&& \qquad  \quad +\,\bar p^{\bar \sigma }\,p^\alpha\, g_{\bar \kappa \kappa}\,
\Big\{ S_{b}^{\bar \tau \bar \kappa  }(\bar p+l)\,S^{\mu \kappa, \tau }_{b}(p+l)
+ S_{b}^{ \bar \tau ,\mu \bar \kappa }(\bar p+l)\,S^{ \kappa \tau}_{b}(p+l)\Big\} \Bigg\}
\,,
\nonumber\\
&& \bar E^{\mu, \alpha \beta}_{ab} (\bar p, p)= - \, i\,\int
\frac{d^4l}{(2\pi)^4}\,\epsilon_{\tilde \alpha \tilde \beta \tau}^{\quad \;\, \beta }\,
\epsilon_{\bar \alpha \bar \beta \bar \sigma \bar \tau }\,S^{ \bar \tau ,\tilde \alpha \tilde \beta}_a(l)\,
S_{b}^{\rho \bar \beta ,\tau  }(p+l)
\nonumber\\
&& \qquad \quad \times \,p^\alpha \,\bar p^{\bar \sigma}\,\Big(
g^{\mu }_{\;\;\rho}\,q^{\bar \alpha }-g^{\mu \bar \alpha }\,q_\rho\Big) \,,
\nonumber\\
&& F^{\mu, \alpha \beta }_{+,ab}(\bar p, p) =+ \,i\,\int
\frac{d^4l}{(2\pi)^4}\,\epsilon_{\tilde \alpha \tilde \beta \sigma }^{\quad \;\, \beta }\,
\epsilon_{\bar \alpha \bar \beta \bar \sigma \bar \tau }\,S^{\bar \tau \sigma}_a(l)\,\Bigg\{
 \bar p^{\bar \sigma }\, g^{\mu \alpha}\,S_{b}^{\bar \alpha \bar \beta,\tilde \alpha \tilde \beta }(\bar p+l)
 \nonumber\\
&& \qquad  \quad + \,g^{\bar \sigma \mu}\,p^\alpha\, S_{b}^{\bar \alpha \bar \beta,\tilde \alpha \tilde \beta}(p+l)
  +  \bar p^{\bar \sigma }\,p^\alpha\,g_{\bar \kappa \kappa}\,
\Big\{ S_{b}^{\bar \alpha \bar \beta,\bar \kappa  }(\bar p+l)\,S^{\mu \kappa, \tilde \alpha \tilde \beta }_{b}(p+l)
\nonumber\\
&& \qquad  \qquad \quad +\, S_{b}^{\bar \alpha \bar \beta ,\mu \bar \kappa }(\bar p+l)\,S^{ \kappa ,\tilde \alpha \tilde \beta}_{b}(p+l)\Big\} \Bigg\}\,,
\nonumber\\
&& F^{\mu, \alpha \beta}_{-,ab}(\bar p, p) = -\, i\,\int
\frac{d^4l}{(2\pi)^4}\,\epsilon_{\tilde \alpha \tilde \beta \sigma}^{\quad\;\, \beta }\,
\epsilon_{\bar \alpha \bar \beta \bar \sigma \bar \tau }\,S^{ \bar \alpha \bar \beta ,\sigma}_a(l)\, \Bigg\{
\bar p^{\bar \sigma }\,g^{\mu \alpha}\, S_{b}^{\bar \tau , \tilde \alpha \tilde \beta }(\bar p+l)
\nonumber\\
&& \qquad  \quad +\,g^{\bar \sigma \mu}\,p^\alpha\,S_{b}^{\bar \tau ,\tilde \alpha \tilde \beta  }(p+l)
+\bar p^{\bar \sigma }\,p^\alpha\, S_{b}^{\mu \bar \tau , \tilde \alpha \tilde \beta }( p+l)
\nonumber\\
&& \qquad  \quad +\,\bar p^{\bar \sigma }\,p^\alpha\,g_{\bar \kappa \kappa}\,
\Big\{ S_{b}^{\bar \tau \bar \kappa  }(\bar p+l)\,S^{\mu \kappa, \tilde \alpha \tilde \beta }_{b}(p+l)
+ S_{b}^{ \bar \tau ,\mu \bar \kappa }(\bar p+l)\,S^{ \kappa ,\tilde \alpha \tilde \beta}_{b}(p+l)\Big\} \Bigg\}
\,, \nonumber\\
&& \hat F^{\mu, \alpha \beta }_{+,ab}(\bar p, p) =+ \,i\,\int
\frac{d^4l}{(2\pi)^4}\,\epsilon_{\tilde \alpha \tilde \beta \sigma }^{\quad \;\, \beta }\,
\epsilon_{\bar \alpha \bar \beta \bar \sigma \bar \tau }\,S^{\bar \tau \sigma}_a(l)\,
\nonumber\\
&& \qquad  \quad \Bigg\{\,\bar p^{\bar \sigma }\,p^\alpha\,q_{ \kappa}\,
\Big\{ S_{b}^{\bar \alpha \bar \beta, \kappa  }(\bar p+l)\,S^{\mu , \tilde \alpha \tilde \beta }_{b}(p+l)
- S_{b}^{\bar \alpha \bar \beta ,\mu  }(\bar p+l)\,S^{ \kappa ,\tilde \alpha \tilde \beta}_{b}(p+l)\Big\} \Bigg\}\,,
\nonumber\\
&& \hat F^{\mu, \alpha \beta}_{-,ab}(\bar p, p) = -\, i\,\int
\frac{d^4l}{(2\pi)^4}\,\epsilon_{\tilde \alpha \tilde \beta \sigma}^{\quad\;\, \beta }\,
\epsilon_{\bar \alpha \bar \beta \bar \sigma \bar \tau }\,S^{ \bar \alpha \bar \beta ,\sigma}_a(l)\,
\nonumber\\
&& \qquad  \quad \Bigg\{\,\bar p^{\bar \sigma }\,p^\alpha\,q_{ \kappa}\,
\Big\{ S_{b}^{\bar \tau  \kappa  }(\bar p+l)\,S^{\mu , \tilde \alpha \tilde \beta }_{b}(p+l)
- S_{b}^{ \bar \tau \mu  }(\bar p+l)\,S^{ \kappa ,\tilde \alpha \tilde \beta}_{b}(p+l)\Big\} \Bigg\}
\,\nonumber\\
&& G^{\mu, \alpha \beta}_{abc}(\bar p, p) = +\,2\,i\,\int
\frac{d^4l}{(2\pi)^4}\,q_\nu\,\epsilon^{\mu \nu}_{\quad  \bar \sigma \bar \tau }\,
p^\alpha\, \epsilon_{\tilde \alpha  \tilde \beta \tau  }^{\quad \;\, \beta}\,S^{ \rho ,\tilde \alpha \tilde \beta}_a(l)\,
\nonumber\\
&& \qquad  \quad \times (l+\bar p)^{\bar \tau }\,\bar p_{\rho }\,
 S_{b}(\bar p+l)\,S^{\bar \sigma \tau }_{c}(p+l)\,,
 \nonumber\\
 &&  H^{\mu, \alpha \beta}_{abc}(\bar p, p) = -\,2\,i\,\int
\frac{d^4l}{(2\pi)^4}\,q_\nu\,\epsilon^{\mu \nu}_{\quad  \bar \sigma \bar \tau }\,
p^\alpha\,\epsilon_{\tilde \alpha  \tilde \beta \sigma }^{\quad \;\, \beta }\,S^{ \rho \sigma}_a(l)\,
\nonumber\\
&& \qquad  \quad \times (l+\bar p)^{\bar \tau }\,\bar p_{\rho }\,
 S_{b}(\bar p+l)\,S^{\bar \sigma ,\tilde \alpha  \tilde \beta }_{c}(p+l)\,.
\label{def-EFGH-1plus-d}
\end{eqnarray}
The  decay constant $d_{1^+\to \gamma 0^-}$ defined in (\ref{result:width-1plus:a}) is
computed by contracting the gauge-invariant tensors
(\ref{def-AB-1plus}- \ref{def-D-1plus}, \ref{def-barB-1plus}) and (\ref{def-EFGH-1plus-d})
with the antisymmetric tensor
\begin{eqnarray}
&&  P^{(0-)}_{\mu, \alpha \beta} =-
\frac{1}{2}\,
\Bigg( \Big\{g_{\mu \alpha}- \frac{p^\mu\,q^\alpha}{p\cdot q}\Big\}\,p_\beta
- \Big\{g_{\mu \beta }- \frac{p^\mu\,q^\beta}{p\cdot q}\Big\}\,p_\alpha \Bigg) \,.
\label{def-projections12}
\end{eqnarray}
We provide the results in terms of the master loop integrals $I_{ab}, \bar I_{ab}, J_{abc}$ and
$\bar J_{abc}$ introduced in (\ref{def-master-loops}). According to the
arguments of Section 3.3 reduced tadpole integrals are dropped. Using $p^2=M_i^2$ and
$(p-q)^2=M_f^2$, we derive
\begin{eqnarray}
&& 16\,(p \cdot q)\,M_f^2\,P^{(0-)}_{\mu, \alpha \beta}\,A^{\mu, \alpha \beta }_{ab} =
2\, M_f^2\, \Big[M_i^6-\left(m_a^2-m_b^2\right)^2 M_i^2
\nonumber\\
&&   \qquad\,-\,\left(m_a^2+3\, m_b^2+M_i^2\right) \left(M_i^2-M_f^2\right)
   M_i^2
\nonumber\\
&&   \qquad\,+\,\left(M_f^2-M_i^2\right)^2 \left(-m_a^2+m_b^2+M_i^2\right)\Big]\,I_{ab}
\nonumber\\
&&   \,+\, \Big[  M_f^8-2 \left(m_a^2+m_b^2\right)
   M_f^6+\left(m_a^2-m_b^2\right)^2 M_f^4
\nonumber\\
&&   \qquad\,+\, 2 \left(\left(m_a^2-m_b^2\right)^2-M_f^4\right) M_i^2 M_f^2
\nonumber\\
&&   \qquad\,-\left(m_a^4-2
   \left(m_b^2+M_f^2\right) m_a^2+\left(m_b^2-M_f^2\right)^2\right) M_i^4 \Big]\,\bar I_{ab}
\nonumber\\
&&   \,+\, 4 \,m_a^2 \,M_f^2\,M_i^2 \left(m_a^2-m_b^2-M_f^2\right)
   \left(M_i^2-M_f^2\right) J_{aab}\,,
\nonumber\\ \nonumber\\
&& 24\,M_i^2\,P^{(0-)}_{\mu, \alpha \beta}\,\bar A^{\mu, \alpha \beta }_{ab} =
\left(M_i^2-M_f^2\right) \Big[3\, M_i^6-3 \left(m_a^2-m_b^2\right)^2 M_i^2
\nonumber\\
&&   \qquad\,-\,\left(M_i^2-M_f^2\right)
   \Big(M_i^4+\left(m_a^2+m_b^2\right) M_i^2-2 \left(m_a^2-m_b^2\right)^2\Big)\Big]\,I_{ab}\,,
\nonumber\\ \nonumber\\
&& 16\,(p \cdot q)\,M_f^2\,P^{(0-)}_{\mu, \alpha \beta}\,B^{\mu, \alpha \beta }_{ab} =
M_f^2 \Big[-\left(m_a^2-m_b^2\right) M_f^4-\left(8\, m_b^2+M_f^2\right) M_i^4
\nonumber\\
&&   \qquad\,-\,\Big(M_f^4-\left(5\, m_a^2+3 \,m_b^2\right)
   M_f^2+2 \left(m_a^2-m_b^2\right)^2\Big) M_i^2\Big]\,I_{ab}
\nonumber\\
&&   \,+\,\Big[
   M_f^8-2 \left(m_a^2+m_b^2\right) M_f^6+\left(m_a^2-m_b^2\right)^2
   M_f^4
   \nonumber\\
&&   \qquad\,-\,\Big(M_f^4+\left(m_a^2-m_b^2\right) M_f^2
   -2 \left(m_a^2-m_b^2\right)^2\Big)\, M_i^2\, M_f^2
   \nonumber\\
&&   \qquad\,-\,\Big(-2\, M_f^4+\left(m_a^2-5\,
   m_b^2\right) M_f^2+\left(m_a^2-m_b^2\right)^2\Big)\, M_i^4 \Big]\,\bar I_{ab}
\nonumber\\
&&   \,-\,2 \,m_b^2\, M_f^2 \,M_i^2 \left(M_f^2-M_i^2\right) \left(-2\, m_a^2+2\,
   m_b^2-M_f^2+3\, M_i^2\right) \bar J_{abb} \,,
\nonumber\\ \nonumber\\
&& 12\,P^{(0-)}_{\mu, \alpha \beta}\,\bar B^{\mu, \alpha \beta }_{ab} =
-\left(M_i^2-M_f^2\right) \Big[m_a^4+\left(m_b^2-2\, M_i^2\right) m_a^2
  \nonumber\\
&&   \qquad\,-\,2 \,m_b^4+M_i^4+7\, m_b^2 \,M_i^2\Big]\,I_{ab}\,,
\nonumber\\ \nonumber\\
&& 12\,M_i^2\,P^{(0-)}_{\mu, \alpha \beta}\,\tilde B^{\mu, \alpha \beta }_{ab} =
\left(M_i^2-M_f^2\right) \Big[-3\, M_i^6+3 \left(m_a^2-m_b^2\right)^2 M_i^2
 \nonumber\\
&&   \qquad\,+\,\left(M_i^2-M_f^2\right)
   \Big(M_i^4+\left(m_a^2+m_b^2\right) M_i^2-2 \left(m_a^2-m_b^2\right)^2\Big)\Big]\,I_{ab}\,,
\nonumber\\ \nonumber\\
&& 16\,(p \cdot q)\,M_f^2\,P^{(0-)}_{\mu, \alpha \beta}\,C^{\mu, \alpha \beta }_{abc} =
-4\, m_c^2\, M_f^2 \,\Big[-2\, M_i^6+3\, M_f^2 \,M_i^4
   \nonumber\\
&&   \qquad\,+\,M_f^2 \left(-3\, m_a^2+2 m_b^2+m_c^2-M_f^2\right) M_i^2
   +\left(m_a^2-m_c^2\right) M_f^4\Big]\,I_{ac}
\nonumber\\
&&   \,-\,  4\, m_c^2 \,M_f^2\, M_i^2\, \Big[\left(2\, m_a^2-3\, m_b^2+m_c^2-M_f^2\right) M_f^2
   \nonumber\\
&&   \qquad\, +\,\left(m_b^2-m_c^2+M_f^2\right) M_i^2\Big]\,\bar I_{bc}
\nonumber\\
&&   \,-\,8\, m_c^2 \,M_f^2\, M_i^2 \,\Big[m_b^2 M_f^4-\left(m_a^2-m_b^2\right)^2 M_f^2-2 m_b^2\, M_i^2\, M_f^2
   +m_b^2\, M_i^4\Big]\,J_{abc}\,,
\nonumber\\ \nonumber\\
&& 12\,(p \cdot q)\,M_f^2\,M_i^2\,P^{(0-)}_{\mu, \alpha \beta}\,\bar C^{\mu, \alpha \beta }_{abc} =
-M_f^2\, \Big[2\, \Big(M_f^3-M_f\, M_i^2\Big)^2 \Big(m_a^2-2\,m_c^2\Big)\, m_a^2
   \nonumber\\
&&   \qquad\, -\,\Big(\left(M_f^2-M_i^2\right)^2 \left(M_f^2+3\, M_i^2\right)-3\,
   m_b^2 \left(M_f^4-3 \,M_f^2\, M_i^2\right)\Big)\, M_i^2\, m_a^2
     \nonumber\\
&&   \qquad\,+\,\left(m_c^2-M_i^2\right) \left(M_f^2-M_i^2\right)^2
       \left(-3\, M_i^4+M_f^2\, M_i^2+2\, m_c^2 \,M_f^2\right)
      \nonumber\\
&&   \qquad\,+\, 6\, m_b^4\, M_f^2\, M_i^4-3\, m_b^2\, M_i^2
   \left(M_i^2-M_f^2\right) \left(2\, M_i^4-M_f^2\, M_i^2-m_c^2\, M_f^2\right)\Big]\,I_{ac}
\nonumber\\
&&   \,-\,   3 \,m_b^2\, M_f^2 \,M_i^4 \Big[\left(2 \,m_a^2-3\,
   m_b^2+m_c^2-M_f^2\right) M_f^2+\left(m_b^2-m_c^2+M_f^2\right) M_i^2\Big]\,\bar I_{bc}
\nonumber\\
&&   \,-\, 6\, m_b^2\, M_f^2\, M_i^4\, \Big[m_b^2\,
   M_f^4-\left(m_a^2-m_b^2\right)^2 M_f^2-2 \,m_b^2 \,M_i^2\, M_f^2+m_b^2\, M_i^4\Big]\,J_{abc}\,,
\nonumber\\ \nonumber\\
&& 48\,M_f^2\,P^{(0-)}_{\mu, \alpha \beta}\,D^{\mu, \alpha \beta }_{ab} =
2\, M_f^2\, \Big[-3 \,m_b^2 \left(-m_a^2+m_b^2+M_i^2\right) M_i^2
     \nonumber\\
&&   \qquad\,-\,\left(M_i^2-M_f^2\right) \left(m_a^4+\left(m_b^2-2\,
   M_i^2\right) m_a^2-2\, m_b^4+M_i^4+m_b^2\, M_i^2\right)\Big]\,I_{ab}
\nonumber\\
&&   \,+\,  \Big[ 6 \,\Big(m_a^4-2 \left(m_b^2+M_f^2\right)
   m_a^2+\left(m_b^2-M_f^2\right)^2\Big) \,M_i^4
      \nonumber\\
&&   \qquad\, -\,6 \,M_f^2 \left(m_a^4-\left(m_b^2+2\, M_f^2\right) m_a^2+M_f^4-3 \,m_b^2 \,M_f^2\right)
   M_i^2 \Big]\,\bar I_{ab}
\nonumber\\
&&   \,+\,   12 \,m_b^4\, M_f^2\, M_i^2 \left(M_i^2-M_f^2\right) \bar J_{abb}\,,
\end{eqnarray}
\begin{eqnarray}
&& 24\,m_b^2\,(p \cdot q)\,M_f^2\,P^{(0-)}_{\mu, \alpha \beta}\,E^{\mu, \alpha \beta }_{+,ab} =
 M_f^2 \,\Big[24\, m_b^2 \,\Big(\left(m_a^2-m_b^2\right)^2-M_i^4\Big)\, M_i^2
         \nonumber\\
&&   \qquad\,  +\,3 \left(M_i^2-M_f^2\right)
   \Big(m_a^4+20 \,m_b^2\, m_a^2+11\, m_b^4+M_i^4
\nonumber\\
   &&   \qquad\,-\,2 \left(m_a^2+2\, m_b^2\right) M_i^2\Big)\, M_i^2
  -2\left(M_f^2-M_i^2\right)^2
   \Big(m_a^4+\left(7 \,m_b^2-2\, M_i^2\right) m_a^2
         \nonumber\\
&&   \qquad\,
   -\,8\, m_b^4+M_i^4+m_b^2\, M_i^2\Big)\Big]\,I_{ab}
\nonumber\\
&&   \,   -3\,
   \Big[-\left(m_a^2+3\, m_b^2\right) M_f^2\, M_i^6
   +4 \left(m_b^3-m_a^2 m_b\right)^2 \Big( 2\,M_f^2\,M_i^2 -M_i^4\Big)
        \nonumber\\
&&   \qquad\,-\left(-M_f^6+10\, m_b^2 \,M_f^4+\left(m_a^4-8\,
   m_b^2 \,m_a^2-9 \,m_b^4\right) M_f^2\right) M_i^4
        \nonumber\\
&&   \qquad\,+\,M_f^2 \left(-M_f^6+\left(m_a^2+m_b^2\right)
   M_f^4+2 \,m_b^2 \left(m_a^2-m_b^2\right) M_f^2\right) M_i^2
        \nonumber\\
&&   \qquad\,+\,4\, m_b^2\, M_f^4 \,\Big(m_a^4-2
   \left(m_b^2+M_f^2\right) m_a^2+\left(m_b^2-M_f^2\right)^2\Big)
        \nonumber\\
&&   \qquad\,+\,\left(m_a^2-m_b^2\right)^2 M_i^6   \Big]\,\bar I_{ab}
\nonumber\\
&&   \,+\, 6\, m_b^2\, M_f^2\, M_i^2 \left(M_i^2-M_f^2\right)
  \Big[-8\,m_b^4+7\, M_f^2\, m_b^2-2 \,M_f^4+M_i^4
       \nonumber\\
&&   \qquad\,+\left(m_b^2+M_f^2\right) M_i^2
  +m_a^2 \left(8\, m_b^2+3\, M_f^2-3\, M_i^2\right)\Big] \,\bar J_{abb}\,,
\nonumber\\ \nonumber\\
&& 24\,m_a^2\,(p \cdot q)\,M_f^2\,P^{(0-)}_{\mu, \alpha \beta}\,E^{\mu, \alpha \beta }_{-,ab} =
M_f^2 \Big[24\, m_b^2 \,\Big(\left(m_a^2-m_b^2\right)^2-M_i^4\Big)\, M_i^2
    \nonumber\\
&&   \qquad\,+\,3 \left(M_i^2-M_f^2\right)
   \Big(m_a^4+20 \,m_b^2 \,m_a^2+11\, m_b^4+M_i^4
    \nonumber\\
&&   \qquad\,-2\, \left(m_a^2+2 \,m_b^2\right) M_i^2\Big)\, M_i^2
    +2\left(M_f^2-M_i^2\right)^2
   \Big(m_a^4+\left(m_b^2-2\, M_i^2\right) m_a^2
       \nonumber\\
&&   \qquad\,-\,2 \,m_b^4+M_i^4+7\, m_b^2 \,M_i^2\Big)\Big]\,I_{ab}
\nonumber\\
&&   \,-\, 3\,
   \Big[-\left(m_a^2+3\, m_b^2\right) M_f^2\, M_i^6
    +4 \left(m_b^3-m_a^2 \,m_b\right)^2 \Big( 2\,M_f^2\,M_i^2 -M_i^4\Big)
     \nonumber\\
&&   \qquad\,-\,\left(-M_f^6+10 \,m_b^2\, M_f^4+\left(m_a^4-8\,
   m_b^2 \,m_a^2-9\, m_b^4\right) M_f^2\right) M_i^4
     \nonumber\\
&&   \qquad\,+\,M_f^2 \left(-M_f^6+\left(m_a^2+m_b^2\right)
   M_f^4+2 \,m_b^2 \left(m_a^2-m_b^2\right) M_f^2\right) M_i^2
     \nonumber\\
&&   \qquad\,+\,4 \,m_b^2\, M_f^4\, \Big(m_a^4-2
   \left(m_b^2+M_f^2\right) m_a^2+\left(m_b^2-M_f^2\right)^2\Big)
          \nonumber\\
&&   \qquad\,+\,\left(m_a^2-m_b^2\right)^2 M_i^6\Big]\, \bar I_{ab}
\nonumber\\
&&   \,+\,6\, m_b^2 \,M_f^2\, M_i^2 \left(M_i^2-M_f^2\right) \Big[-8\,
   m_b^4+7\, M_f^2 \,m_b^2-2\, M_f^4+M_i^4
         \nonumber\\
&&   \qquad\,+\,\left(m_b^2+M_f^2\right) M_i^2
   +m_a^2 \left(8\, m_b^2+3\, M_f^2-3\, M_i^2\right)\Big]\,\bar J_{abb}\,,
\nonumber\\ \nonumber\\
&& 12\,M_i^2\,P^{(0-)}_{\mu, \alpha \beta}\,\bar E^{\mu, \alpha \beta }_{ab} =
\left(M_i^2-M_f^2\right) \Big[-7\, M_i^6+\left(5 \left(m_a^2+m_b^2\right)-M_f^2\right) M_i^4
 \nonumber\\
&&   \qquad\,+\,\Big(2
   \left(m_a^2-m_b^2\right)^2-\left(m_a^2+m_b^2\right) M_f^2\Big)\, M_i^2
    \nonumber\\
&&   \qquad\,+\,2 \left(m_a^2-m_b^2\right)^2 M_f^2\Big]\,I_{ab}\,,
\end{eqnarray}
\begin{eqnarray}
&& 24\,m_b^2\,(p \cdot q)\,M_f^2\,P^{(0-)}_{\mu, \alpha \beta}\,F^{\mu, \alpha \beta }_{+,ab} =
2 \,M_f^2 \,\Big[12\, m_a^2 \left(-m_a^2+m_b^2+M_i^2\right)^2 M_i^2
\nonumber\\
&&   \qquad\,+\,3\, m_a^2 \left(7\, m_a^2+9\, m_b^2-7\, M_i^2\right)
   \left(M_i^2-M_f^2\right) M_i^2
   \nonumber\\
&&   \qquad\,+\,\left(M_f^2-M_i^2\right)^2
   \Big(-7 \,m_a^4+\left(5 \,m_b^2+11\, M_i^2\right) m_a^2
 \nonumber\\
&&   \qquad\, +\, 2 \left(m_b^2-M_i^2\right)^2\Big)\Big]\,I_{ab}
\nonumber\\
&&   \,+\, 6\, m_a^2 \Big[-2 \,M_f^8+4 \left(m_a^2+m_b^2\right) M_f^6-2 \left(m_a^2-m_b^2\right)^2 M_f^4
\nonumber\\
&&   \qquad\,+\,\left(4 \,m_a^2-4\, m_b^2+M_f^2\right) \left(-m_a^2+m_b^2+M_f^2\right) M_i^2\, M_f^2
\nonumber\\
&&   \qquad\,+\,\Big(-3 M_f^4+\left(m_a^2-9\, m_b^2\right)
   M_f^2+2 \left(m_a^2-m_b^2\right)^2\Big) M_i^4\Big]\, \bar I_{ab}
\nonumber\\
&&   \,-\,12 \,m_a^2\, m_b^2 \,M_f^2 \,M_i^2 \left(4\, m_a^2-4\, m_b^2+M_f^2-5 \,M_i^2\right)
   \left(M_f^2-M_i^2\right) \bar J_{abb}\,,
\nonumber\\ \nonumber\\
&& 24\,m_b^2\,(p \cdot q)\,M_f^2\,P^{(0-)}_{\mu, \alpha \beta}\,F^{\mu, \alpha \beta }_{-,ab} =
6\, m_b^2\, M_f^2  \Big[ 4\left(-m_a^2+m_b^2+M_i^2\right)^2 M_i^2
\nonumber\\
&&   \qquad\,+\, \left(M_i^2-M_f^2\right)\left(7\, m_a^2+9\, m_b^2-7\, M_i^2\right) M_i^2
\nonumber\\
&&   \qquad\,+\,\left(m_a^2-m_b^2+M_i^2\right) \left(M_i^2-M_f^2\right)^2\Big]\,I_{ab}
\nonumber\\
&&   \,-\, 6\, m_b^2  \Big[2\, \Big(m_a^4-2 \left(m_b^2+M_f^2\right) m_a^2+\left(m_b^2-M_f^2\right)^2\Big)
   M_f^4
   \nonumber\\
&&   \qquad\,-\,\left(4 \,m_a^2-4\, m_b^2+M_f^2\right) \left(-m_a^2+m_b^2+M_f^2\right) M_i^2 \,M_f^2
\nonumber\\
&&   \qquad\,+\,  \Big(3\, M_f^4-\left(m_a^2-9\, m_b^2\right)
   M_f^2-2 \left(m_a^2-m_b^2\right)^2\Big)\, M_i^4 \Big]\, \bar I_{ab}
\nonumber\\
&&   \,+\,   12\, m_b^4\, M_f^2\, M_i^2 \left(M_f^2-M_i^2\right) \left(-4\,
   m_a^2+4\, m_b^2-M_f^2+5\, M_i^2\right)\, \bar J_{abb}\,,
\end{eqnarray}
\begin{eqnarray}
&& 2\,P^{(0-)}_{\mu, \alpha \beta}\,\hat F^{\mu, \alpha \beta }_{+,ab} =
m_a^2 \left(-2\, M_i^4+M_f^2\, M_i^2+\left(m_a^2-m_b^2 \right) M_f^2\right) I_{ab}
\nonumber\\
&&   \,+\,m_a^2 \left(-m_a^2+m_b^2+M_f^2\right) M_i^2\,\bar I_{ab}
+2\, m_a^2\, m_b^2\, M_i^2 \left(M_i^2-M_f^2\right) \bar J_{abb}\,,
\nonumber\\ \nonumber\\
&& 6\,M_i^2\,P^{(0-)}_{\mu, \alpha \beta}\,\hat F^{\mu, \alpha \beta }_{-,ab} =
\Big[3\, M_i^8-\left(3 \,m_a^2+9\, m_b^2+4 \,M_f^2\right) M_i^6
 \nonumber\\
&&   \qquad\,+\,M_f^2 \left(2\, m_a^2+5\, m_b^2+M_f^2\right) M_i^4 -2 \left(m_a^2-m_b^2\right)^2 M_f^4
 \nonumber\\
&&   \qquad\,+\,M_f^2 \left(2\,
   m_a^4-m_b^2\, m_a^2-m_b^4+\left(m_a^2+m_b^2\right) M_f^2\right) M_i^2
    \Big]\,I_{ab}
\nonumber\\
&&   \,+\, 3\, m_b^2   \left(-m_a^2+m_b^2+M_f^2\right) M_i^4\,\bar I_{ab}
\nonumber\\
&&   \,+\,   6 \,m_b^4\, M_i^4 \left(M_i^2-M_f^2\right)\bar J_{abb}\,
\end{eqnarray}
\begin{eqnarray}
&& 12\,(p \cdot q)\,M_f^2\,P^{(0-)}_{\mu, \alpha \beta}\,G^{\mu, \alpha \beta }_{abc} =
3\, m_c^2\, M_f^2 \,\Big[2\, M_i^6-3\, M_f^2\, M_i^4
\nonumber\\
&&   \qquad\,+\,M_f^2 \left(-m_a^2-2 m_b^2+3 \,m_c^2+M_f^2\right) M_i^2
   +\left(m_a^2-m_c^2\right)
   M_f^4\Big]\,I_{ac}
\nonumber\\
&&   \,+\,M_i^2 \Big[\Big(3\, m_c^2 \left(3 M_f^2-M_i^2\right) M_f^2
+4   \left(m_a^2+M_f^2\right) \left(M_f^2-M_i^2\right)^2\Big)\, m_b^2
\nonumber\\
&&   \qquad\,-\,2\left(M_f^2-M_i^2\right)^2 m_b^4-6 \,m_c^4 \,M_f^4-2\left(m_a^2-M_f^2\right)^2
   \left(M_f^2-M_i^2\right)^2
   \nonumber\\
&&   \qquad\,+\,3\, m_c^2 \,M_f^2 \left(M_f^2-m_a^2\right) \left(M_f^2-M_i^2\right)\Big]\,\bar I_{ab}
\nonumber\\
&&   \,+\, 6\, m_c^2\, M_f^2\,
   M_i^2 \,\Big[ M_f^2 \,m_b^4-\Big(2\, m_c^2\, M_f^2+\left(M_f^2-M_i^2\right)^2\Big)\, m_b^2+m_c^4\, M_f^2\Big]\,\bar J_{abc}\,,
\nonumber\\ \nonumber\\
&& 48\,(p \cdot q)\,M_f^2\,P^{(0-)}_{\mu, \alpha \beta}\,H^{\mu, \alpha \beta }_{abc} =
-2\, M_f^2\, \Big[6 \left(m_b^2-m_c^2\right)  \left(m_a^2+m_b^2-M_i^2\right) M_i^4
 \nonumber\\
&&   \qquad\,+\,3
   \left(M_i^2-M_f^2\right) \Big(m_a^4+\left(-3 \,m_b^2+3\, m_c^2-2\, M_i^2\right) m_a^2
   \nonumber\\
&&   \qquad\,-\,\left(m_c^2-M_i^2\right) \left(m_b^2+M_i^2\right)\Big)\, M_i^2
 - \left(M_f^2-M_i^2\right)^2 \Big(4 \,m_a^4
    \nonumber\\
&&   \qquad\,-\,\left(2\, m_c^2+5\, M_i^2\right)
   m_a^2-2\, m_c^4+M_i^4+7\, m_c^2 M_i^2\Big)\Big]\,I_{ac}
\nonumber\\
&&   \,-\, 6 \,M_i^2 \Big[\left(2\, M_f^4-3\, M_i^2 M_f^2+M_i^4\right) m_a^4
 -2 \,\Big(m_b^2
   \left(3 \,M_f^4-3\, M_i^2\, M_f^2+M_i^4\right)
     \nonumber\\
&&   \qquad\,-\,M_f^2 \left(2\, M_f^2-M_i^2\right) \left(m_c^2-M_f^2+M_i^2\right)\Big)
   m_a^2
-\left(m_b^2-M_f^2\right) \Big(2 \,M_f^6
      \nonumber\\
&&   \qquad\,-\,\left(-3 \,m_b^2+2 \,m_c^2+3 \,M_f^2\right) M_i^2\, M_f^2
   +\left(M_f^2-m_b^2\right) M_i^4\Big)\Big]\,\bar I_{ab}
\nonumber\\
&&   \,-\,12\, M_f^2 \,M_i^2 \Big[m_c^2 \left(M_i^2-M_f^2\right)^3+\left(m_b^2-m_c^2\right)^2 \left(2\,
   m_a^2-M_i^2\right) \left(M_i^2-M_f^2\right)
    \nonumber\\
&&   \qquad\,+\,m_c^2 \left(m_a^2+m_b^2-M_i^2\right)
   \left(M_f^2-M_i^2\right)^2
      \nonumber\\
&&   \qquad\,-\,\left(m_b^2-m_c^2\right)^2 M_i^2 \left(m_a^2+m_b^2-M_i^2\right)\Big]\,\bar J_{abc} \,.
\end{eqnarray}


\section*{Appendix F}

We derive the contribution of the $\phi\,D_s^*$ and $K^* D^*$ channels to the
$1^+ \to \gamma \,0^+$ process.
According to (\ref{result:width-1plus:b}) it suffices to evaluate the contractions of the four loop tensors
introduced in (\ref{def-A-1plus:0plus}, \ref{def-D-1plus:0plus}, \ref{def-BC-1plus:0plus}) with the projector
\begin{eqnarray}
P^{(0+)}_{\mu, \alpha \beta} = -\frac{1}{2}\,q^\tau\,p^\sigma\,\Big\{
\epsilon_{\mu \tau \sigma \alpha} \,p_\beta -\epsilon_{\mu \tau \sigma \beta} \,p_\alpha\Big\}\,.
\end{eqnarray}
We have
\begin{eqnarray}
&&\frac{2\,M_f^2}{(p \cdot q)\,M_i^2}\,P^{(0+)}_{\mu, \alpha \beta }\,A^{\mu, \alpha \beta}_{+,abc} =
2\, m_c^2\, M_f^2\, I_{ac}+ 2\, m_c^2 \left(m_c^2-m_b^2\right) M_f^2\, \bar J_{abc}
\nonumber\\
&& \,-\,  \Big[\left(2\,m_c^2+\left(M_i^2-M_f^2\right)\right) M_f^2+
   \left(m_a^2-m_b^2\right) \left(M_f^2-M_i^2\right)\Big]\,\bar I_{ab}\,,
\nonumber\\ \nonumber\\
&&\,\frac{24\,m_b^2\,M_f^2}{(p \cdot q)}\,P^{(0+)}_{\mu, \alpha \beta }\,B^{\mu, \alpha \beta}_{+,ab} =
M_f^2 \Big[-2 M_i^6+\left(m_a^2-11\, m_b^2-M_f^2\right) M_i^4
\nonumber\\
&& \qquad \,+\,\left(m_a^4+5 \left(8 \,m_b^2+M_f^2\right) m_a^2+7
   \left(m_b^4-m_b^2\, M_f^2\right)\right) M_i^2\nonumber\\
&& \qquad \,-\,2 \left(2\, m_a^4-m_b^2\, m_a^2-m_b^4\right) M_f^2\Big]\, I_{ab}
\nonumber\\
&& \, +\, \Big[3 \left(m_a^4-m_b^4+M_f^4-2
   \left(m_a^2-2 \,m_b^2\right) M_f^2\right) M_i^4
   \nonumber\\
&& \qquad \,-\,6\, m_b^2 \,M_f^2 \left(7 \,m_a^2+m_b^2-M_f^2\right) M_i^2\Big]\,\bar I_{ab}
\nonumber\\
&& \,+\,   6\, m_b^2\, M_f^2\, M_i^2 \left(3\,
   m_a^2+m_b^2-M_f^2-2\, M_i^2\right) \left(M_f^2-M_i^2\right) \bar J_{abb}\,,
\nonumber\\ \nonumber\\
&&\,\frac{4\,M_f^2}{(p \cdot q)}\,P^{(0+)}_{\mu, \alpha \beta }\,C^{\mu, \alpha \beta}_{+,ab} =
M_f^2\, \Big[\left(M_f^2+7\, M_i^2\right) m_b^2
 \nonumber\\
&& \qquad \,+\,\left(m_a^2+M_i^2\right) \left(M_i^2-M_f^2\right)\Big]\,I_{ab}
\nonumber\\
&& \,+\,M_i^2\,\Big[\left(m_a^2+M_f^2\right) \left(M_i^2-M_f^2\right)-m_b^2 \left(7\, M_f^2+M_i^2\right)\Big]\,\bar I_{ab}
\nonumber\\
&& \,+\,8\, m_b^2\, M_f^2\, M_i^2   \left(M_f^2-M_i^2\right) \bar J_{abb}\,,
\nonumber\\ \nonumber\\
&&24\,P^{(0+)}_{\mu, \alpha \beta }\,D^{\mu, \alpha \beta}_{+,abc} =
 \Big[3\, m_b^2 \left(m_a^2+3\, m_c^2+M_i^2\right) \left(M_i^2-M_f^2\right)    M_i^2
   \nonumber\\
&& \qquad\,-\, \left(M_f^2-M_i^2\right)^2 \left(m_a^4+\left(4\, m_c^2-2 \,M_i^2\right) m_a^2
   -5\, m_c^4+M_i^4+4\, m_c^2\, M_i^2\right)
 \nonumber\\
&& \qquad\,+\,6\, m_b^2 \left(m_b^2-m_a^2\right) M_i^4 \Big]\,I_{ac}
\nonumber\\
&&\,+\,   3\, m_b^2\, M_i^2\, \Big[  \left(\left(3\, m_c^2+M_f^2\right)
   \left(M_f^2-M_i^2\right)-m_b^2 \left(M_f^2+M_i^2\right)\right)
\nonumber\\
&& \qquad\,+\,2\,m_a^2\, M_f^2\Big]\, \bar I_{bc}
\nonumber\\
&& \,+\,   6 \,m_b^2\, M_i^2\, \Big[-\left(m_a^2-m_b^2\right)^2
   M_i^2-m_a^2 \left(M_f^2-M_i^2\right)^2
\nonumber\\
&& \qquad\,+\,\left(m_a^2-m_b^2\right) \left(m_a^2+m_c^2+M_i^2\right)
   \left(M_i^2-M_f^2\right)\Big]\, J_{abc}\,,
\label{}
\end{eqnarray}
where we use $p^2=M_i^2$ and $(p-q)^2=M_f^2$.


\section*{Appendix G}

We collect explicit expressions for the contributions to the decay amplitude
(\ref{def-transition-amplitude-axial-b}) which are linear in coupling constant $\hat g_H$ (see
(\ref{def-gH})). We have
\begin{eqnarray}
&& -i\,M^{\bar \alpha \bar \beta, \mu, \alpha \beta }_{1^+ \to \gamma\,1^-}=
\frac{e\,g_T\,\hat g_H}{4\,f^2}\, \Big\{
F_{+,\phi D_s^*}^{ \bar \alpha \bar \beta, \mu , \alpha \beta }(\bar p, p)
-E_{-, D^*K^*}^{ \bar \alpha \bar \beta, \mu , \alpha \beta }(\bar p, p)
+F_{+, K^*D^*}^{ \bar \alpha \bar \beta, \mu , \alpha \beta }(\bar p, p)
 \Big\}
\nonumber\\
&& \quad  \;\;\;
+\,\frac{e\,\tilde g_V\,\hat g_H}{4\,f^2}\, \Big\{
F_{-,\phi D_s^*}^{ \bar \alpha \bar \beta, \mu , \alpha \beta }(\bar p, p)
-E_{+, D^*K^*}^{ \bar \alpha \bar \beta, \mu , \alpha \beta }(\bar p, p)
+F_{-, K^*D^*}^{ \bar \alpha \bar \beta, \mu , \alpha \beta }(\bar p, p)
 \Big\}
\nonumber\\
&& \quad  \;\;\;
+\,\frac{e\,\tilde g_V\,\hat g_H}{4\,f^2}\, \Big\{
F_{\phi D_s^*}^{ \bar \alpha \bar \beta, \mu , \alpha \beta }(\bar p, p)
-E_{ D^*K^*}^{ \bar \alpha \bar \beta, \mu , \alpha \beta }(\bar p, p)
+F_{ K^*D^*}^{ \bar \alpha \bar \beta, \mu , \alpha \beta }(\bar p, p)
 \Big\}
\nonumber\\
&& \quad  \;\;\;
+\,\frac{\hat g_H}{8\,f^2\,M_V^2}\,\Big[ \tilde e_C-e+ \frac{\tilde e_Q}{3}\Big]\, \Big\{
g_T\,\hat F_{+,\phi D_s^*}^{ \bar \alpha \bar \beta, \mu , \alpha \beta }(\bar p, p)
\nonumber\\
&& \qquad \qquad \quad  \;\;\;+ \tilde g_V\,(\hat F_{\phi D_s^*}^{ \bar \alpha \bar \beta, \mu , \alpha \beta }(\bar p, p)
+\hat F_{-,\phi D_s^*}^{ \bar \alpha \bar \beta, \mu , \alpha \beta }(\bar p, p))
 \Big\}
 \nonumber\\
&& \quad  \;\;\;
+\,\frac{\hat g_H}{4\,f^2\,M_V^2}\,\Big[ \tilde e_C-\frac{e}{2}- \frac{\tilde e_Q}{6}\Big]\, \Big\{
g_T\,\hat F_{+, K^*D^*}^{ \bar \alpha \bar \beta, \mu , \alpha \beta }(\bar p, p)
\nonumber\\
&& \qquad \qquad \quad  \;\;\;+ \tilde g_V\,(\hat F_{ K^*D^*}^{ \bar \alpha \bar \beta, \mu , \alpha \beta }(\bar p, p)
+ \hat F_{-, K^*D^*}^{ \bar \alpha \bar \beta, \mu , \alpha \beta }(\bar p, p)) \Big\}
\nonumber\\
&& \quad  \;\;\;
-\,\frac{\hat g_H}{4\,f^2\,m_V^2}\, \Big\{
e_T\, \bar E_{-, D^*K^*}^{ \bar \alpha \bar \beta, \mu , \alpha \beta }(\bar p, p)
+\tilde e_V\,\Big( \bar E_{+, D^*K^*}^{ \bar \alpha \bar \beta, \mu , \alpha \beta }(\bar p, p)
+\bar E_{D^*K^*}^{ \bar \alpha \bar \beta, \mu , \alpha \beta }(\bar p, p)\Big) \Big\}
\nonumber\\
&& \quad  \;\;\; +\,
\frac{\tilde e_E\,\hat g_H }{6\,f^2\,m_V^2}\, \Big\{ \tilde E_{ D^*K^*}^{ \bar \alpha \bar \beta, \mu,\alpha \beta }(\bar p, p) +
2\,\tilde E_{D_s^* \phi}^{ \bar \alpha \bar \beta, \mu ,\alpha \beta}(\bar p, p) \Big\}
\nonumber\\
&& \quad  \;\;\;
+\,\frac{(e_C+e_Q/6)\,\hat g_H}{8\,f^2\,M_V^2}\, \Big\{ \tilde g_T \,H_{K^*D D^*}^{ \bar \alpha \bar \beta, \mu , \alpha \beta }(\bar p, p)
+ g_E\,\bar H_{K^*DD^*}^{ \bar \alpha \bar \beta, \mu , \alpha \beta }(\bar p, p) \Big\}
\nonumber\\
&& \quad  \;\;\;
+\,\frac{(e_C-e_Q/3)\,\hat g_H}{16\,f^2\,M_V^2}\,\Big\{ \tilde g_T \,H_{\phi D_s D_s^*}^{ \bar \alpha \bar \beta, \mu , \alpha \beta }(\bar p, p)
+ g_E\,\bar H_{\phi D_s D_s^*}^{ \bar \alpha \bar \beta, \mu , \alpha \beta }(\bar p, p) \Big\}
\nonumber\\
&& \quad  \;\;\;
-\,\frac{e_A\,\tilde g_P\,\hat g_H}{96\,f^3\,m_V}\, \Big\{ G_{D^*K K^*}^{ \bar \alpha \bar \beta, \mu , \alpha \beta }(\bar p, p)
+ \bar G_{D^*K K^*}^{ \bar \alpha \bar \beta, \mu , \alpha \beta }(\bar p, p) \Big\}
\nonumber\\
&& \quad  \;\;\;
-\,\frac{e_A\,\tilde g_P\,\hat g_H}{64\,f^3\,m_V}\, \Big\{ G_{D_s^*\eta \phi}^{ \bar \alpha \bar \beta, \mu , \alpha \beta }(\bar p, p)
+ \bar G_{D_s^*\eta \phi}^{ \bar \alpha \bar \beta, \mu , \alpha \beta }(\bar p, p) \Big\} \,,
\label{decay-amplitude-1plus-AppendixG}
\end{eqnarray}
with
\begin{eqnarray}
&& E^{\bar \alpha \bar \beta, \mu, \alpha \beta }_{ab}(\bar p, p) =+\, i\,\int
\frac{d^4l}{(2\pi)^4}\,\epsilon_{\tilde \alpha \tilde \beta \rho}^{\quad \;\, \beta}\,
S^{ \bar \sigma ,\tilde \alpha \tilde \beta}_a(l)\, \Bigg\{
 \bar p^{\bar \alpha}\,p^{\alpha}\,
S^{\mu \bar \beta}_{\bar p,\;\, \bar \sigma \tau}(p)\,S_b^{\tau \rho} (p+l)
\nonumber\\
&& \qquad  \quad +\,g^{\bar \alpha}_{\;\bar \sigma}\,p^\alpha \,S_{b}^{\bar \beta ,\mu \rho  }(\bar p+l)
+g^{\bar \alpha}_{\;\bar \sigma}\,g^{\mu \alpha} \,S_{b}^{\bar \beta \rho  }(\bar p+l)
+g^{\bar \alpha}_{\;\bar \sigma}\,p^\alpha \, S_{b}^{\mu \bar \beta ,\rho  }(p+l)
\nonumber\\
&& \qquad  \quad +\,
g^{\bar \alpha}_{\;\bar \sigma}\,g_{\bar \kappa \kappa}\,p^\alpha \,\Big\{ S_{b}^{\bar \beta \bar \kappa  }(\bar p+l)\,S^{\mu \kappa, \rho }_{b}(p+l)
+ S_{b}^{ \bar \beta ,\mu \bar \kappa }(\bar p+l)\,S^{ \kappa \rho}_{b}(p+l)\Big\} \Bigg\}
\,,
\nonumber\\
&& E^{\bar \alpha \bar \beta, \mu, \alpha \beta}_{+,ab}(\bar p, p) = +\, i\,\int
\frac{d^4l}{(2\pi)^4}\,\epsilon_{\tilde \alpha \tilde \beta  \rho}^{\quad \;\,\beta}\,g_{\bar \sigma \sigma}\,
S^{ \bar \sigma \bar \beta  ,\tilde \alpha \tilde \beta}_a(l)\,\Bigg\{
\bar p^{\bar \alpha }\, p^\alpha \,S_{b}^{\sigma ,\mu \rho }(\bar p+l)
\nonumber\\
&& \qquad  \quad +\,\bar p^{\bar \alpha }\, g^{\mu \alpha } \,S_{b}^{\sigma  \rho }(\bar p+l)
+\bar p^{\bar \alpha }\,p^\alpha \, S_{b}^{\mu \sigma ,\rho }(p+l)
+g^{\bar \alpha \mu}\, p^\alpha \,S_{b}^{\sigma \rho  }( p+l)
\nonumber\\
&& \qquad  \quad + \,\bar p^{\bar \alpha }\,p^\alpha \,g_{\bar \kappa \kappa}\,
\Big\{ S_{b}^{\sigma \bar \kappa  }(\bar p+l)\,S^{\mu \kappa, \rho }_{b}(p+l)
+ S_{b}^{ \sigma ,\mu \bar \kappa }(\bar p+l)\,S^{ \kappa \rho}_{b}(p+l)\Big\} \Bigg\}
\,,
\nonumber\\
&& E^{\bar \alpha \bar \beta, \mu, \alpha \beta}_{-,ab}(\bar p, p) =+\, i\,\int
\frac{d^4l}{(2\pi)^4}\,\epsilon_{\tilde \alpha \tilde \beta \rho}^{\quad \; \, \beta}\,g_{\bar \sigma \sigma}\,
S_{a}^{\bar \sigma ,\tilde \alpha \tilde \beta} (l)\, \Bigg\{
g^{\mu \bar \alpha }\,p^\alpha \,S_{b}^{\sigma \bar \beta,\rho  }(p+l)
\nonumber\\
&& \qquad  \quad  +\, \bar p^{\bar \alpha }\,p^\alpha \,g_{\bar \kappa \kappa}\,
\Big\{ S_{b}^{\sigma \bar \beta,\bar \kappa  }(\bar p+l)\,S^{\mu \kappa, \rho }_{b}(p+l)
+ S_{b}^{ \sigma \bar \beta ,\mu \bar \kappa }(\bar p+l)\,S^{ \kappa \rho}_{b}(p+l)\Big\}
\nonumber\\
&& \qquad  \quad +\,\bar p^{\bar \alpha}\,g^{\mu \alpha} \,S_{b}^{\sigma \bar \beta,\rho }(\bar p+l)
+\bar p^{\bar \alpha}\,p^\alpha \,S_{b}^{\sigma \bar \beta,\mu \rho }(\bar p+l)
 \Bigg\}\,,
\nonumber\\
&& \bar E^{\bar \alpha\bar \beta, \mu, \alpha \beta}_{ab}(\bar p, p) =\,i\,\int
\frac{d^4l}{(2\pi)^4}\,\epsilon_{\tilde \alpha \tilde \beta \rho}^{\quad \;\, \beta}\,p^\alpha\,S^{ \bar \alpha,
\tilde \alpha \tilde \beta }_a(l)\, S_{b}^{\tau \rho  }(p+l)\,\Big( g^{\mu
}_{\;\;\tau}\,q^{\bar \beta} - g^{\mu \bar \beta}\,q_\tau \Big) \,,
\nonumber\\
&& \tilde E^{\bar \alpha\bar \beta, \mu, \alpha \beta}_{ab}(\bar p, p) =
\,2\,i\,\int \frac{d^4l}{(2\pi)^4}\,\epsilon_{\tilde \alpha \tilde \beta \rho}^{\quad \;\, \beta}\,p^\alpha\,
S_{a}^{\bar \beta ,\tilde \alpha \tilde \beta } (l)\, \bar
p^{\bar \alpha }\,S_{b}^{\mu \tau ,\rho  }(p+l) \,q_\tau \,,
\nonumber\\
&& \bar E^{\bar \alpha\bar \beta, \mu, \alpha \beta}_{+,ab}(\bar p, p) =
\, i\,\int \frac{d^4l}{(2\pi)^4}\,\epsilon_{\tilde \alpha \tilde \beta \rho}^{\quad \;\, \beta}\,p^\alpha\,
S^{ \sigma \bar \beta  ,\tilde \alpha \tilde \beta}_a(l)\, \bar p^{\bar \alpha }\,
S_{b}^{\tau \rho }(p+l)\,
\nonumber\\
&& \qquad  \quad \times \Big( g^{\mu}_{\;\;\tau}\,q_\sigma -
g^{\mu}_{\;\;\sigma}\,q_\tau \Big) \,,
\nonumber\\
&& \bar E^{\bar \alpha\bar \beta, \mu, \alpha \beta}_{-,ab}(\bar p, p) =
\,i\,\int \frac{d^4l}{(2\pi)^4}\,\epsilon_{\tilde \alpha \tilde \beta \rho}^{\quad \;\, \beta}\,p^\alpha\,
S_{a}^{\sigma ,\tilde \alpha \tilde \beta } (l)\, \bar p^{\bar \alpha}\,S_{b}^{\tau
\bar \beta ,\rho  }(p+l) \,
\nonumber\\
&& \qquad  \quad \times \Big( g^{\mu \tau}\,q_\sigma - g^{\mu
\sigma}\,q_\tau \Big)\,,
\nonumber\\
&& F^{\bar \alpha \bar \beta, \mu, \alpha \beta }_{ab}(\bar p, p) =- i\,\int
\frac{d^4l}{(2\pi)^4}\,\epsilon_{\tilde \alpha \tilde \beta \bar \rho}^{\quad \;\, \beta}\,S^{ \bar \sigma \bar \rho}_a(l)\,
\Bigg\{
\bar p^{\bar \alpha}\,p^{\alpha}\,S^{\mu \bar \beta}_{\bar p,\;\,\bar  \sigma \tau}(p)\,S_b^{\tau , \tilde \alpha \tilde \beta} (p+l)
\nonumber\\
&& \qquad  \quad
+\, g^{\bar \alpha}_{\; \bar \sigma}\,g^{\mu \alpha}\,S_{b}^{\bar \beta ,\tilde \alpha \tilde \beta   }(\bar p+l)
+
g^{\bar \alpha }_{\;\bar \sigma }\,p^\alpha\,S_{b}^{\mu \bar \beta ,\tilde \alpha \tilde \beta  }(p+l)
\nonumber\\
&& \qquad  \quad +\,
g^{\bar \alpha }_{\;\bar \sigma }\,g_{\bar \kappa \kappa}\,p^\alpha\,\Big\{ S_{b}^{\bar \beta \bar \kappa  }(\bar p+l)\,S^{\mu \kappa, \tilde \alpha \tilde \beta }_{b}(p+l)
+ S_{b}^{ \bar \beta ,\mu \bar \kappa }(\bar p+l)\,S^{ \kappa ,\tilde \alpha \tilde \beta}_{b}(p+l)\Big\} \Bigg\}
\,,
\nonumber\\
&& F^{\bar \alpha \bar \beta, \mu, \alpha \beta}_{+,ab}(\bar p, p) = -\,  i\,\int
\frac{d^4l}{(2\pi)^4}\,\epsilon_{\tilde \alpha \tilde \beta \bar \rho}^{\quad \;\,\beta}\,
g_{\bar \sigma \sigma}\,S^{ \bar \sigma \bar \beta  ,\bar \rho}_a(l)\,\Bigg\{
\bar p^{\bar \alpha }\,g^{\mu \alpha } S_{b}^{\sigma ,\tilde \alpha \tilde \beta  }(\bar p+l)
\nonumber\\
&& \qquad  \quad +\,\bar p^{\bar \alpha }\,p^\alpha\, S_{b}^{\mu \sigma ,\tilde \alpha \tilde \beta }(p+l)
+g^{\bar \alpha \mu}\, p^\alpha\,S_{b}^{\sigma ,\tilde \alpha \tilde \beta  }( p+l)
\nonumber\\
&& \qquad  \quad +\, \bar p^{\bar \alpha }\,p^\alpha\,g_{\bar \kappa \kappa}\,
\Big\{ S_{b}^{\sigma \bar \kappa  }(\bar p+l)\,S^{\mu \kappa, \tilde \alpha \tilde \beta }_{b}(p+l)
+ S_{b}^{ \sigma ,\mu \bar \kappa }(\bar p+l)\,S^{ \kappa,\tilde \alpha \tilde \beta}_{b}(p+l)\Big\} \Bigg\}
\,,
\nonumber\\
&& F^{\bar \alpha \bar \beta, \mu, \alpha \beta}_{-,ab}(\bar p, p) =- i\,\int
\frac{d^4l}{(2\pi)^4}\,\epsilon_{\tilde \alpha \tilde \beta \bar \rho}^{\quad \;\,\beta}\,g_{\bar \sigma \sigma}\,
S_{a}^{\bar \sigma  \bar \rho} (l)\, \Bigg\{
\bar p^{\bar \alpha}\,g^{\mu \alpha}\,S_{b}^{\sigma \bar \beta,\tilde \alpha \tilde \beta  }(\bar p+l)
\nonumber\\
&& \qquad  \quad  +\,g^{\mu \bar \alpha }\,p^\alpha\,S_{b}^{\sigma \bar \beta, \tilde \alpha \tilde \beta  }(p+l)
  + \,\bar p^{\bar \alpha }\,p^\alpha\,g_{\bar \kappa \kappa}\,
\Big\{ S_{b}^{\sigma \bar \beta,\bar \kappa  }(\bar p+l)\,S^{\mu \kappa, \tilde \alpha \tilde \beta }_{b}(p+l)
\nonumber\\
&& \qquad  \qquad \qquad \qquad \quad + S_{b}^{ \sigma \bar \beta ,\mu \bar \kappa }(\bar p+l)\,S^{ \kappa ,\tilde \alpha \tilde \beta}_{b}(p+l)\Big\} \Bigg\}\,,
\nonumber\\
&& \hat F^{\bar \alpha \bar \beta, \mu, \alpha \beta }_{ab}(\bar p, p) =- i\,\int
\frac{d^4l}{(2\pi)^4}\,\epsilon_{\tilde \alpha \tilde \beta \bar \rho}^{\quad \;\, \beta}\,S^{ \bar \sigma \bar \rho}_a(l)\,
\nonumber\\
&& \qquad  \quad \Bigg\{\,
g^{\bar \alpha }_{\;\bar \sigma }\,q_{ \kappa}\,p^\alpha\,\Big\{ S_{b}^{\bar \beta  \kappa  }(\bar p+l)\,S^{\mu , \tilde \alpha \tilde \beta }_{b}(p+l)
- S_{b}^{ \bar \beta ,\mu \bar \kappa }(\bar p+l)\,S^{ \kappa ,\tilde \alpha \tilde \beta}_{b}(p+l)\Big\} \Bigg\}
\,,
\nonumber\\
&& \hat F^{\bar \alpha \bar \beta, \mu, \alpha \beta}_{-,ab}(\bar p, p) =- i\,\int
\frac{d^4l}{(2\pi)^4}\,\epsilon_{\tilde \alpha \tilde \beta \bar \rho}^{\quad \;\,\beta}\,g_{\bar \sigma \sigma}\,
S_{a}^{\bar \sigma  \bar \rho} (l)\,
\nonumber\\
&& \qquad  \quad
  \Bigg\{\,\bar p^{\bar \alpha }\,p^\alpha\,q_{ \kappa}\,
\Big\{ S_{b}^{\sigma \bar \beta, \kappa  }(\bar p+l)\,S^{\mu , \tilde \alpha \tilde \beta }_{b}(p+l)
- S_{b}^{ \sigma \bar \beta ,\mu }(\bar p+l)\,S^{ \kappa ,\tilde \alpha \tilde \beta}_{b}(p+l)\Big\} \Bigg\}\,,
\nonumber\\
&& \hat F^{\bar \alpha \bar \beta, \mu, \alpha \beta}_{+,ab}(\bar p, p) = -\,  i\,\int
\frac{d^4l}{(2\pi)^4}\,\epsilon_{\tilde \alpha \tilde \beta \bar \rho}^{\quad \;\,\beta}\,
g_{\bar \sigma \sigma}\,S^{ \bar \sigma \bar \beta  ,\bar \rho}_a(l)\,
\nonumber\\
&& \qquad  \quad \Bigg\{\, \bar p^{\bar \alpha }\,p^\alpha\,q_{ \kappa}\,
\Big\{ S_{b}^{\sigma \kappa  }(\bar p+l)\,S^{\mu , \tilde \alpha \tilde \beta }_{b}(p+l)
- S_{b}^{ \sigma \mu }(\bar p+l)\,S^{ \kappa,\tilde \alpha \tilde \beta}_{b}(p+l)\Big\} \Bigg\}\,
\nonumber\\
&& \bar G^{\bar \alpha \bar \beta, \mu, \alpha \beta}_{abc}(\bar p, p) = +\,2\, i\,\int
\frac{d^4l}{(2\pi)^4}\,p^\alpha\,\epsilon_{\tilde \alpha \tilde \beta \rho}^{\quad \;\, \beta}\,
S^{ \bar \tau,\tilde \alpha \tilde \beta}_a(l)\,q_\nu\, \epsilon^{\mu \nu }_{\quad \sigma \tau}
\,\epsilon^{\bar \alpha \bar \beta}_{\quad  \bar \sigma \bar \tau}
\nonumber\\
&& \qquad  \quad \times  (\bar p+l)^\tau\,(\bar p+l)^{\bar \sigma}\,
 S_{b}(\bar p+l)\,S^{\sigma \rho}_{c}(p+l)\,,
\nonumber\\
&& G^{\bar \alpha \bar \beta, \mu, \alpha \beta}_{abc}(\bar p, p) =+\,2\, i\,\int
\frac{d^4l}{(2\pi)^4}\,p^\alpha\,\epsilon_{\tilde \alpha \tilde \beta \rho}^{\quad \;\,\beta}\,
S^{\bar \mu \bar \nu,\tilde \alpha \tilde \beta}_a(l)\,q_\nu\, \epsilon^{\mu \nu}_{ \quad \sigma \tau}
\,\epsilon_{\bar \mu \bar \nu}^{\quad \bar \sigma \bar \beta}
\nonumber\\
&& \qquad  \quad  \times  (\bar p+l)^\tau\,(\bar p+l)_{\bar \sigma}\,\bar p^{\bar \alpha}\,
S_{b}(\bar p+l)\,S^{\sigma  \rho}_{c}(p+l)\,,
\nonumber\\
&& \bar H^{\bar \alpha \bar \beta, \mu, \alpha \beta}_{abc}(\bar p, p) = -\,2\, i\,\int
\frac{d^4l}{(2\pi)^4}\,p^\alpha\,\epsilon_{\tilde \alpha \tilde \beta \bar \rho}^{\quad\; \beta}\, S^{ \bar \tau \bar \rho}_a(l)\,
q_\nu\, \epsilon^{\mu \nu }_{\quad \sigma \tau}
\,\epsilon^{\bar \alpha \bar \beta}_{\quad  \bar \sigma \bar \tau}
\nonumber\\
&& \qquad  \quad \times  (\bar p+l)^\tau\,(\bar p+l)^{\bar \sigma}\,
 S_{b}(\bar p+l)\,S^{\sigma ,\tilde \alpha \tilde \beta}_{c}(p+l)\,,
\nonumber\\
&& H^{\bar \alpha \bar \beta, \mu, \alpha \beta}_{abc}(\bar p, p) =-\,2\, i\,\int
\frac{d^4l}{(2\pi)^4}\,p^\alpha\,\epsilon_{\tilde \alpha \tilde \beta \bar \rho }^{\quad \;\beta}\,
S^{\bar \mu \bar \nu,\bar \rho}_a(l)\,q_\nu\,
\epsilon^{\mu \nu}_{ \quad \sigma \tau}
\,\epsilon_{\bar \mu \bar \nu}^{\quad \bar \sigma \bar \beta}
\nonumber\\
&& \qquad  \quad  \times  (\bar p+l)^\tau\,(\bar p+l)_{\bar \sigma}\,\bar p^{\bar \alpha}\,
S_{b}(\bar p+l)\,S^{\sigma ,\tilde \alpha \tilde \beta }_{c}(p+l)\,.
\label{AppendixF5}
\end{eqnarray}

According to (\ref{result-projection-1plus}) it is enough to compute the
contractions of the tensor integrals
(\ref{AppendixF5})
 with two tensors
\begin{eqnarray}
&&P^{(1)}_{\bar \alpha \bar \beta, \mu, \alpha \beta } = -\frac{1}{4}\,p^{\sigma}\,q^{\tau}\,\Big\{
 \epsilon_{\sigma  \tau  \bar \alpha \mu }\,\bar p_{\bar \beta}\,q_{\alpha}\,p_\beta
-\epsilon_{\sigma  \tau  \bar \alpha \mu }\,\bar p_{\bar \beta}\,q_{\beta}\,p_\alpha
\nonumber\\
&& \qquad \qquad \qquad \qquad \quad -\,\epsilon_{\sigma  \tau  \bar \beta \mu }\,\bar p_{\bar \alpha}\,q_{\alpha}\,p_\beta
+\epsilon_{\sigma  \tau  \bar \beta \mu }\,\bar p_{\bar \alpha}\,q_{\beta}\,p_\alpha
\Big\}\,,
\nonumber\\
&&P^{(2)}_{\bar \alpha \bar \beta, \mu, \alpha \beta } = -\frac{1}{4}\,p^{\sigma}\,q^{\tau}\,\Big\{
 \epsilon_{\sigma  \tau\mu \alpha }\,q_{\bar \alpha}\,\bar p_{\bar \beta}\,p_\beta
-\epsilon_{\sigma  \tau\mu \beta }\,q_{\bar \alpha}\,\bar p_{\bar \beta}\,p_\alpha
\nonumber\\
&& \qquad \qquad \qquad \qquad \quad -\,\epsilon_{\sigma  \tau\mu \alpha }\,q_{\bar \beta}\,\bar p_{\bar \alpha}\,p_\beta
+\epsilon_{\sigma  \tau\mu \beta }\,q_{\bar \beta}\,\bar p_{\bar \alpha}\,p_\alpha
\Big\} \,.
\end{eqnarray}
We establish  the results


We turn to the contractions with the tensor $P^{(2)}_{\bar \alpha \bar \beta, \mu, \alpha \beta }$ of
(\ref{def-projections12}). We obtain
%



\begin{thebibliography}{99}
\bibitem{private-communication-Hofmann}
In Eq. (8) of \cite{Hofmann-Lutz-2004}
there are two misprints. In the expression for $V_{0+}$ ,the sign of the s-channel contribution has to be reversed.
In the expression for $V_{1+}$, the r.h.s. has to be multiplied by a factor $3/2$. As a consequence the numerical
values given in Table 3 change slightly. The conclusions of the work are not affected.
\end{thebibliography}
\end{document}